\documentclass[useAMS,usenatbib]{mn2e}
\usepackage{graphicx}
\usepackage{color}

\usepackage{longtable}
\usepackage{lscape}
\usepackage{subfigure}

\title[A far-infrared molecular and atomic line survey of the Orion KL region]
{A far-infrared molecular and atomic line survey of the Orion KL
region.}
\author[M.R. Lerate et al.]
{M.R. Lerate,$^{1,2}$
M.J. Barlow,$^1$
  B.M. Swinyard,$^2$
  J.R. Goicoechea,$^3$
  \newauthor
  J. Cernicharo,$^4$
  T.W. Grundy,$^2$
  T.L. Lim,$^2$
  E.T. Polehampton,$^2$
  J.P. Baluteau, $^5$
  \newauthor
  S. Viti,$^1$ and J. Yates$^1$ \\
  $^1$University College London, Gower Street, London WC1E 6BT,
  U.K\\
  $^2$Rutherford Appleton Laboratory, Chilton, Didcot OX11 0QX,
  U.K\\
  $^3$LERMA, UMR 8112, CNRS, Observatoire de Paris and Ecole Normale Superieure, 24 Rue Lhomond 75231 Paris Cedex 05, France\\
  $^4$CSIC, Instituto de Estructura de la Materia, Serrano 121, 28006 Madrid, Spain.\\
    $^5$ Laboratoire d'Astrophysique de Marseille, CNRS \& Universite
de Provence, BP-8, F-13376 Marseille Cedex, France }

\date{Released 2005 Xxxxx XX}

\pagerange{\pageref{firstpage}--\pageref{lastpage}} \pubyear{2005}

\def\LaTeX{L\kern-.36em\raise.3ex\hbox{a}\kern-.15em
    T\kern-.1667em\lower.7ex\hbox{E}\kern-.125emX}

\begin{document}

\label{firstpage}

\maketitle

   \begin{abstract}
  {We have carried out a high spectral resolution ($\lambda/\Delta\lambda \sim 6800-9700$) line survey towards the Orion Kleinmann-Low (KL)
cluster from 44--188 $\mu$m. The observations were taken with the
Long Wavelength Spectrometer (LWS) in Fabry-P\'erot mode, on board
the {\em Infrared Space Observatory (ISO)}. A total of 152 lines
are clearly detected and a further 34 features are present as
possible detections. The spectrum is dominated by the molecular
species H$_{2}$O, OH and CO, along with [O~{\sc i}] and [C~{\sc
ii}] lines from PDR or shocked gas and [O~{\sc iii}], [N~{\sc
iii}] lines from the foreground M42 H~{\sc ii} region. Several
isotopic species, as well as NH$_{3}$, are also detected. HDO and
H$_{3}$O$^{+}$ are tentatively detected for the first time in the
far-infrared range towards Orion-KL. A basic analysis of the line
observations is carried out, by comparing with previous
measurements and published models and deriving rotational
temperatures and column densities in the case of the molecular
species.

Analysis of the [O~{\sc i}] and [C~{\sc ii}] fine structure lines
indicates that although a shock model can reproduce the observed [O~{\sc i}]
 surface brightness levels, it falls short of the observed [C~{\sc ii}] level by
 more than a factor of 30. A PDR model can reproduce the [O~{\sc i}] 63.2 $\mu$m and
  [C~{\sc ii}] surface brightness levels within 35\%, although over-predicting the LWS [O~{\sc i}] 145.5 $\mu$m
  emission by a factor of 2.7.  The 70 water lines and 22 OH lines detected by the
  survey appear with mainly P Cygni profiles at the shortest survey wavelengths
  and with mainly pure emission profiles at the longest survey wavelengths.
  The emission and absorption velocity peaks of the water and OH lines indicate
  that they are associated with gas expanding in the outflow from the KL cluster.
  The estimated column densities are (2-5)$\times$10$^{14}$ cm$^{-2}$ for H$_{2}$O
   and (2.5-5.1)$\times$10$^{16}$ cm$^{-2}$ for OH. The 26 detected
   CO lines confirm the presence of three distinct components, with
   temperature and column density combinations ranging from 660~K,
   $6\times10^{17}$~cm$^{-2}$ to 360~K, $2\times10^{19}$~cm$^{-2}$.
Comparison of the surface brightnesses and integrated fluxes of
the CO lines measured in the 80-arcsec LWS beam with those
measured previously by the {\em Kuiper Airborne Observatory (KAO)}
in a 44 arcsec beam shows similar surface brightnesses in the
different beams for the lowest-J CO lines and similar integrated
fluxes in the different beams for the highest-J CO lines,
indicating that emission from the former lines has a uniform
surface brightness within the LWS beam, while the latter lines
originate from a region less than 44 arcsec in diameter. The
complexity of the region requires more sophisticated models for
the interpretation of all the line observations.}
 \end{abstract}


\section{Introduction}

The formation of a star is often associated with outflows which
can compress and heat the surrounding quiescent gas,
 altering the chemistry of the parent molecular cloud in which the star is forming.  Many of the  molecules responsible for cooling the gas in these
molecular clouds, such as CO (J $\geq$ 14), OH and H$_{2}$O, emit
at far-infrared wavelengths ($>$ 40$\mu$m) which are difficult to
detect from the ground due to the constraints of the Earth's
atmosphere.  The launch of ESA's {\em Infrared Space Observatory}
({\em ISO}) in 1995 marked a revolution in our ability to observe
these transitions, and in particular, opened a window for
large-scale high resolution far-infrared spectroscopy of star
forming regions such as Orion. In this paper we report
 on the first, and to date only, unbiased high spectral resolution spectral survey in the far-infrared waveband of
 the Kleinmann-Low (KL) region of Orion and discuss the
 implications of the results for the interpretation of the chemical evolution of the region.\\

\begin{figure}
    \centering
   \includegraphics[width=8cm,height=8cm]{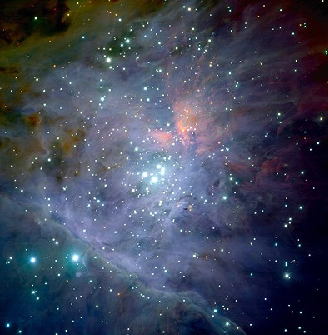}
    \caption{Colour composite
    mosaic image of the central part of the Orion Nebula, M42 , based on 81 images
    obtained with the infrared multi-mode ISAAC instrument on the ESO Very Large
    Telescope (VLT) at the Paranal Observatory. The Trapezium stars are seen near the centre (ESO PR Photo 03a/01 2001).}
        \label{M42}
   \end{figure}

The Orion star-forming complex, at a distance of 450 pc, is our
nearest
 region of high mass star formation. It mainly consists
of two giant molecular clouds: Orion molecular clouds 1 and 2
(OMC1 and OMC2), which were revealed via large scale mapping in
the J=1$\rightarrow$0 $^{12}$CO rotational transition  (Tucker,
Kutner \& Thaddeus, 1973).  OMC1 is associated with the dark
clouds L1640, L1641, L1647 (Lynds 1962), extending over
6$^{\circ}$ southwards from the Orion nebula. The OMC1 ridge is
associated with regions of prominent molecular emission, with the
region near the Orion Nebula being the best studied example. OMC1
contains the visible nebula M42, a blister of hot, photo-ionised
luminous gas around the hot Trapezium stars (see Figure~\ref{M42})
and also contains a number of IR-emitting regions such as the
Kleinmann-Low nebula, which is the brightest far-infrared region
in the complex (Kleinmann \& Low 1967). The KL nebula includes an
infrared cluster of massive stars at an early evolutionary stage,
e.g. the Becklin-Neugebauer object (BN, Becklin \& Neugebauer
1967) and IRc2, which dominate the mid-infrared radiation (Downes
et al. 1981). There are at least two molecular outflows, one
associated with material ejected at hundreds of km s$^{-1}$
forming a bipolar cone of molecular fingers, whose axis is
perpendicular to the NH$_{3}$ emission associated with the hot
core (Wilson et al. 2000), and a second low-velocity outflow first
detected by H$_{2}$O maser observations (Genzel et al. 1981). IRc2
was originally believed to be the major source of the KL
luminosity (L$\sim$ 10$^{5}$L$_{\odot}$, Genzel \& Stutzki, 1989).
However, the infrared observations of Dougados et al.(1993)
resolved IRc2 into four components which may not even be
self-luminous, and radio studies located the origin of the outflow
as being offset by 0.5 arcsec from IRc2-A; at the position of
radio source I (Menten
\& Reid 1995, Gezari et al. 1998, Chandler \& Greenhill 2002).\\
Within the KL region, molecular emission arises from several
physically distinct regions:
\begin{itemize}
\item The Hot Core is composed of very dense material with $n$
$\sim$ 10$^{7}$ cm$^{-3}$ and $T$ $\sim$
 200K in clumps located $\sim$ 2$^{\prime\prime}$ south of IRc2, 1$^{\prime\prime}$ offset from radio source I
 (Wright et al. 1996, Schilke et al. 2001).
\item The Compact Ridge is a compact region (14$^{\prime\prime}$)
of dense gas ($n$ $\sim$ 10$^{6}$ cm$^{-3}$) surrounded by
outflows.
\item The Plateau; a region of inhomogeneous density
containing both low--velocity
  ($\sim$ 18 km s$^{-1}$) and high--velocity ($\sim$ 100 km s$^{-1}$) flows with an approximate diameter of
 50$^{\prime\prime}$ (Blake et al. 1987).
\item Finally the whole region is embedded in the Extended Ridge,
a region of quiescent and cooler gas from which many narrow lines
($\Delta v$ $\sim$ 3--5 km s$^{-1}$ FWHM) are detected (Greaves \&
White 1991).
\end{itemize}
The proximity and complexity of the KL region have made it the
target of many astronomical studies, and it is one of the best
studied examples of the interaction between massive stars in their
earliest stages and their parental molecular cloud.  In this paper
we report the results of an unbiased spectral survey between
44--188$\mu$m (1612 to 6813 GHz) at a resolving power of $\approx$
8000, which represents the first systematic study of the molecular
spectrum in this wavelength range. A number of the detected lines
from this spectral survey and from a series of targeted
observations have already been reported (e.g. the OH lines;
Goicoechea et al. 2006, some H$_{2}$O lines; Harwit et al. 1998,
Cernicharo et al. 1999, and some CO lines Sempere et al. 2000),
however, this is the first analysis of the entire spectrum.\\
Our spectral survey provides an insight into the physical
conditions and dynamical processes in the Orion KL region and
complements studies performed at lower frequencies, e.g. from 72
to 91 GHz (Johansson et al. 1984), 70 to 115 GHz (Turner 1989),
150 to 160 GHz (Ziurys \& McGonagle 1993), 215 to 247 GHz (Sutton
et al. 1985), 247 to 263 GHz (Blake et al. 1986) 325 to 360 GHz
(Schilke et al. 1997), 455 to 507 GHz (White et al. 2003), 607 to
725 GHz (Schilke et al. 2001)
and 795 to 903 GHz (Comito et al. 2005).  \\
The aim of this work is to present the data, describe its
reduction and calibration and discuss some basic results. Section
2 describes
 the observations and data
reduction, including a brief summary of the instrumental
operation. Section 3 presents the results, while the individual
species are discussed in Section 4. Our results are summarised in
Section 5.

\section{Observations and Data reduction}
\subsection{Instrumental
operation} The Long Wavelength Spectrometer (LWS; Clegg et al.
1996) was one of the four instruments on board the {\em Infrared
Space Observatory} ({\em ISO}), which operated between November
1995 and April 1998, when the superfluid helium used to cool the
instruments ran out. The LWS covered the spectral range between 43
and 197$\mu$m at medium resolution ($\lambda$/$\Delta\lambda$
$\sim$ 150--200) using a diffraction grating and at high
resolution ($\lambda$/$\Delta$$\lambda$ $\sim$ 6800--9700) with
one of
two Fabry-P\'erot (FP) etalons, order sorted by the grating.\\
Four different observing modes were available to users, via the
so-called `Astronomical Observation Templates' (AOTs), which
allowed the observer a choice of wavelength ranges, sampling
intervals and exposure times. These observing modes were: medium
resolution, full wavelength range spectrum (AOT L01), medium
resolution and narrow band photometry, selected wavelength range
spectrum (AOT L02), high-resolution, full wavelength range (AOT
L03) and high-resolution, narrow wavelength range (AOT L04). Also,
a parallel mode was carried out while an AOT was active with
another instrument and a serendipity mode was carried out during
the slews with no prime instrument. Since a full spectral scan
with the LWS FP (AOT L03) took a long time, a significant part of
the spectrum was recorded at high spectral resolution for only
four
objects: Orion BN/KL, Sgr B2 (Polehampton  2002, Polehampton et al. 2006 in prep.), Sgr A and Jupiter.\\

In L03 mode, data for a given wavelength range were recorded by
scanning the FP gap over a small range with the grating held
stationary, thus shifting the wavelength of the selected FP order
across the grating response function.  To record data for another
wavelength range the grating was moved by a portion of the grating
resolution element
and the FP scanned again: the data set recorded at each grating setting is termed a $\it{mini}$--$\it{scan}$.\\
During each observation the LWS FP and grating settings were
optimised for the detector whose band pass filter included the
wavelength range of interest. This was denoted as the `prime'
detector.  However, all ten LWS detectors recorded data
simultaneously in their own spectral ranges; the other detectors
are termed `non-prime'.

\subsection{Observations}
\begin{figure}
    \centering
    \includegraphics[width=8cm,height=18cm]{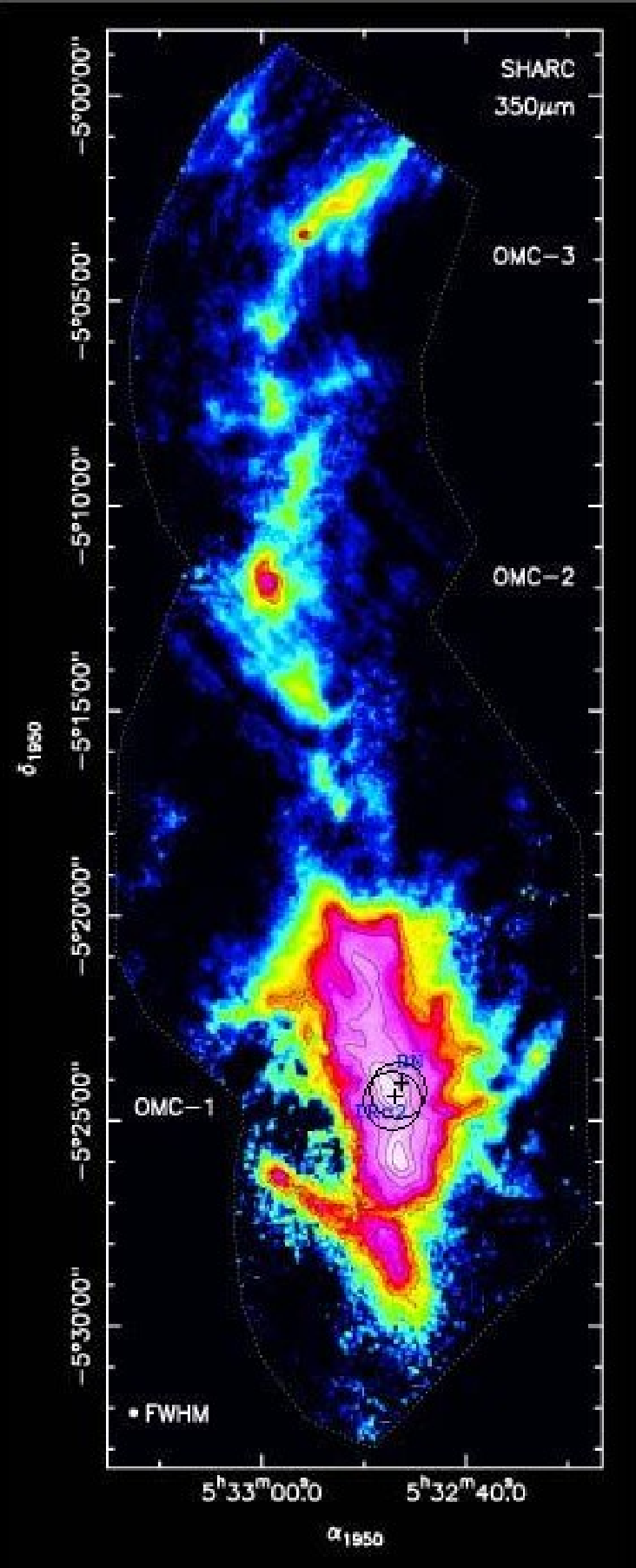}
    \caption{Approximate LWS observation center (cross points) at the positions of IRc2 and 10.5$^{\prime\prime}$ offset from the BN object on a false colour-image from Lis et al. (1998)
    showing the 350$\mu$m continuum
    emission observed with SHARC. (Courtesy of D. Lis). The circles
    show the LWS beam positions.}
    \label{milimiter}
   \end{figure}

The observations were carried out between September 1997 and April
1998. The dataset consists of 26 individual observations making up
a total of 27.9 hours of {\em ISO} LWS observing time in L03 mode,
16 observations making up 13.1 hours in L04 mode and 1 observation
in the lower resolution L01 grating mode. The instrumental field
of view for all L03 observations was either centred on a position
offset by 10.5$^{\prime\prime}$ from the BN object (which is at
05$^{\rm h}$ 35$^{\rm m}$ 14.12$^{\rm s}$ -- 05$^{\circ}$
22$^{\prime}$ 22.9$^{\prime\prime}$ J2000), or a position offset
by 5.4$^{\prime\prime}$ from IRc2 (which is at 05$^{\rm h}$
35$^{\rm m}$ 14.45$^{\rm s}$ -- 05$^{\circ}$ 22$^{\prime}$
30.0$^{\prime\prime}$ J2000) while most of the L04 observations
were centred on IRc2 (see Table~\ref{tdts} and
Figure~\ref{milimiter}). The LWS beam had a diameter $\approx$
80$^{\prime\prime}$ (Gry et al. 2003). Table~\ref{tdts} lists the
observations: column one is the Target Dedicated Time
identification number (TDT). The TDT is an eight digit number
unique to each observation; the first three digits are the {\em
ISO} revolution number, the next three digits are the sequence
number of the observation in the revolution and the last two
numbers were assigned by the observer. Column two lists the
wavelength range covered by the prime detector. The length of the
observation in seconds can be found in column three while the
observation date (d.m.y) is in column four. The J2000 coordinates
of the pointing position for each observation can be found in
columns five
and six respectively. \\

\begin{table*}
\centering                          
\begin{tabular}{c c c c c c}        
\hline\hline
TDT  &  Wavelength($\mu$m)  &  length(s) &  Date & J2000 RA & J2000 Dec \\    
\hline
 &  & AOT = L03 \\
 \hline
66302402& 47--52 &   5586  & 9.9.97 & 5$^{\rm h}$ 35$^{\rm m}$
14$^{\rm s}$.17 & -5$^{\circ}$ 22$^{\prime}$
33$^{\prime\prime}$.5 \\
66302406&  47--52 &   5586  & 9.9.97 & 5$^{\rm h}$ 35$^{\rm m}$
14$^{\rm s}$.17 & -5$^{\circ}$ 22$^{\prime}$
33$^{\prime\prime}$.5 \\
66002003&  52--57 &   4434  & 6.9.97 & 5$^{\rm h}$ 35$^{\rm m}$
14$^{\rm s}$.17 & -5$^{\circ}$ 22$^{\prime}$
33$^{\prime\prime}$.4 \\
66002007& 52--57 &   4434  & 6.9.97 & 5$^{\rm h}$ 35$^{\rm m}$
14$^{\rm s}$.17 & -5$^{\circ}$ 22$^{\prime}$
33$^{\prime\prime}$.4 \\
70101704&  57--63 &   5726 &17.10.97 & 5$^{\rm h}$ 35$^{\rm m}$
14$^{\rm s}$.17 & -5$^{\circ}$ 22$^{\prime}$
33$^{\prime\prime}$.8 \\
 70101708&  57--63 & 5726  & 17.10.97 & 5$^{\rm h}$ 35$^{\rm m}$ 14$^{\rm s}$.17 & -5$^{\circ}$ 22$^{\prime}$
33$^{\prime\prime}$.8 \\
 70001205&  63--70 &   5641 & 16.10.97 & 5$^{\rm h}$ 35$^{\rm m}$ 14$^{\rm s}$.17 & -5$^{\circ}$ 22$^{\prime}$
33$^{\prime\prime}$.8 \\
 70001209& 63--70 &  5642 & 16.10.97 & 5$^{\rm h}$ 35$^{\rm m}$ 14$^{\rm s}$.17 & -5$^{\circ}$ 22$^{\prime}$
33$^{\prime\prime}$.8 \\
 69901510&  70--73 & 2592 &  15.10.97 & 5$^{\rm h}$ 35$^{\rm m}$ 14$^{\rm s}$.17 & -5$^{\circ}$ 22$^{\prime}$
33$^{\prime\prime}$.8 \\
 69901514 & 70--73 &   2591& 15.10.97 & 5$^{\rm h}$ 35$^{\rm m}$ 14$^{\rm s}$.17 & -5$^{\circ}$ 22$^{\prime}$
33$^{\prime\prime}$.8 \\
69901312& 77--81  &  4106 &  15.10.97 & 5$^{\rm h}$ 35$^{\rm m}$
14$^{\rm s}$.17 & -5$^{\circ}$ 22$^{\prime}$
33$^{\prime\prime}$.8  \\
69901316&  77--81  &  4106  & 15.10.97 & 5$^{\rm h}$ 35$^{\rm m}$
14$^{\rm s}$.17 & -5$^{\circ}$ 22$^{\prime}$
33$^{\prime\prime}$.8  \\
 69901413&  81--85 &   3842  & 15.10.97 & 5$^{\rm h}$ 35$^{\rm m}$ 14$^{\rm s}$.17 & -5$^{\circ}$ 22$^{\prime}$
33$^{\prime\prime}$.8 \\
 69901417&  81--85 &   3842 & 15.10.97 & 5$^{\rm h}$ 35$^{\rm m}$ 14$^{\rm s}$.17 & -5$^{\circ}$ 22$^{\prime}$
33$^{\prime\prime}$.8 \\
 70001105&  85--89 &   2530 & 15.10.97 & 5$^{\rm h}$ 35$^{\rm m}$ 14$^{\rm s}$.16 & -5$^{\circ}$ 22$^{\prime}$
33$^{\prime\prime}$.7 \\
 70001127&  85--89 & 2530 & 16.10.97 & 5$^{\rm h}$ 35$^{\rm m}$ 14$^{\rm s}$.17 & -5$^{\circ}$ 22$^{\prime}$
33$^{\prime\prime}$.8 \\
87301008 &  99--104 &  2924 &  6.4.98 & 5$^{\rm h}$ 35$^{\rm m}$
14$^{\rm s}$.15 & -5$^{\circ}$ 22$^{\prime}$
33$^{\prime\prime}$.4  \\
87301030& 99--104 &  2924 &  6.4.98  & 5$^{\rm h}$ 35$^{\rm m}$
14$^{\rm s}$.15 & -5$^{\circ}$ 22$^{\prime}$
33$^{\prime\prime}$.4 \\
87300909&   104--109 & 2748 &  6.4.98 & 5$^{\rm h}$ 35$^{\rm m}$
14$^{\rm s}$.15 & -5$^{\circ}$ 22$^{\prime}$
33$^{\prime\prime}$.4 \\
 87300931 & 104--109 &  2748 &  6.4.98 & 5$^{\rm h}$ 35$^{\rm m}$ 14$^{\rm s}$.15 & -5$^{\circ}$ 22$^{\prime}$
33$^{\prime\prime}$.4 \\
 82702210 & 109--115 & 3226 & 19.2.98 & 5$^{\rm h}$ 35$^{\rm m}$ 14$^{\rm s}$.15 & -5$^{\circ}$ 22$^{\prime}$
33$^{\prime\prime}$.5 \\
 82702232  &  109--115 & 3226 &  20.2.98 & 5$^{\rm h}$ 35$^{\rm m}$ 14$^{\rm s}$.15 & -5$^{\circ}$ 22$^{\prime}$
33$^{\prime\prime}$.5 \\
 84101911 &   115--121 & 3140  & 5.3.98 & 5$^{\rm h}$ 35$^{\rm m}$ 14$^{\rm s}$.15 & -5$^{\circ}$ 22$^{\prime}$
33$^{\prime\prime}$.4 \\
 84101933 & 115--121 &  3140 &  5.3.98 & 5$^{\rm h}$ 35$^{\rm m}$ 14$^{\rm s}$.15 & -5$^{\circ}$ 22$^{\prime}$
33$^{\prime\prime}$.5 \\
82602316  &  147--154 & 3830 &  18.2.98 & 5$^{\rm h}$ 35$^{\rm m}$
14$^{\rm s}$.15 & -5$^{\circ}$ 22$^{\prime}$
33$^{\prime\prime}$.5 \\
 82602338 &  147--154 & 3830  & 18.2.98  & 5$^{\rm h}$ 35$^{\rm m}$ 14$^{\rm s}$.15 & -5$^{\circ}$ 22$^{\prime}$
33$^{\prime\prime}$.5 \\
 \hline
  &  & AOT = L04 \\
 \hline
83101203 & 51--101 & 2166 & 23.02.98 & 5$^{\rm h}$ 35$^{\rm m}$ 14$^{\rm s}$.45 & -5$^{\circ}$ 22$^{\prime}$ 30$^{\prime\prime}$.0\\
69000714 & 51--125 & 6918 & 06.10.95 & 5$^{\rm h}$ 35$^{\rm m}$ 14$^{\rm s}$.21 & -5$^{\circ}$ 22$^{\prime}$ 23$^{\prime\prime}$.5\\
82901207 & 54--75 & 1818 & 22.02.98 & 5$^{\rm h}$ 35$^{\rm m}$ 14$^{\rm s}$.45 & -5$^{\circ}$ 22$^{\prime}$ 30$^{\prime\prime}$.0\\
82901101 & 68--129 & 3858 & 21.02.98 & 5$^{\rm h}$ 35$^{\rm m}$ 14$^{\rm s}$.45 & -5$^{\circ}$ 22$^{\prime}$ 29$^{\prime\prime}$.5 \\
69602606 & 70--124 & 2334 & 12.10.97 & 5$^{\rm h}$ 35$^{\rm m}$ 14$^{\rm s}$.47 & -5$^{\circ}$ 22$^{\prime}$ 29$^{\prime\prime}$.8\\
69602605 & 73--180 & 2028 & 12.10.9 & 5$^{\rm h}$ 35$^{\rm m}$ 14$^{\rm s}$.47 & -5$^{\circ}$ 22$^{\prime}$ 29$^{\prime\prime}$.8\\
82901206 & 75--94 & 1772 & 21.02.98 & 5$^{\rm h}$ 35$^{\rm m}$ 14$^{\rm s}$.45 & -5$^{\circ}$ 22$^{\prime}$ 30$^{\prime\prime}$.1\\
83101202 & 83--180 & 2282 & 23.02.98 & 5$^{\rm h}$ 35$^{\rm m}$ 14$^{\rm s}$.45 & -5$^{\circ}$ 22$^{\prime}$ 30$^{\prime\prime}$.1\\
70101216 & 84--179 & 3028 & 17.10.97 & 5$^{\rm h}$ 35$^{\rm m}$14$^{\rm s}$.47 & -5$^{\circ}$ 22$^{\prime}$ 30$^{\prime\prime}$.4\\
83101201 & 96--180 & 2416 & 23.02.98 & 5$^{\rm h}$ 35$^{\rm m}$ 14$^{\rm s}$.45 & -5$^{\circ}$ 22$^{\prime}$ 30$^{\prime\prime}$.1\\
69602317 & 100--180 & 1980 & 12.10.97 & 5$^{\rm h}$ 35$^{\rm m}$14$^{\rm s}$.47 & -5$^{\circ}$ 22$^{\prime}$ 30$^{\prime\prime}$.4\\
82901204 & 105--146 & 1630 & 21.02.98 & 5$^{\rm h}$ 35$^{\rm m}$ 14$^{\rm s}$.45 & -5$^{\circ}$ 22$^{\prime}$ 30$^{\prime\prime}$.1\\
83101301 & 119--120 & 3380 & 23.02.98 & 5$^{\rm h}$ 35$^{\rm m}$ 14$^{\rm s}$.49 & -5$^{\circ}$ 22$^{\prime}$ 30$^{\prime\prime}$.9\\
68701716 & 119--165 & 7661 & 03.10.97 & 5$^{\rm h}$ 35$^{\rm m}$ 14$^{\rm s}$.21 & -5$^{\circ}$ 22$^{\prime}$ 23$^{\prime\prime}$.5\\
69602318 & 119--179 & 2492 & 12.10.97 & 5$^{\rm h}$ 35$^{\rm m}$ 14$^{\rm s}$.48 & -5$^{\circ}$ 22$^{\prime}$ 30$^{\prime\prime}$.4\\
68701301 & 148--152 & 1294 & 03.10.97 & 5$^{\rm h}$ 35$^{\rm m}$ 14$^{\rm s}$.47 & -5$^{\circ}$ 22$^{\prime}$ 29$^{\prime\prime}$.8\\

\hline
\end{tabular}
\caption{List of all L03 and L04 observations centred on Orion
IRc2-BN/KL analysed in this work. Observations are sorted by
wavelength range covered by the prime detector. Note that for L04
spectra, only certain targeted observations were covered in the
quoted wavelength range.} \label{tdts}
\end{table*}

\begin{figure}
    \centering
    \includegraphics[width=6cm,height=5cm]{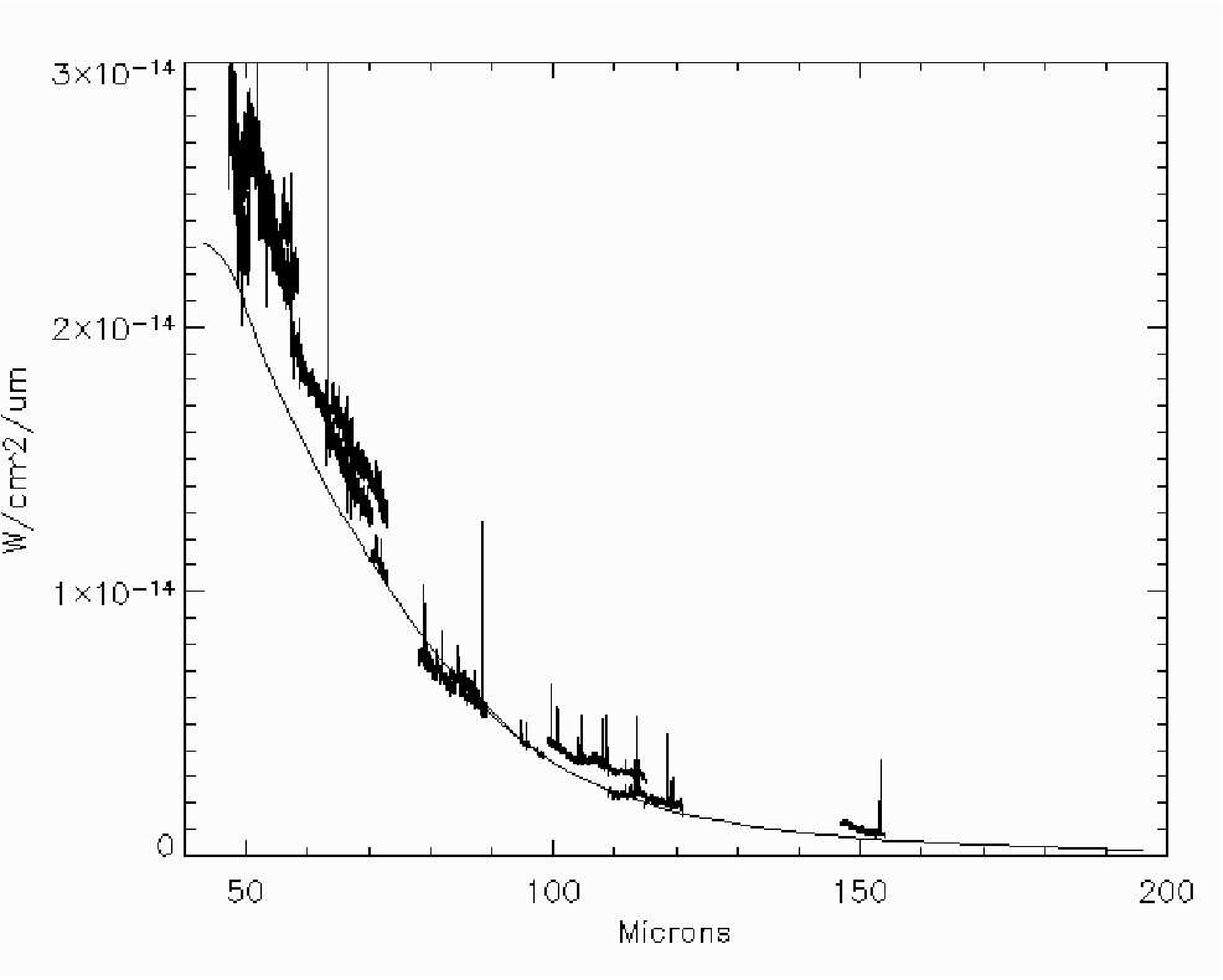}
    \includegraphics[width=6cm,height=5cm]{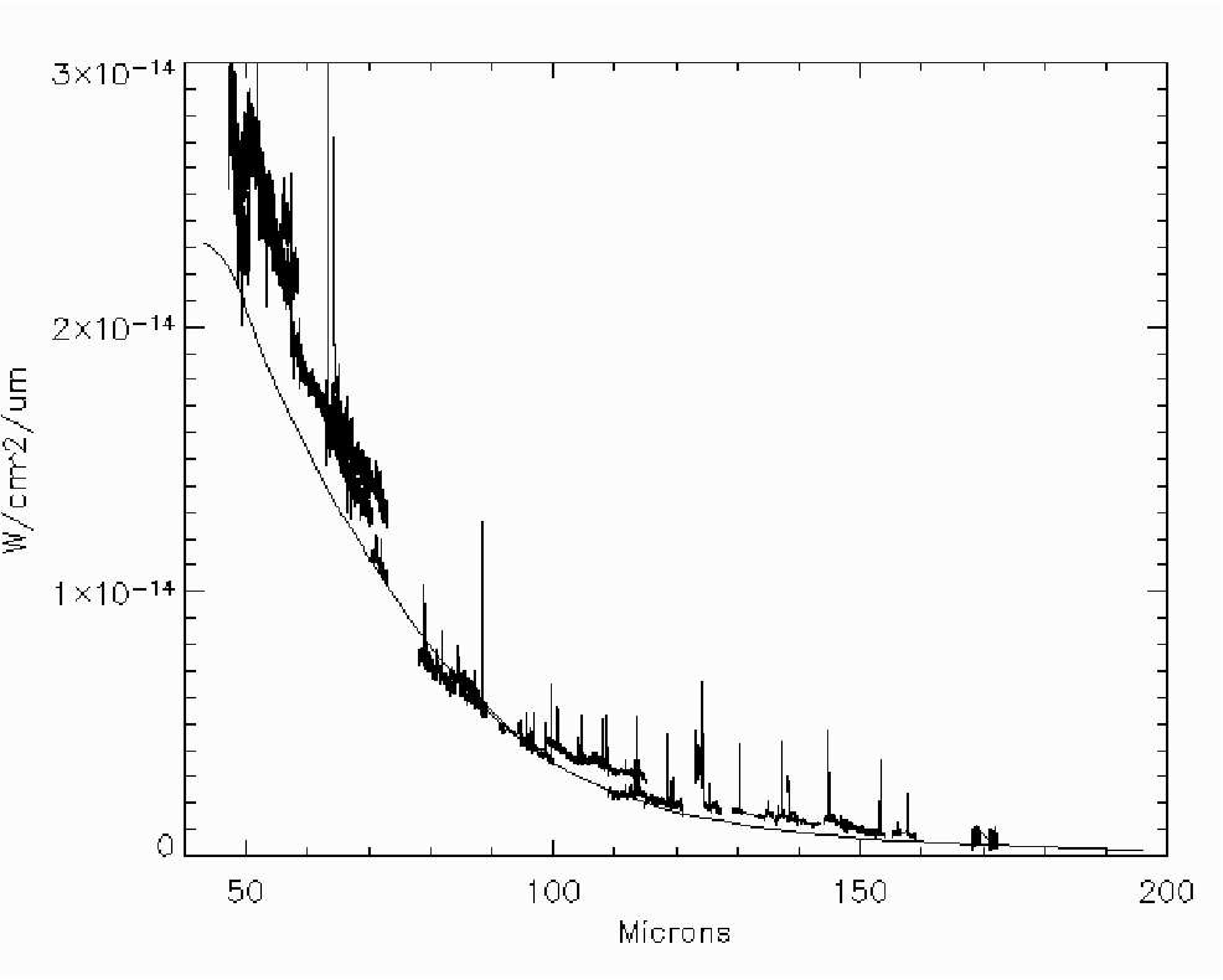}
    \caption{All L03 prime observations (upper plot) and non-prime observations (down plot) together with
    a continuum fit to the grating observations. Note that for wavelengths $\geq$ 120 $\mu$m the spectrum is mainly
    covered by non-prime data.}
    \label{fit_continuo}
\end{figure}

\subsection{Data reduction}

Processing of the LWS FP data was carried out using the Offline
Processing (OLP) pipeline and the LWS Interactive Analysis (LIA)
package version 10. The basic calibration is fully described in
the LWS handbook (Gry et al. 2003). Further processing, including
dark current optimisation, de-glitching and removal of the LWS
grating profile
 was then carried out interactively using the LWS Interactive Analysis package version 10
(LIA10; Lim et al. 2002) and the {\em ISO} Spectral Analysis
Package (ISAP; Sturm 1998). The reduction process was performed in
the same way as that for the {\em ISO} L03 Sagittarius B2 data
(Polehampton et al. 2006 in prep) with the only exception being
the $\it{mini}$--$\it{scan}$ shape removal (see below). The
dataset contains data from
both prime and non-prime detectors.\\
Detector dark currents were determined during the mission by three
different methods (Swinyard et al. 1998). Table~\ref{darkcurrent}
lists the nominal dark current, which is the average of these
measurements. However, scattering of light into the detectors may
have occurred during observations of extended or off-axis sources.
This could contribute an additional component to the dark signal
which varied from source to source and is not easy to distinguish
from the detector dark current or the source signal.  A method to
determine both the dark current and stray light corrections for a
particular observation was developed by Polehampton (2002) using
non-prime data from the observation of Sagittarius B2 (see also
Polehampton et al. 2006). The method is based on the fact that for
some $\it{mini}$--$\it{scans}$ the transmitted orders from the FP
do not correspond to the wavelength range transmitted by the
grating. In these circumstances no light was transmitted and the
detector will have recorded only the dark current and the stray
light that finds its way to the detectors
from outside of the main beam - the dark signal. \\
This method was used on the Orion data and dark signal corrections
were applied via the graphical interface available in the LIA
interactive reduction package. Table~\ref{darkcurrent} lists the
 values of the dark signal that were subtracted from the observations for each detector
 compared to the `nominal' dark current.\\
Basic removal of the $\it{mini}$--$\it{scan}$ shape was also
performed interactively in the LIA. Any remaining
$\it{mini}$--$\it{scan}$ residual shape is due to incorrect
substraction of the grating response profile as a consequence of a
positional uncertainty in the absolute grating position. The
effect can be seen as periodic structures in regions where
$\it{mini}$--$\it{scans}$ overlap. These features are very
difficult to remove being still present in the final LIA product.
Therefore, a further step was
performed for the Orion data by using smooth functions and removing the residuals individually for every $\it{mini}$--$\it{scan}$ (see also following section).\\
 The Orion KL region was also observed using the grating
in L01 mode in revolution 695. Due to the strength of the source,
the grating data had to be corrected for the effects of detector
saturation (Leeks et al. 1999). Channel fringing and spurious
features introduced by the Relative Spectral Response Function
(RSRF) were also corrected by using the Highly Processed Data
Products (HPDP) available in the {\em ISO} data archive (Lloyd et
al. 2003). The final L01 product has a superior flux calibration
accuracy compared to the FP data and was therefore used to derive
the continuum level. As part of the normalisation process, FP data
were first smoothed to the same resolution as the grating, using a
bin size of 0.07 $\mu$m for the SW detectors and 0.15 $\mu$m for
the LW detectors. Finally, each FP spectrum was corrected to the
Local Standard of Rest velocity (LSR) using the Starlink routine
RV (Wallace \& Clayton 1996).

\begin{table*}
\begin{center}
\begin{tabular}[!h]{lcccc}
\hline \\[-5pt]
Detector &   Nominal  & FPS & FPL \\[+5pt]
         &   dark current ($\times$ 10$^{-16}$ A) & ($\times$ 10$^{-16}$ A) & ($\times$ 10$^{-16}$ A) \\
\hline \\[-5pt]

SW1 & 4.96  $\pm$ 0.55  \\
SW2 &  2.08  $\pm$ 0.43  & 6.10  $\pm$ 0.20 \\
SW3 & 2.20   $\pm$ 0.21  & 4.97  $\pm$ 0.19 \\
SW4 & 1.18   $\pm$ 0.34 & 2.41  $\pm$ 0.11  & 5.48  $\pm$ 0.49 \\
SW5 &  1.56  $\pm$ 0.24 & 2.11  $\pm$ 0.06  & 5.75  $\pm$ 0.56 \\
LW1 &  2.50  $\pm$ 0.29  & 3.80  $\pm$ 0.15  & 4.20  $\pm$  0.07 \\
LW2 & 0.07  $\pm$ 0.27  & 1.03  $\pm$ 0.05  & 4.64 $\pm$ 0.06 \\
LW3 & 0.53  $\pm$ 0.39  & 2.01  $\pm$ 0.42  & 4.83  $\pm$ 0.35  \\
LW4 & 1.76  $\pm$ 0.42  & 4.41  $\pm$ 0.65  & 5.28  $\pm$ 0.28 \\
LW5 & 1.21  $\pm$ 0.25  & 1.49  $\pm$ 0.14  & 1.47  $\pm$  0.06 \\
\hline \\
\end{tabular}
\end{center}
\caption{Nominal dark currents adopted in the standard pipeline
processing for OLP version 10 (column one) and averaged values of
the dark current and stray light correction determined for the
Orion dataset for the two FPs.} \label{darkcurrent}
\end{table*}

\subsection{Spectrum reliability, signal to noise achieved }
After the dark current optimization and $\it{mini}$--$\it{scan}$
shifting many spurious features were observed in the data. These
structures can be easily recognized and divided into two different
types: features due to $\it{mini}$--$\it{scans}$ shifting, and
features transferred via the Relative Spectral Response Function
(RSRF) in the calibration stage. The first are introduced by the
fact that each observation is built up of
$\it{mini}$--$\it{scans}$ and interactively adjusted as part of
the reduction process. \\
This residual shape is difficult to remove, especially for short
wavelength detectors and an alternative method was used to
minimize transferred features. For each $\it{mini}$--$\it{scan}$ a
smooth function was fitted and used to remove the
$\it{mini}$--$\it{scan}$ shape. This method significantly improved
the signal to noise achieved but introduced an error due to the
fact that each observation is divided by the flux level achieved
in each particular $\it{mini}$--$\it{scan}$. This means that the
same lines recorded in different observations could present a
slight difference in flux level in the final reduced file.
However, the difference in flux when the same spectral range was
observed by two adjacent detectors was found to be relatively
small (see Figure~\ref{fit_continuo}), introducing errors of less
than
20$\%$.\\
The second group of features are transferred from the calibration
via the RSRF and are probably caused by transient effects in the
detectors. They can be
seen as large scale features.\\

\begin{figure}
   \centering

    \includegraphics[width=6cm,height=5cm]{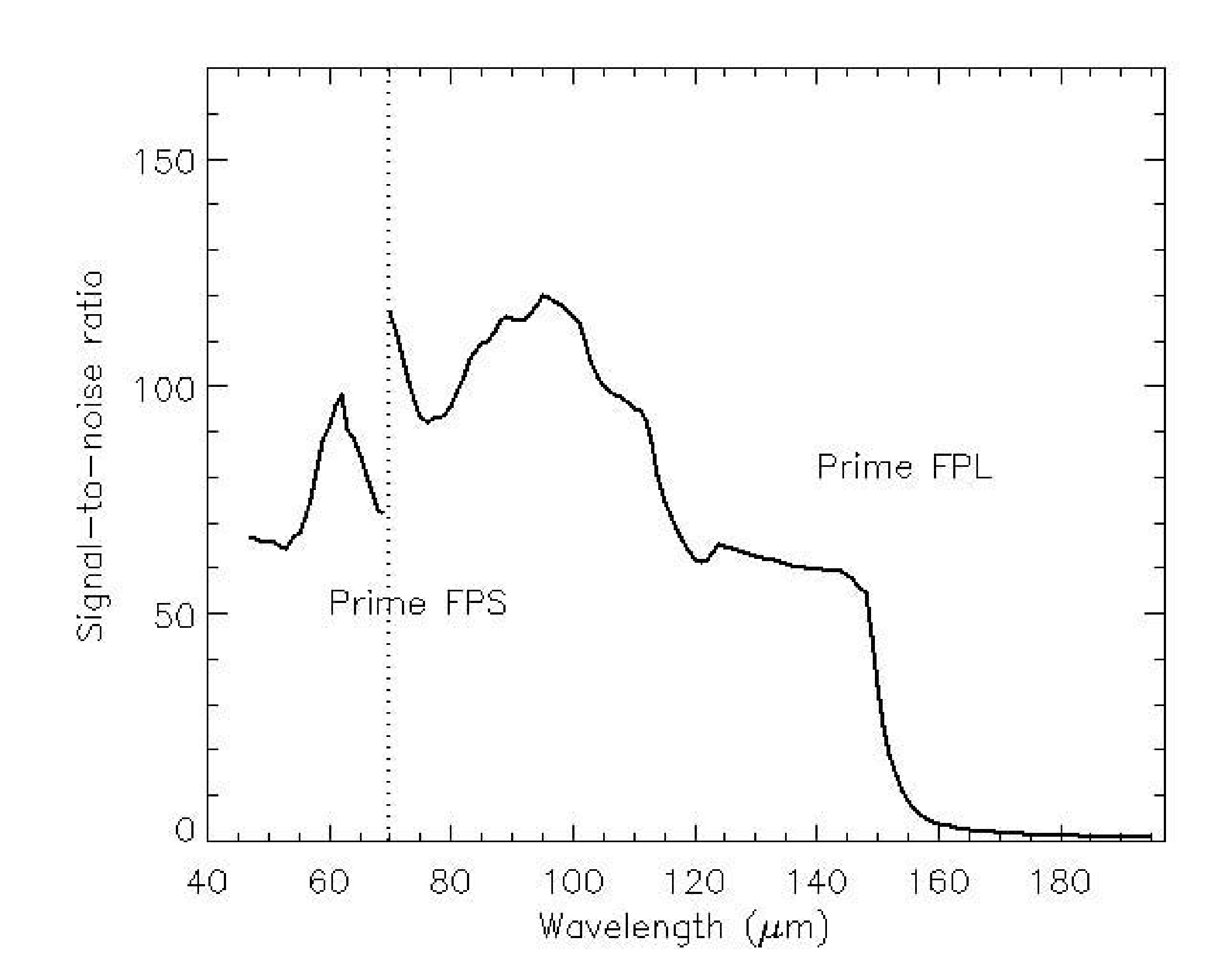}
    \caption{The signal to noise achieved for the L03 prime-detector spectra. Peaks in the curve represent the
    spectral ranges with the highest data reliability. The vertical dotted line
    divides the spectral ranges observed by the short wavelength
    FP (FPS) from those observed by the long wavelength FP (FPL)}
    \label{primes}
\end{figure}

\begin{figure}
    \centering
    \includegraphics {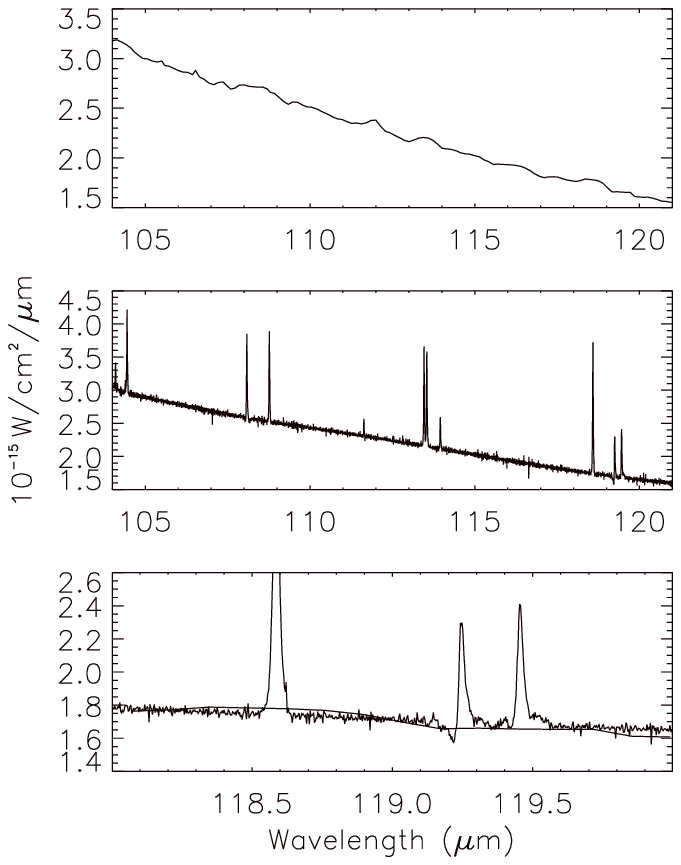}
    \caption{Comparison of a section of the low resolution L01 spectrum of Orion KL (top plot) with the high resolution L03 spectrum (middle plot).
    The bottom plot shows a direct comparison of the two spectra in a narrower spectral range.}
    \label{resolution}
\end{figure}

 The overall signal to noise achieved in the survey as a function of
wavelength is shown in Figure~\ref{primes}. The inclusion of
non-prime data fills in some gaps in spectral coverage that were
present in the prime data as well as overlapping in some places
with the prime observations. This improves the signal to noise
achieved by a factor that depends on the detector. Polehampton et
al. (2006 in prep.) give a detailed explanation of the effect of
including non-prime data on the overall signal to noise, based on
the throughput of
the LWS FP etalons.\\
High spectral resolution can provide detailed information as a
function of velocity for atomic and molecular transitions as well
as unique information about the temperature and density regimes in
which the transitions occur. Information on dynamical processes
can be inferred by comparing line shapes; this can be easily seen
for Orion KL, with H$_{2}$O and OH lines showing pure absorption, P-Cygni and pure emission profiles (see Section 4).\\
The advantages of high resolution spectroscopy can be seen by
comparing the same wavelength region observed in the low
resolution (L01) and high resolution (L03) mode. Figure
\ref{resolution} shows a section of the low spectral resolution
observation (L01 mode) compared with the high spectral resolution
(L03 mode) scan of the same region. At least 80\% of the lines
detected in the L03 scan are missed in the lower resolution
spectrum.

\subsection{The LWS Beam profile}

 The LWS beam profile is the convolution of the telescope point
spread function (PSF) with the aperture of each detector. Ideally,
the telescope PSF should be an Airy profile but
 the central obscuration, secondary supports and any optical imperfections redistribute
 power from the core of the profile to the Airy rings (Gry et al. 2003).
 The latest model of the telescope PSF (Gry et al. 2003) includes the effects of the central obscuration and its supporting structure,
 and indicates that the power in the Airy rings is increased and that the wings of the profile contain 2D asymmetric structure.
 The asymmetry introduced into the profile is due to the three-legged secondary support. The aperture was assumed
 to be circular with a top hat profile (Gry et al. 2003, Lloyd et al. 2003).
Table~\ref{beams} lists the effective diameter of the aperture of
each detector used to derive the effective solid angle. The solid
angle is defined as $2\pi(1-\cos{\theta})$, where $\theta$ is the
effective beam angular radius.

\begin{table}
\begin{center}
    \leavevmode
    \footnotesize
    \begin{tabular}[h]{lccc}
      \hline \\[-5pt]
      Detector &  Wavelength &Effective & Effective  \\[+5pt]
       & range ($\mu$m) & radius($^{\prime\prime}$) & solid angle (sr) \\
      \hline \\[-5pt]
      SW1 & 43-50.5  & 39.4  &  1.15 $\times$ 10$^{-7}$\\
      SW2 & 49.5-64 & 42.3    & 1.32 $\times$ 10$^{-7}$\\
     SW3  & 57-70 &  43.5  &  1.40 $\times$ 10$^{-7}$ \\
     SW4  & 67-82 & 40.9  &  1.23 $\times$ 10$^{-7}$ \\
     SW5 & 76-93 &  39.5 &   1.15 $\times$ 10$^{-7}$ \\
      LW1 & 84-110 &  38.6 &1.10 $\times$ 10$^{-7}$ \\
      LW2 & 103-128 & 38.9  &  1.12 $\times$ 10$^{-7}$\\
       LW3 & 123-152 & 35.5  &  9.30 $\times$ 10$^{-8}$\\
      LW4 & 142-171 & 34.7 &   8.89 $\times$ 10$^{-8}$\\
       LW5 & 167-197 & 33.2 &   8.14 $\times$ 10$^{-8}$\\
   \hline \\
      \end{tabular}
  \end{center}
   \caption{The effective beam radius and effective solid angle subtended by each detector. Differences in
   effective apertures are due to asymmetries introduced into the profile by the three-legged secondary instrumental
   support (Gry et al. 2003) }
  \label{beams}
\end{table}

\subsection{Line flux measurements}
The shape of the FP spectral response function was determined from
dedicated calibration observations of narrow, unresolved spectral
lines in the spectra of planetary nebula. These observations show
that the response function shape can be accurately described by
the predicted (Hecht and Zajac, 1974) Airy profile. The observed
line profile is then the convolution of the source intrinsic
profile with the Airy function. Line measurements were done by
fitting a Lorentzian profile using the interactive line-fitting
routine in ISAP. The Lorentzian function was found to approximate
the line shape very well, however, for unidentified features and
lines with poor signal, a Gaussian fitting function was selected.
Data were averaged over detectors and scans. The ISAP fitting
procedure gives the line flux, line centre, FWHM, and an estimate
of the statistical error on the line flux.

\section{Data analysis and results}

Line identification was carried out based on the JPL line
catalogue (Pickett et al. 1998) and the molecular catalogue
created and maintained by one of us (JC), which includes more than
1200 molecular species. The spectrum is dominated by molecular
lines of CO, OH and H$_{2}$O and by forbidden lines of [O~{\sc
iii}], [O~{\sc i}], [N~{\sc iii}], [N~{\sc ii}] and [C~{\sc ii}].
Isotopic variants and the lowest transitions of HDO, NH$_{3}$ and
H$_{3}$O$^{+}$ are also detected. Table~\ref{resultado1} and
Figure \ref{survey}
 show the identified
lines.
\section{Discussion of individual species}
A basic analysis and discussion of the individual detected species
is presented in the following sub-sections.\\
From the analysis of the atomic forbidden lines, we derive
physical parameters such as temperatures and densities by
comparing our measurements with models. When conditions are
appropriate, we use the `rotation diagram' method to determine
molecular parameters such as rotation temperatures and column
densities. A problem with using the rotational diagram method is
that it underestimates the total column density if the energy
distribution is far from Local Thermodynamic Equilibrium (LTE). In
order to decide whether the use of this method is appropriate, we
estimated line optical depths using the statistical equilibrium
radiative transfer code RADEX. (Schoier et al. 2005). RADEX is a
one-dimensional non-LTE radiative transfer code that uses the
escape probability formulation assuming an isothermal and
homogeneous medium without large-scale velocity fields.

\subsection{Atomic forbidden Lines}
Table~\ref{atomic} lists several detected forbidden transitions
observed by the survey: [O~{\sc iii}] 52 $\mu$m, 88 $\mu$m,
[N~{\sc iii}] 57 $\mu$m, [N~{\sc ii}] 122 $\mu$m, [O~{\sc i}] 63
$\mu$m, 145 $\mu$m and [C~{\sc ii}] 158 $\mu$m. The flux errors
were deduced from Lorentzian profile fitting (except the [N~{\sc
ii}] 122 $\mu$m line) and do not include systematic errors caused
by uncertainties in the flux calibration, see Section 2.4. The
[N~{\sc ii}] 122 $\mu$m line was fitted using a Gaussian profile
due to its width and low intensity. This broadness is due to
hyperfine components in the transition as can be seen in the
detection of the [N~{\sc ii}] 122 $\mu$m line in Sgr B2
(Polehampton et al. 2006, in prep.).
\subsubsection{Lines from ionised gas: {\rm [O~{\sc iii}], [N~{\sc iii}] and [N~{\sc ii}]}}
 The [O~{\sc iii}] 52, 88 $\mu$m, [N~{\sc iii}] 57 $\mu$m and [N~{\sc ii}] 122
$\mu$m lines originate from the foreground M42 H~{\sc ii} region
which is excited by the $\theta^{1}$ Ori OB stars, known as `The
Trapezium' (see Figure~\ref{M42}); these lines provide an
important tool for probing physical conditions and elemental
abundances in the ionised gas (Simpson et al. 1986, Rubin et al.
1994).
\subsubsection{Electron densities from the {\rm[O~{\sc iii}]} lines}
 The intensity ratio of the
[O~{\sc iii}] 52 $\mu$m and 88 $\mu$m lines provides an electron
density diagnostic. The line ratio has almost no dependence on
electron temperature because of the very low excitation energies
of the transitions. The measured line fluxes give an 88.4
$\mu$m/51.8 $\mu$m flux ratio of $\sim$ 0.41 $\pm$ 0.01. Adopting
the same atomic parameters as Liu et al. (2001), we obtain
an electron density $N_{e}$(O~{\sc iii})=830$\pm^{200}_{140}$ cm$^{-3}$ calculated for $T_e$=8000 K. \\
Another important electron density indicator is the [S~{\sc iii}]
33.5/18.7 $\mu$m ratio.

\onecolumn

\begin{figure*}
 \caption{The complete Orion KL spectral survey obtained by the {\em ISO} LWS in Fabry-P\'erot modes L03 and L04. The grating
 observation (L01 mode) is superimposed in regions where high resolution data are not
available. }
         \label{survey}
          \centering
\includegraphics[width=16cm,height=7cm]{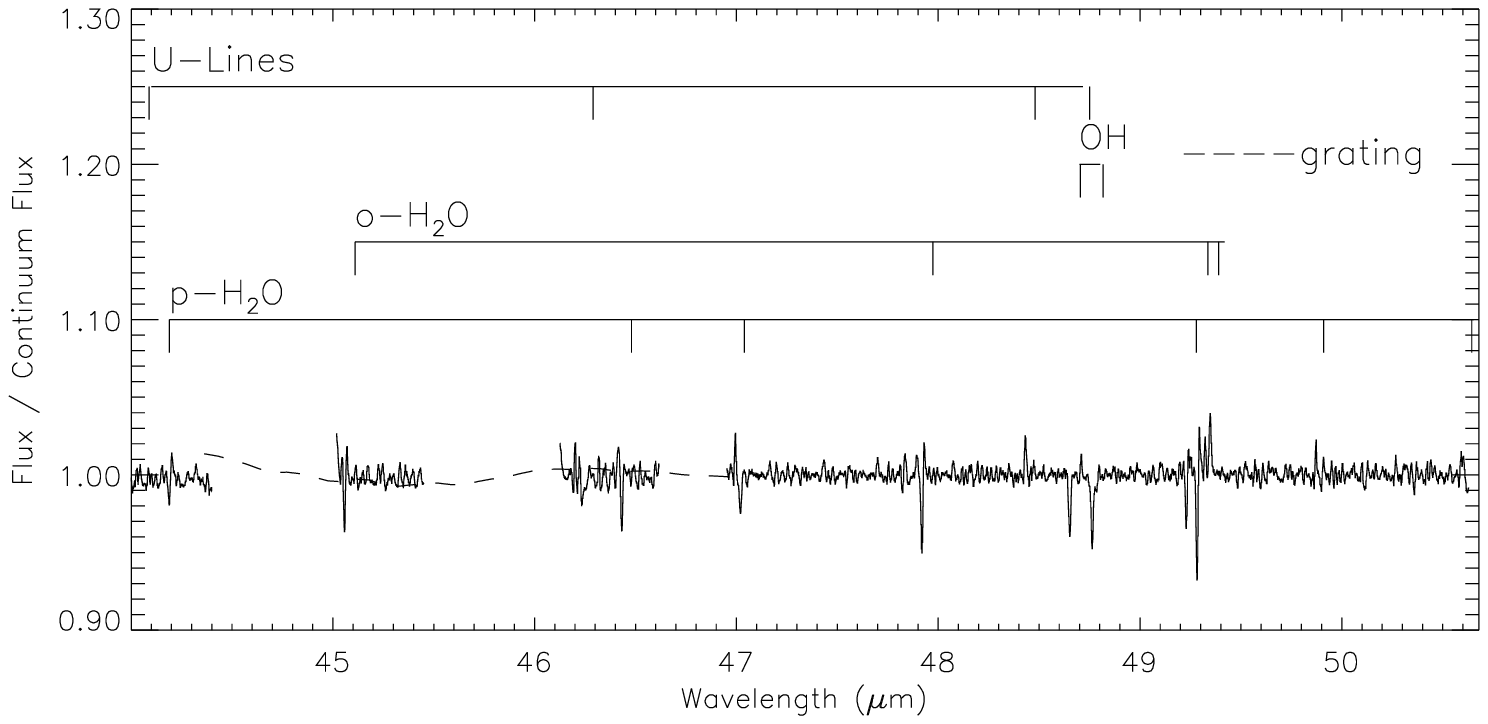}
\includegraphics[width=16cm,height=7cm]{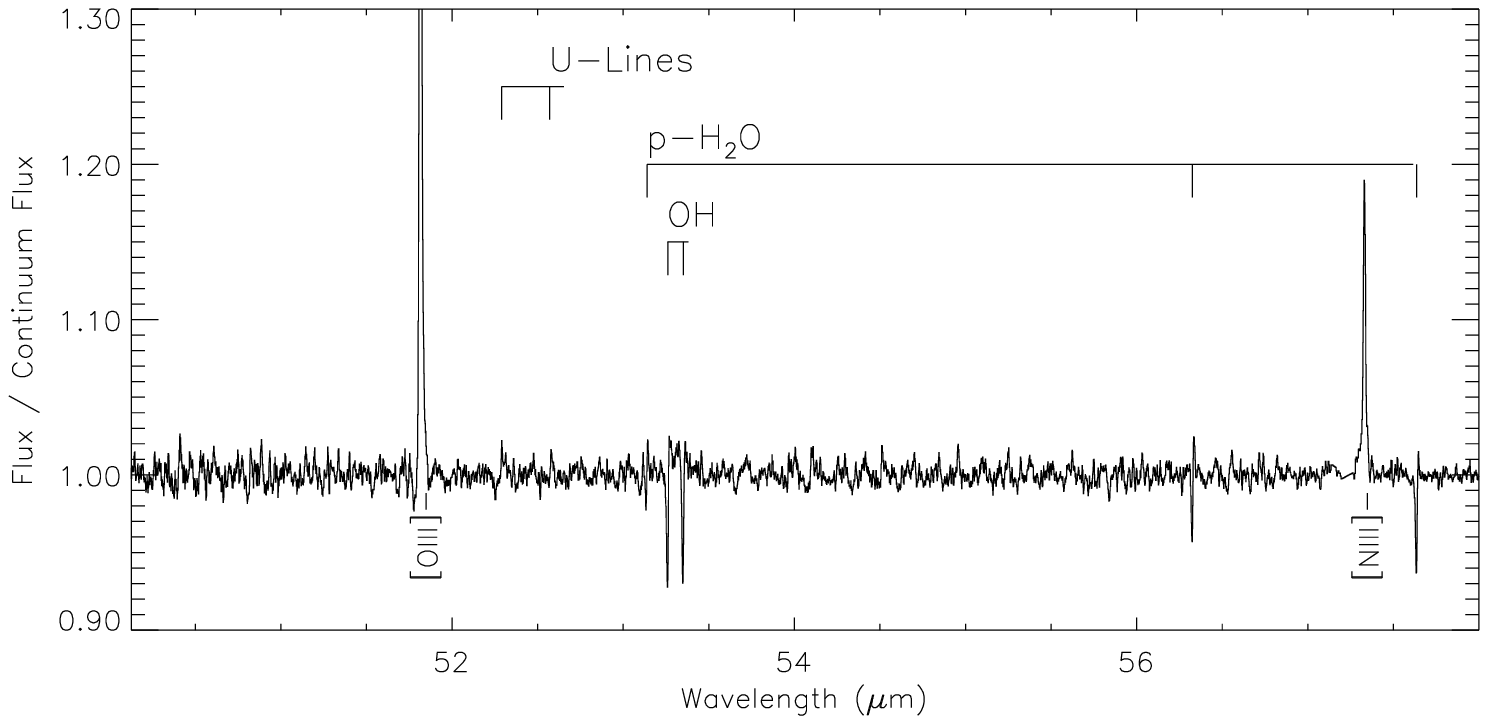}
\includegraphics[width=16cm,height=7cm]{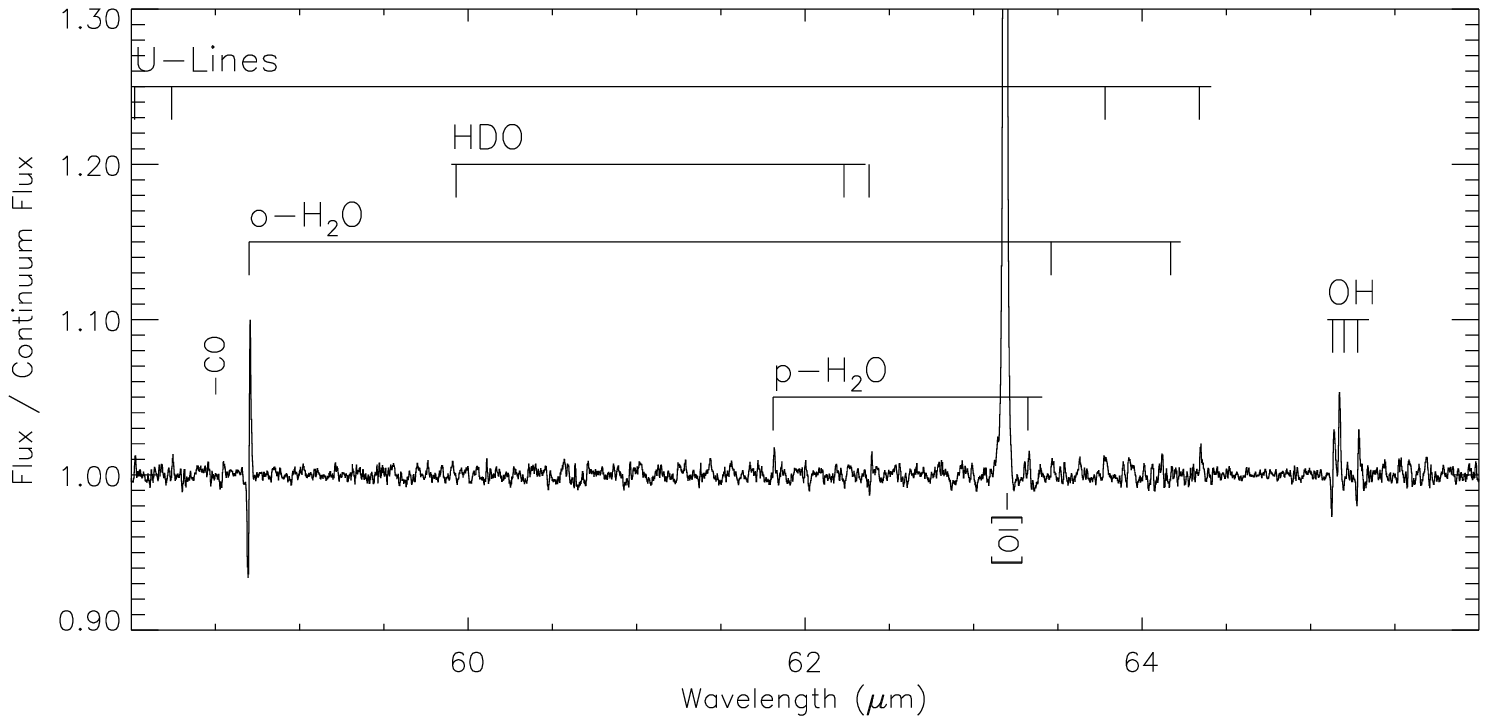}

\end{figure*}
\addtocounter{figure}{-1}
\begin{figure*}
\centering

\includegraphics[width=16cm,height=7cm]{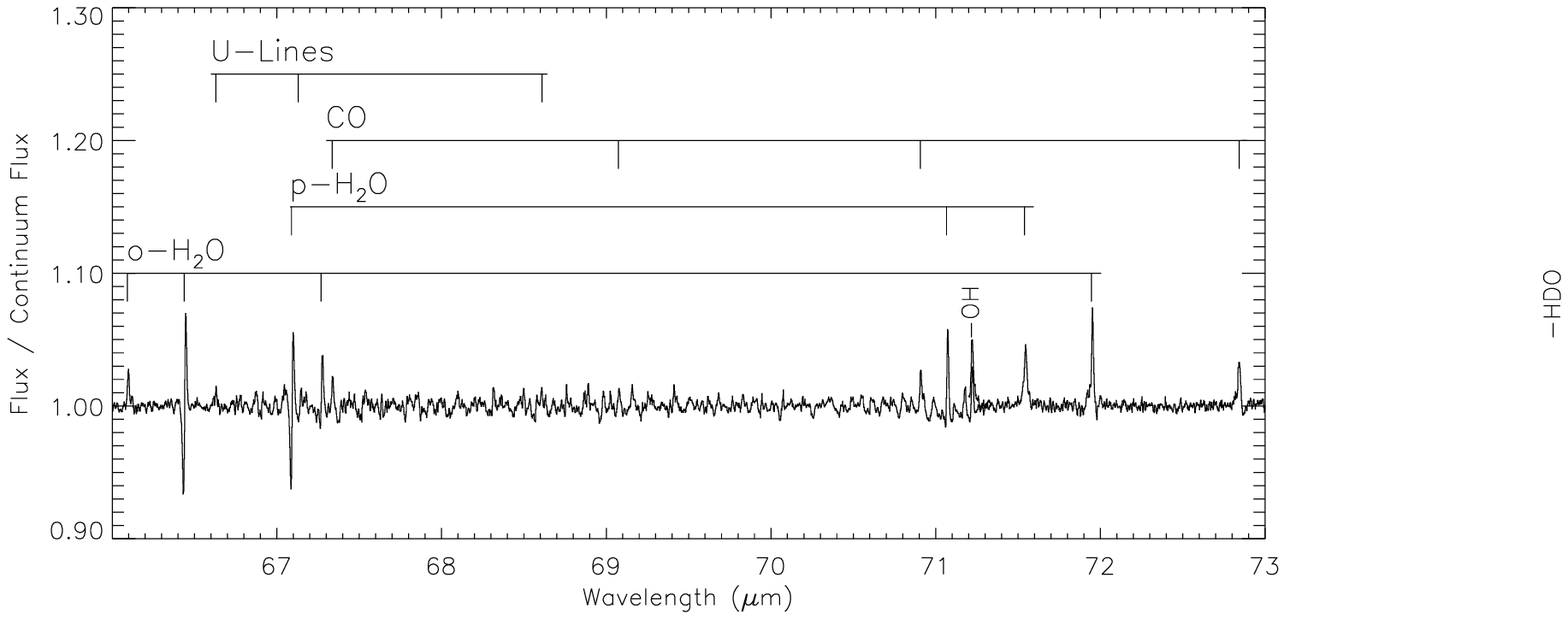}

     \includegraphics[width=16cm,height=7cm]{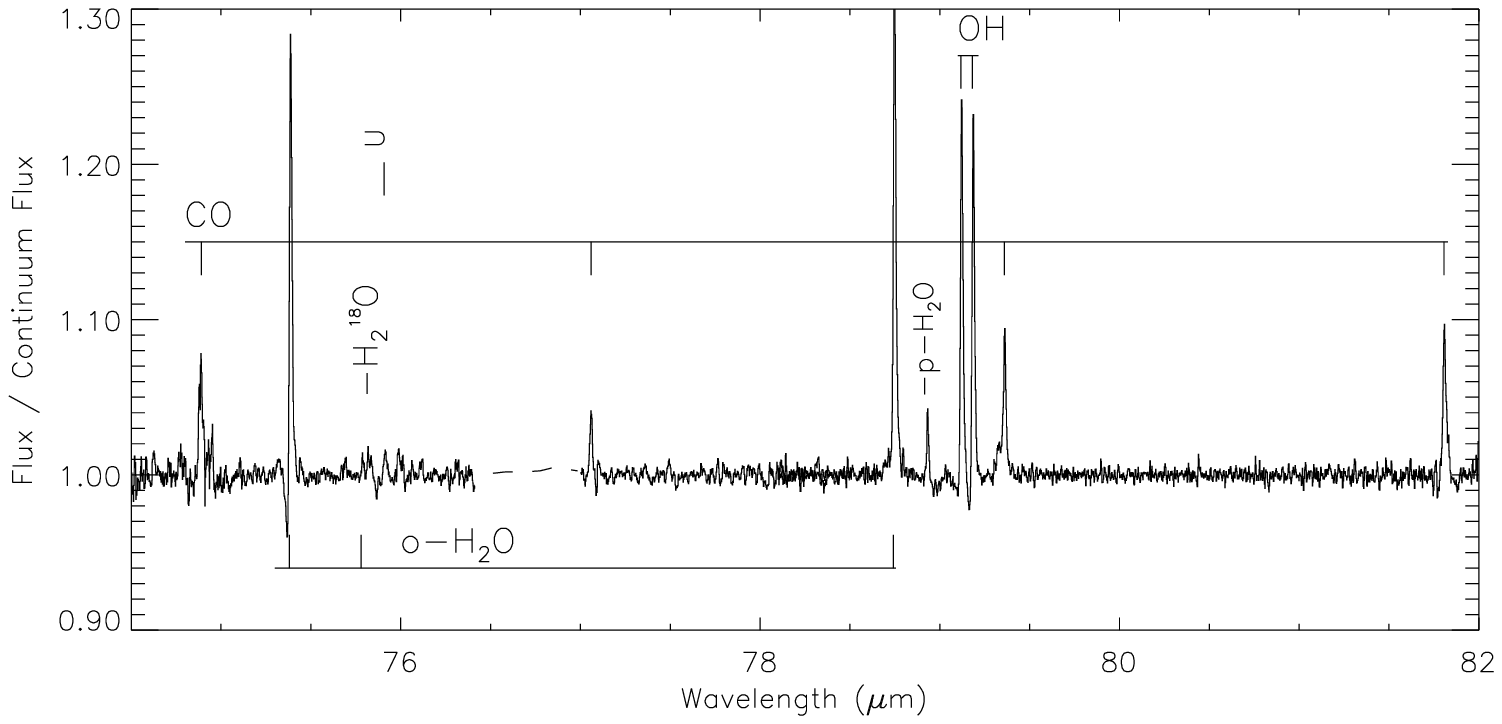}

    \includegraphics[width=16cm,height=7cm]{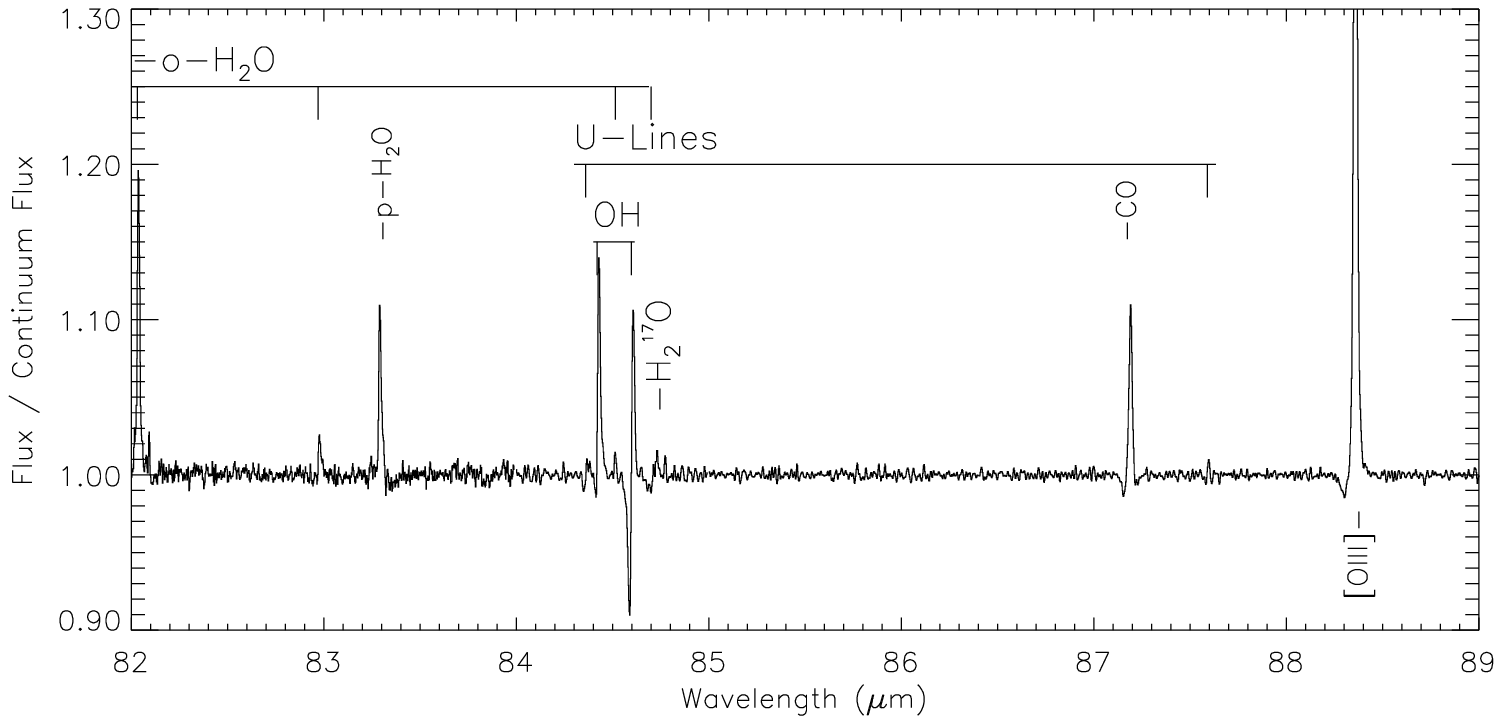}

 \caption{Continued}
   \end{figure*}
   \addtocounter{figure}{-1}
   \begin{figure*}
    \centering
    \includegraphics[width=16cm,height=7cm]{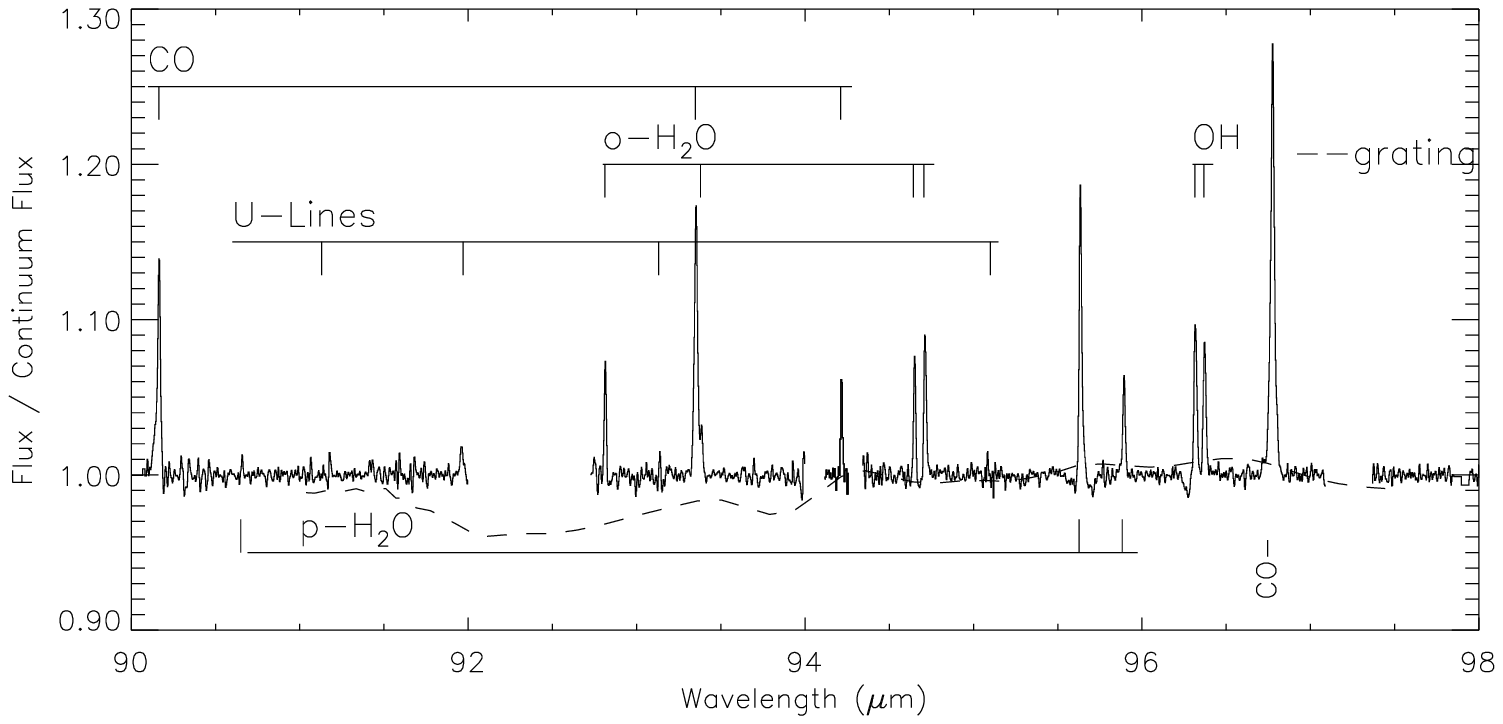}
    \includegraphics[width=16cm,height=7cm]{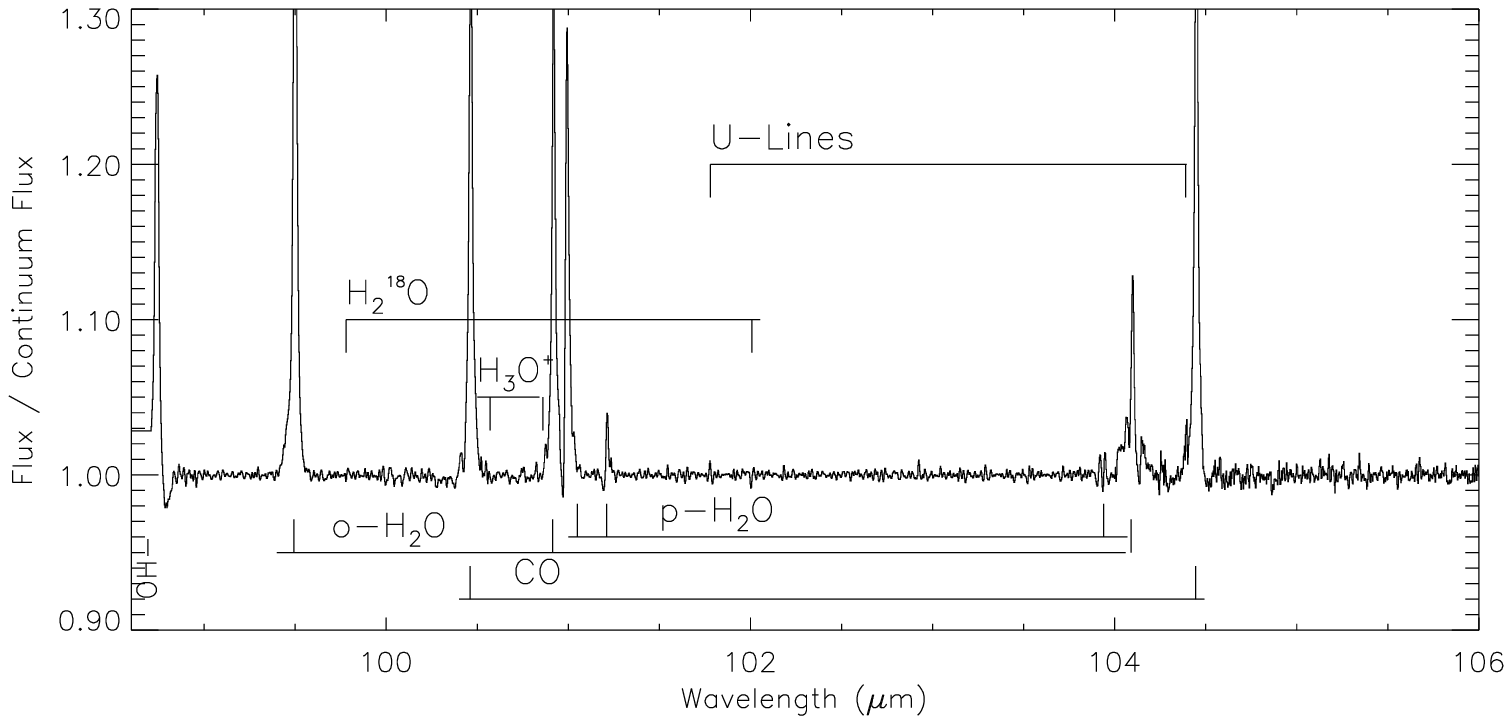}
    \includegraphics[width=16cm,height=7cm]{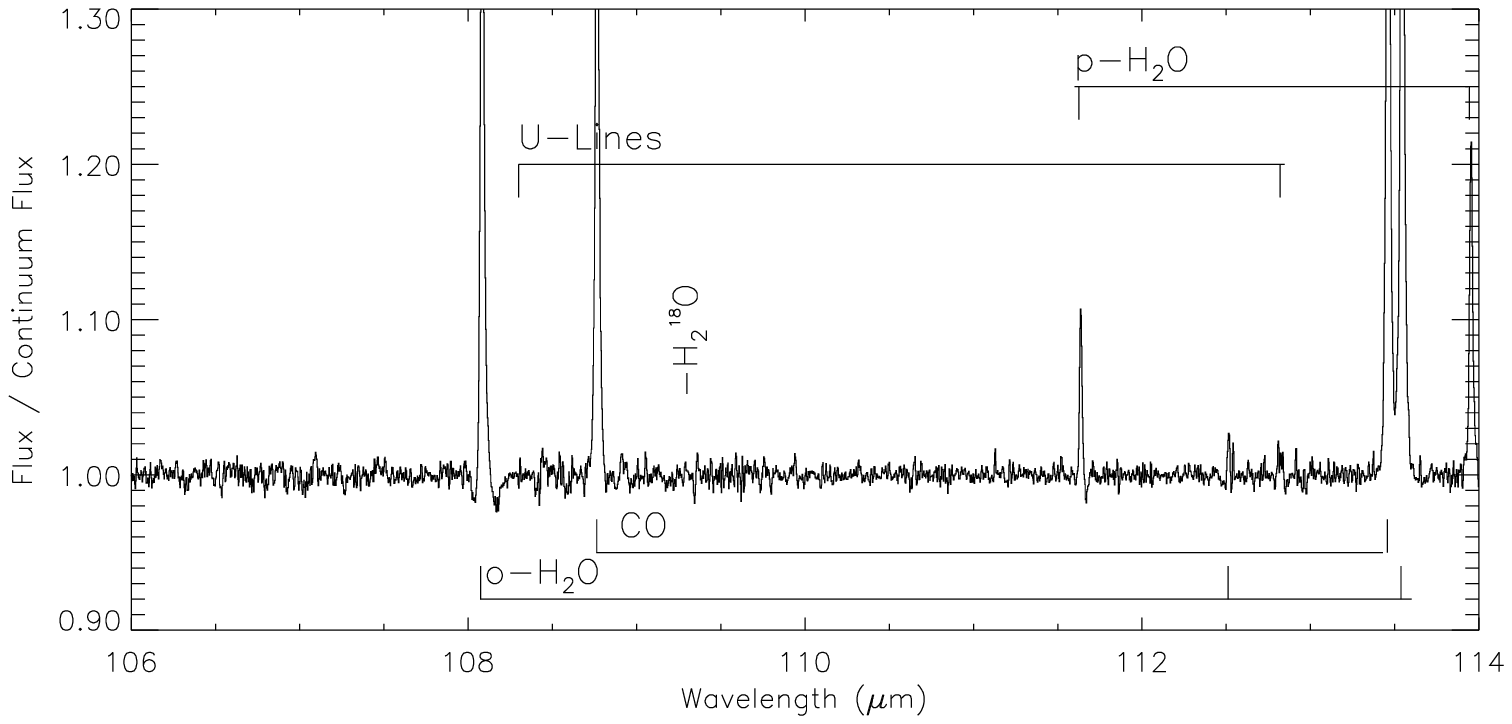}

 \caption{Continued}
   \end{figure*}
    \addtocounter{figure}{-1}
    \begin{figure*}
    \centering

  \includegraphics[width=16cm,height=7cm]{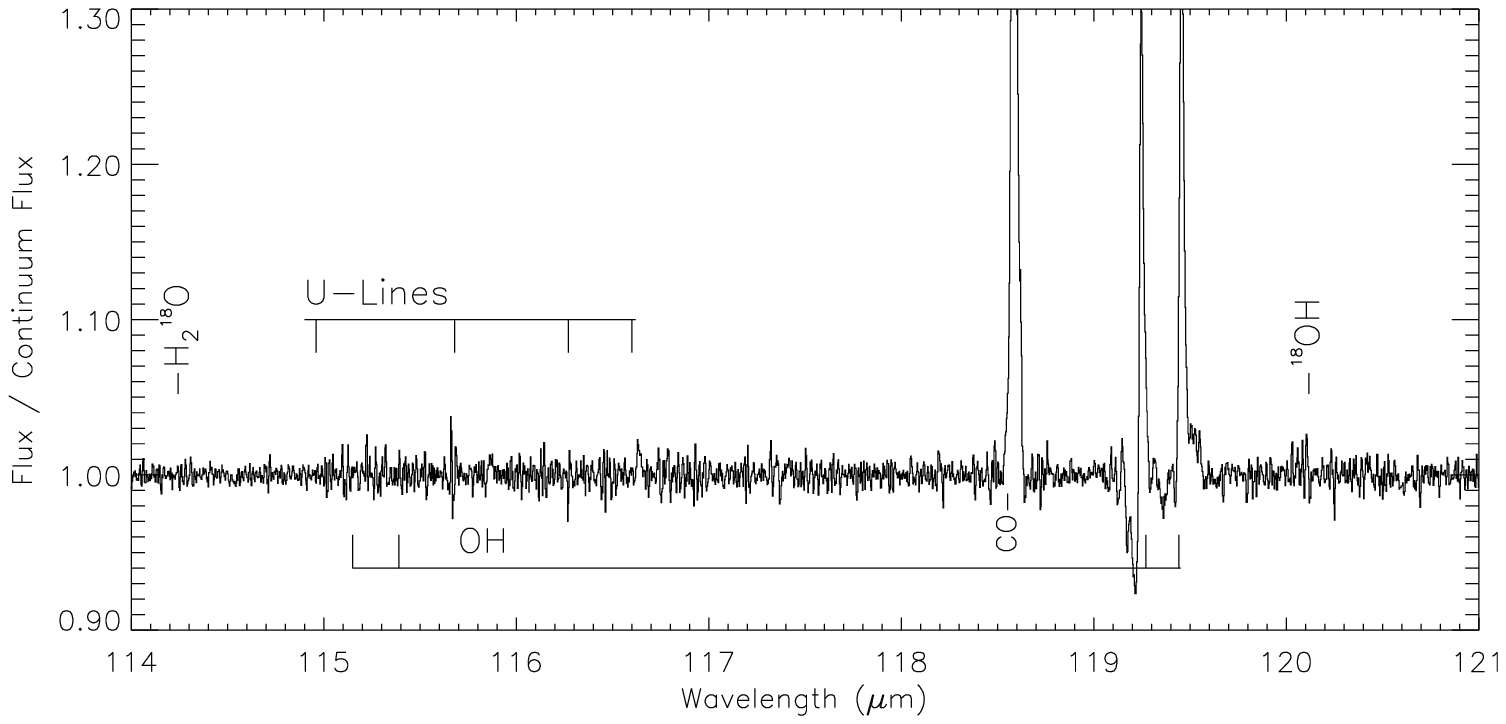}
    \includegraphics[width=16cm,height=7cm]{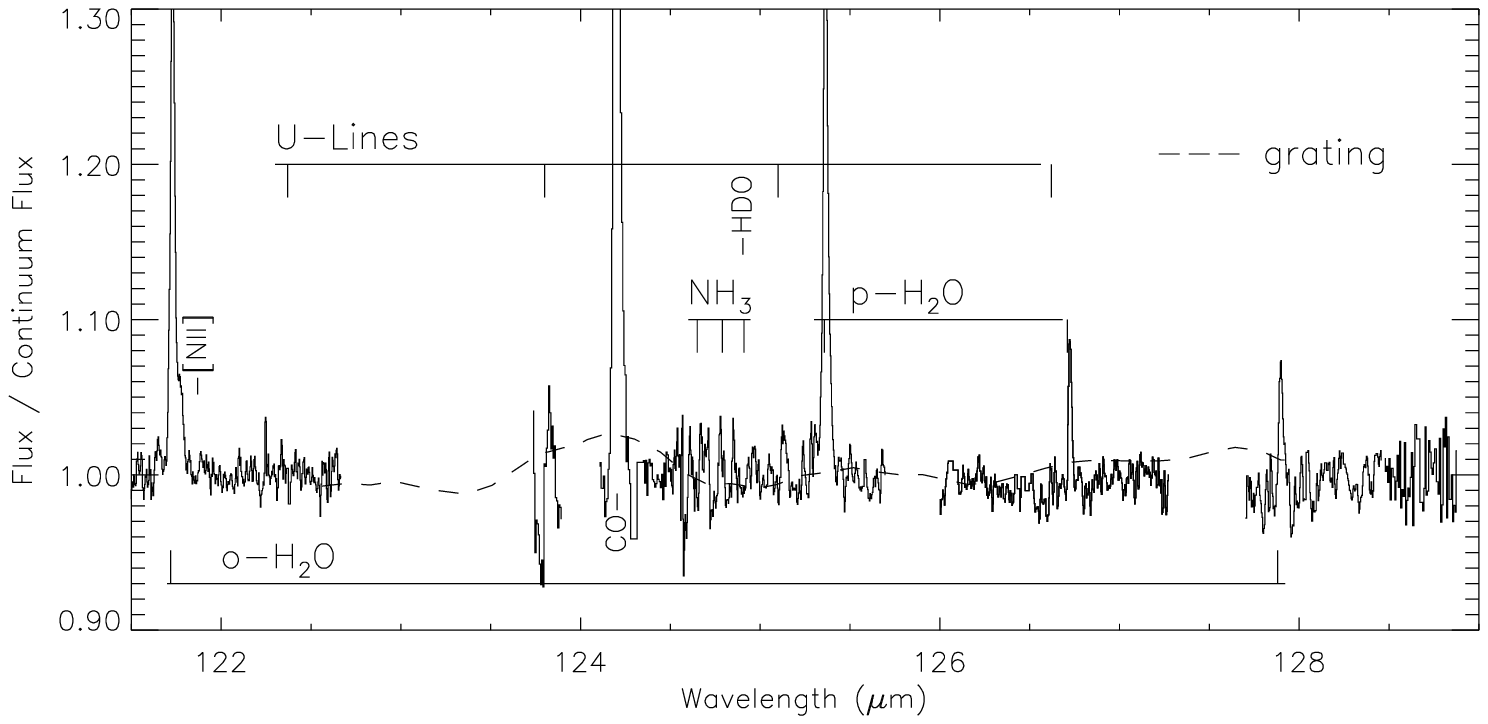}
    \includegraphics[width=16cm,height=7cm]{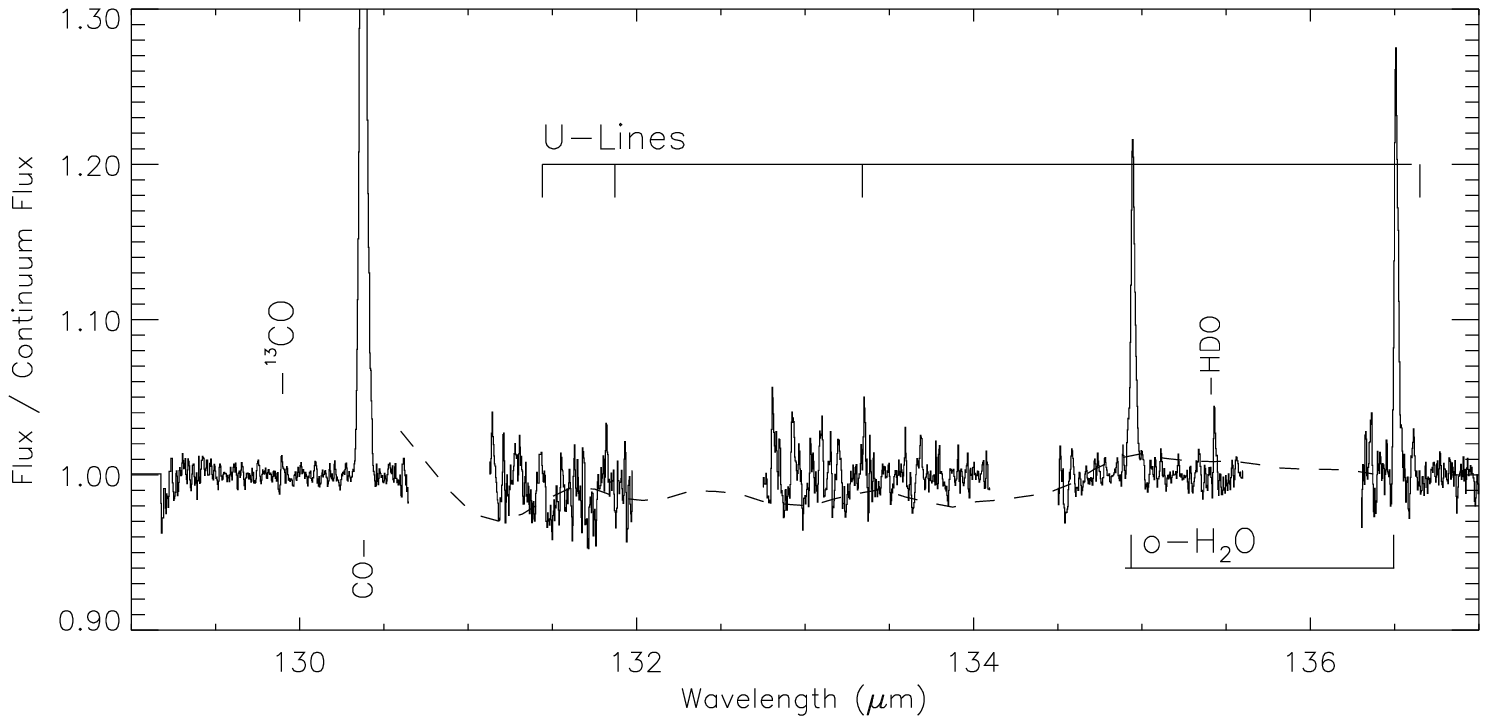}

 \caption{Continued}
   \end{figure*}
   \addtocounter{figure}{-1}
 \begin{figure*}
    \centering
     \includegraphics[width=16cm,height=7cm]{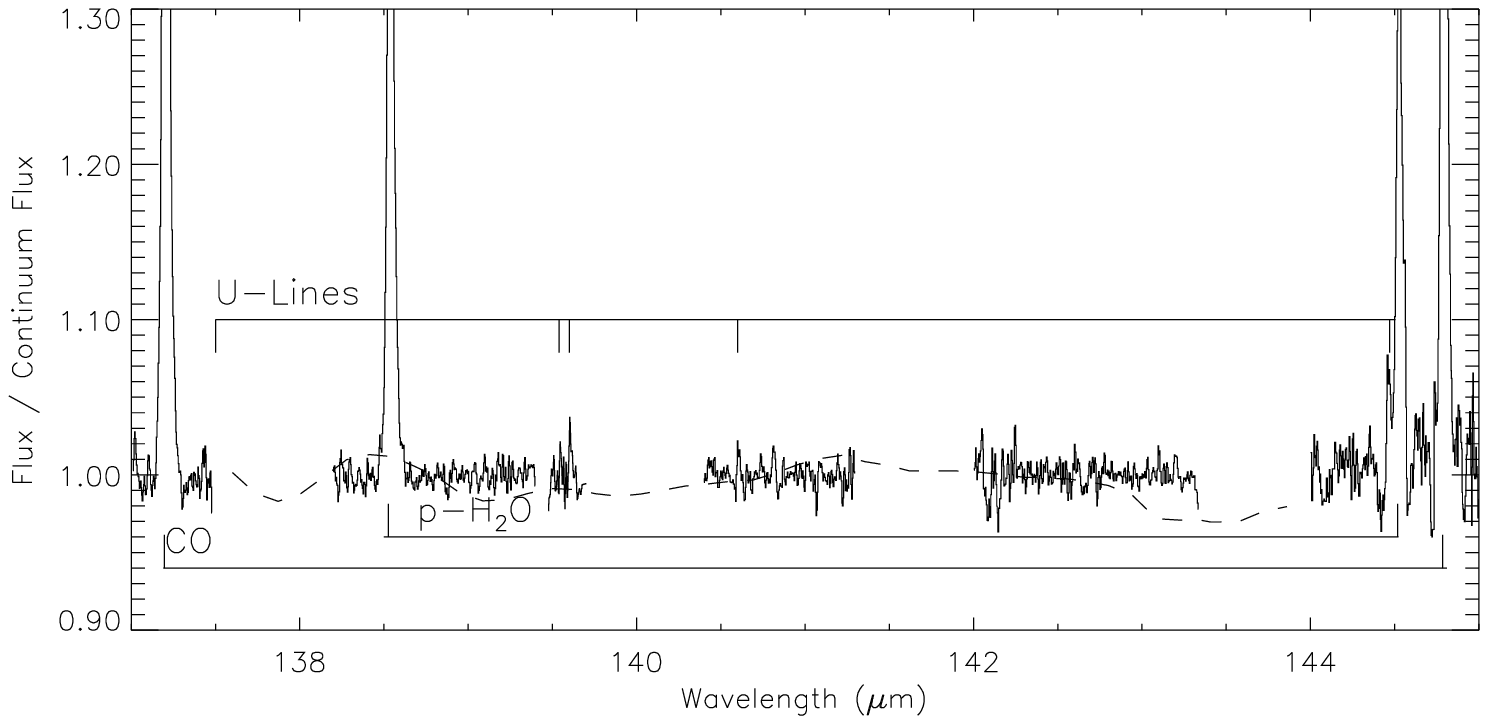}

    \includegraphics[width=16cm,height=7cm]{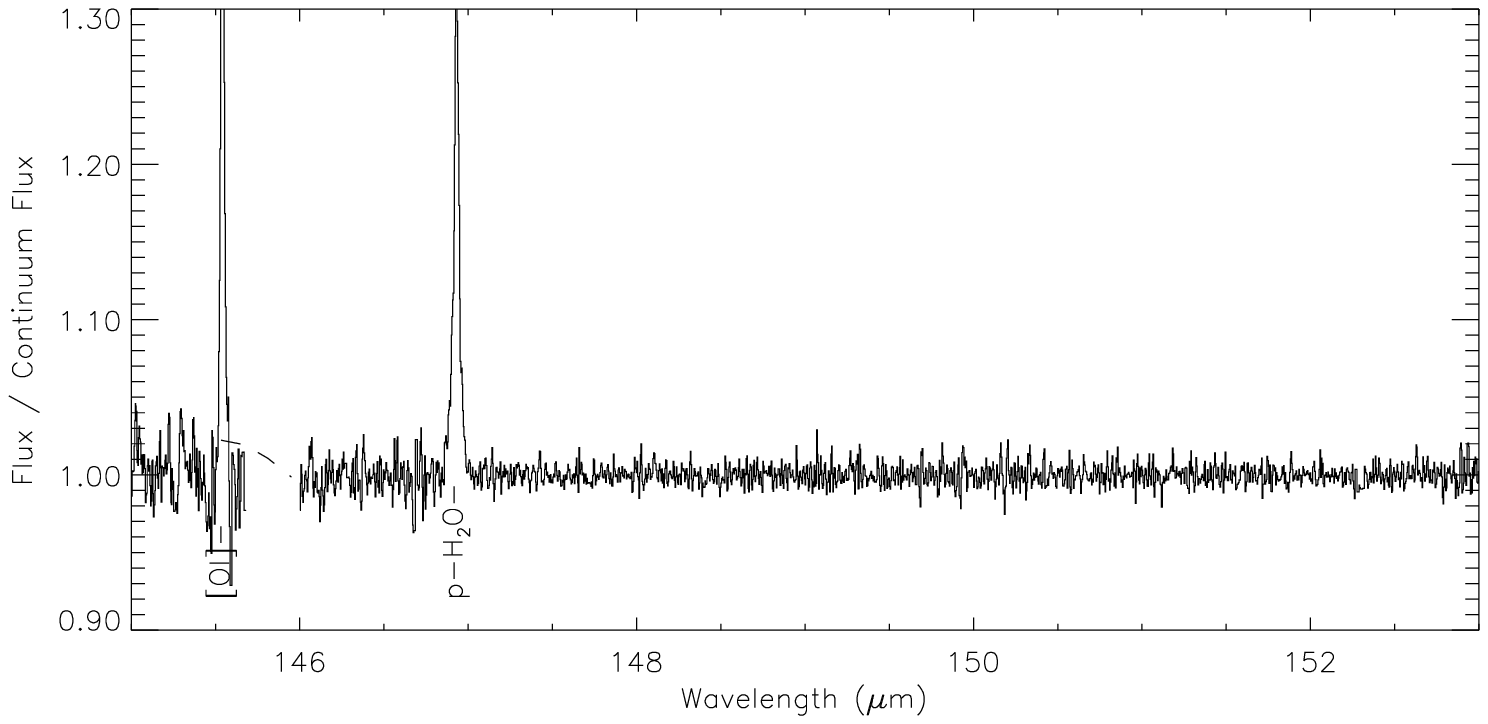}
     \includegraphics[width=16cm,height=7cm]{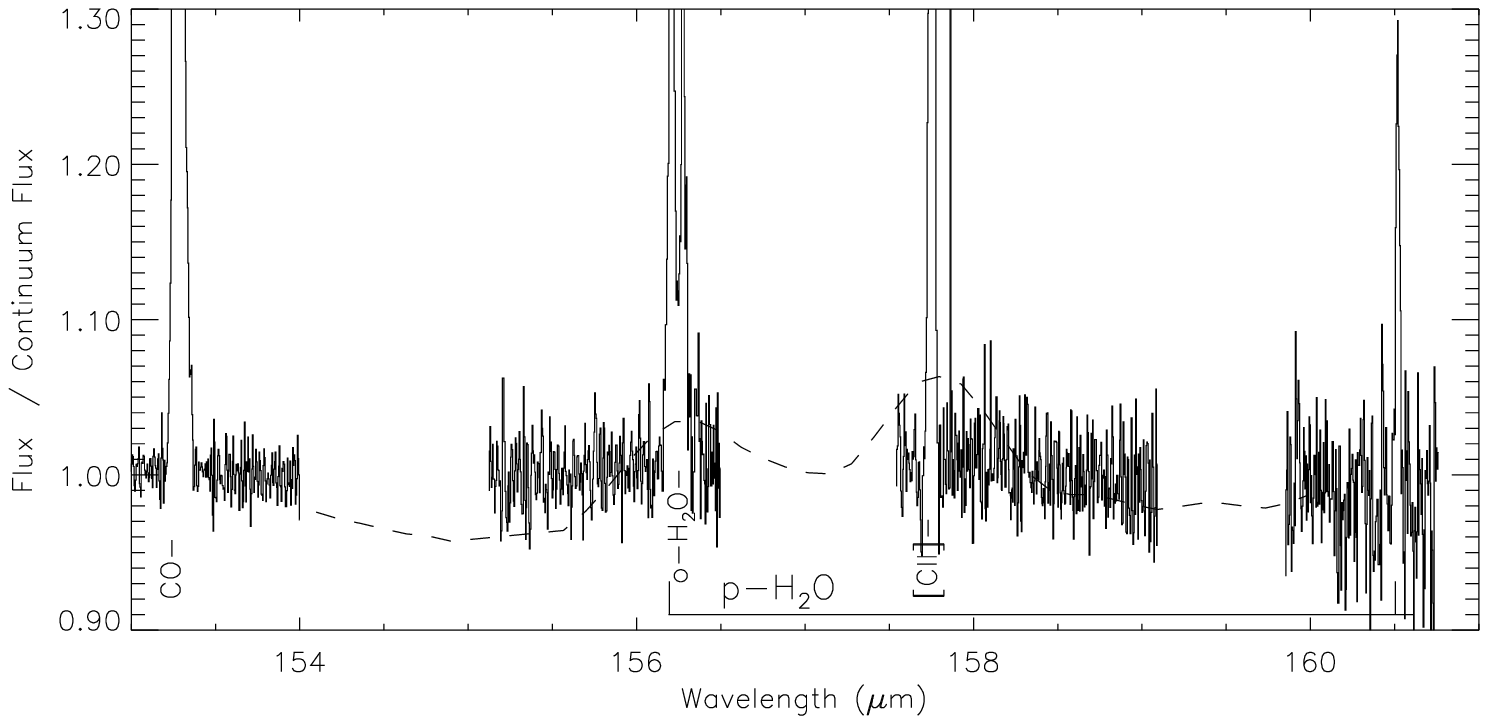}

 \caption{Continued}
   \end{figure*}
   \addtocounter{figure}{-1}
    \begin{figure*}
    \centering
     \includegraphics[width=16cm,height=7cm]{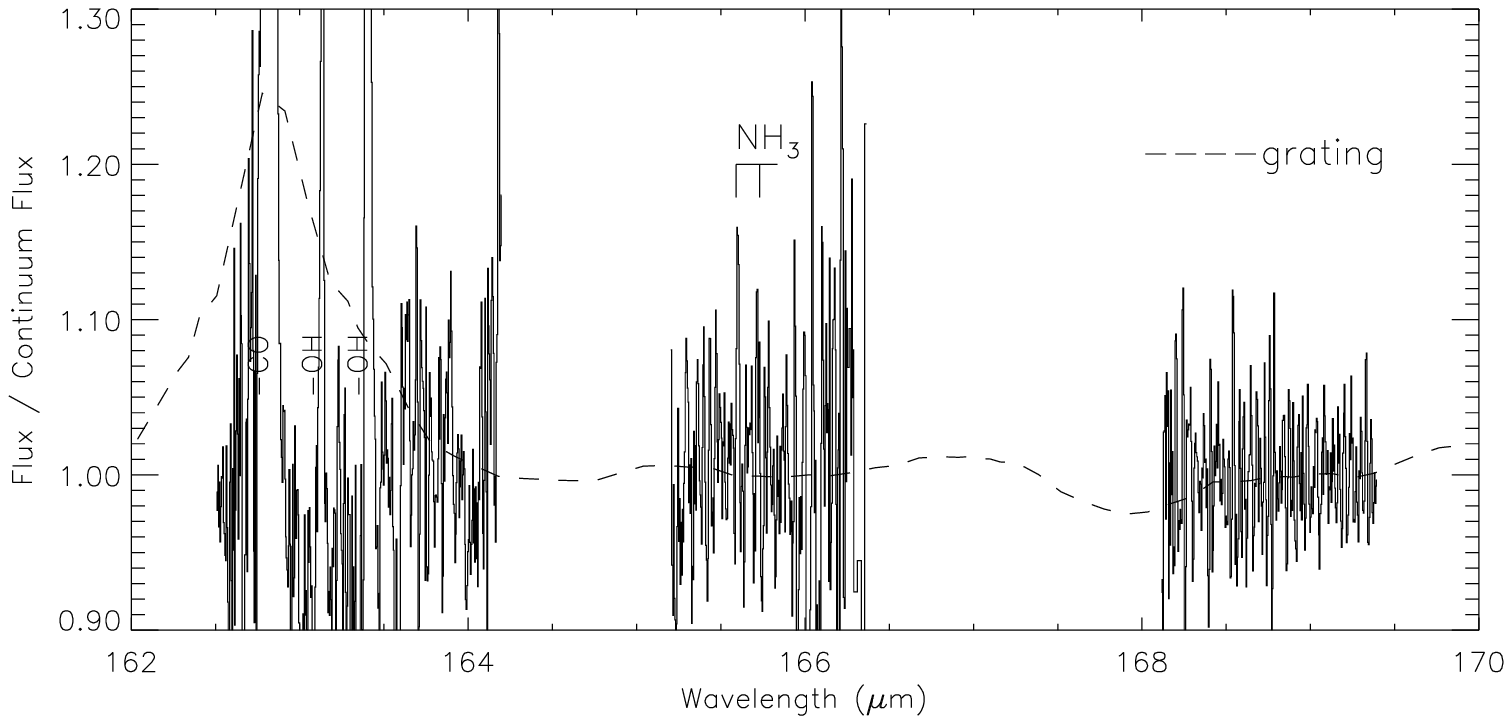}
    \includegraphics[width=16cm,height=7cm]{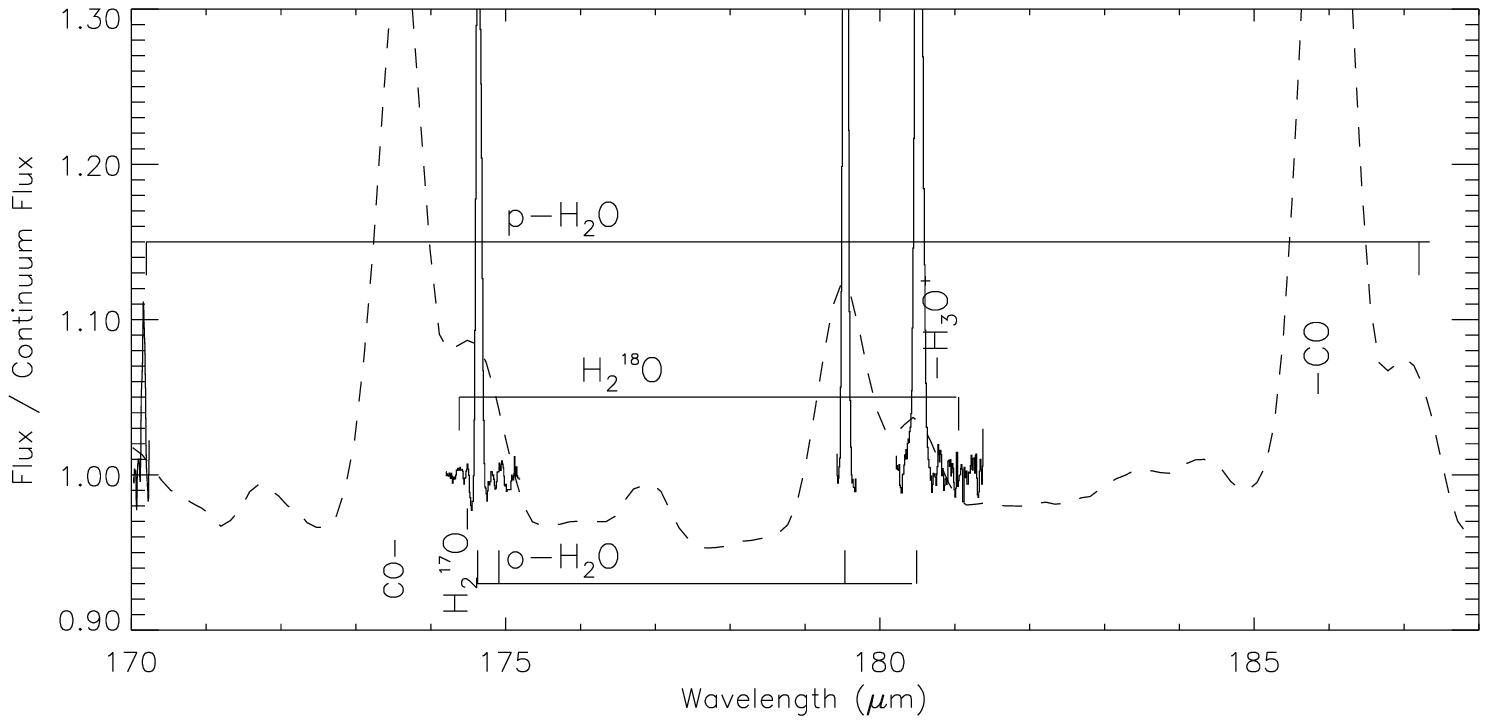}
 \caption{Continued. Note that features with no label on the L01 spectrum are non real features transferred by the calibration stage.}
   \end{figure*}

\begin{center}
\scriptsize

 \centering
 \setlength{\arrayrulewidth}{0.1pt}
 \setlength{\doublerulesep}{0mm}
\setlength\LTleft{0cm} \setlength \LTright{17cm}
    \begin{longtable}{@{\extracolsep{0.0mm}\extracolsep{0.0mm}\extracolsep{0.0mm}\extracolsep{0.0mm}
    \extracolsep{0.0mm}\extracolsep{0.0mm}\extracolsep{0.0mm}}ccccccc}

\caption[ Line measurements obtained from L03 and L04 high
resolution observations towards the Orion KL region. Line fitting
measurements were performed using a Lorentzian function. Quoted
errors are based on goodness of fit estimations. Note that
$^{(1)}$ corresponds to a tentative detection with flux below
3$\sigma$; $^{(2)}$ are blended lines and $^{(3)}$ are lines
detected in the lower resolution L01 spectrum where no velocity
peak information is given.]{\footnotesize Line measurements
obtained from L03 and L04 high resolution observations towards the
Orion KL region. Line fitting measurements were performed using a
Lorentzian function. Quoted errors are based on goodness of fit
estimations. Note that $^{(1)}$ corresponds to a tentative
detection with flux below 3$\sigma$; $^{(2)}$ are blended lines
and $^{(3)}$ are lines detected in the lower resolution L01
spectrum where no velocity peak information is given.}
 \label{resultado1}\\

\hline \hline \\[-2ex]

  \multicolumn{1}{c}{\textbf{Species}} &
  \multicolumn{1}{c}{ \textbf{Transition}} &
    \multicolumn{1}{c}{ \textbf{Rest}} &
    \multicolumn{1}{c}{ \textbf{Flux}} &
    \multicolumn{1}{c}{ \textbf{Flux}} &
\multicolumn{1}{c}{ \textbf{Peak}} &
\multicolumn{1}{c}{ \textbf{Peak}} \\

\multicolumn{1}{c}{\textbf{}} &
  \multicolumn{1}{c}{\textbf{}} &
    \multicolumn{1}{c}{ \textbf{
     Wavelength}} &
    \multicolumn{1}{c}{\textbf{{absorption} }} &
    \multicolumn{1}{c}{\textbf{{emission}}} &
\multicolumn{1}{c}{\textbf{{absorption}}} &
\multicolumn{1}{c}{\textbf{{emission} }}\\

\multicolumn{1}{c}{\textbf{}} &
  \multicolumn{1}{c}{\textbf{}} &
    \multicolumn{1}{c}{ \textbf{($\mu$m)}} &
    \multicolumn{1}{c}{ \textbf{(10$^{-17}$W cm$^{-2}$)}} &
    \multicolumn{1}{c}{ \textbf{(10$^{-17}$W cm$^{-2}$)}} &
  \multicolumn{1}{c}{ \textbf{(km s$^{-1}$)}} &
  \multicolumn{1}{c}{ \textbf{(km s$^{-1}$)}}

\\[3ex] \hline
  \\[-2ex]

\endfirsthead

\multicolumn{3}{c}{{\tablename} \thetable{} -- \footnotesize Continued} \\[2ex]
  \hline \hline \\[-2ex]
\multicolumn{1}{c}{\footnotesize \textbf{Species}} &
  \multicolumn{1}{c}{\footnotesize \textbf{Transition}} &
    \multicolumn{1}{c}{\small \textbf{Rest}} &
    \multicolumn{1}{c}{\small \textbf{Flux}} &
    \multicolumn{1}{c}{\small \textbf{Flux}} &
\multicolumn{1}{c}{\small \textbf{Peak}} &
\multicolumn{1}{c}{\small \textbf{Peak}} \\

\multicolumn{1}{c}{\small \textbf{}} &
  \multicolumn{1}{c}{\small \textbf{}} &
    \multicolumn{1}{c}{\small \textbf{
     Wavelength}} &
    \multicolumn{1}{c}{\small \textbf{\small{absorption} }} &
    \multicolumn{1}{c}{\small \textbf{\small{emission}}} &
\multicolumn{1}{c}{\textbf{\small{absorption}}} &
\multicolumn{1}{c}{\textbf{\small{emission} }}\\

\multicolumn{1}{c}{\textbf{}} &
  \multicolumn{1}{c}{\textbf{}} &
    \multicolumn{1}{c}{\small \textbf{($\mu$m)}} &
    \multicolumn{1}{c}{\small \textbf{(10$^{-17}$W cm$^{-2}$)}} &
    \multicolumn{1}{c}{\small \textbf{(10$^{-17}$W cm$^{-2}$)}} &
\multicolumn{1}{c}{\small \textbf{(km s$^{-1}$)}} &
\multicolumn{1}{c}{\small \textbf{(km s$^{-1}$)}}
  \\[0.5ex] \hline
  \\[-1.8ex]
\endhead

\hline
  \multicolumn{7}{l}{{$^{(1)}$ Tentative
detections $^{(2)}$ Blended
lines} $^{(3)}$ L01 detection } \\

\endfoot

\hline
  \multicolumn{7}{l}{{$^{(1)}$ Tentative
detections $^{(2)}$ Blended
lines} $^{(3)}$ L01 detection } \\

  \\[-1.8ex] \hline \hline
\endlastfoot

\hline
p-H$_{2}$O  & $5_{42}$ -- $4_{31}$  & 44.195 & 0.48 $\pm$ 0.091 & 0.39 $\pm$ 0.20  & -28.9 $\pm$ 5.5 & 53.6 $\pm$ 27.4\\
o-H$_{2}$O   &   $5_{23}$ -- $4_{14}$ & 45.111 & 1.11 $\pm$ 0.016 & 0.39 $\pm$ 0.091 & -22.2 $\pm$ 2.1& 39.7 $\pm$ 9.2\\
p-H$_{2}$O   & $3_{31}$ -- $2_{02}$ & 46.484 & 1.05 $\pm$ 0.12 & & -10.6 $\pm$ 2.1 \\
p-H$_{2}$O   & $10_{19}$ -- $9_{28}$ & 47.039 & 0.19 $\pm$ 0.041 & 0.49 $\pm$ 0.19 & -3.1 $\pm$ 1.1 & 23.6 $\pm$ 9.1 \\
o-H$_{2}$O   &   $5_{32}$ -- $4_{23}$ & 47.972& 1.13 $\pm$ 0.14 &0.55 $\pm$ 0.21  & -16.3 $\pm$ 2.1& 38.9 $\pm$ 14.8 \\
OH  &  $^{2}\Pi_{1/2}$ -- $^{2}\Pi_{3/2}$ J=5/2$^{-}$ -- 5/2$^{+}$ & 48.704 & 1.09 $\pm$ 0.13  & & -12.4 $\pm$ 1.4\\
OH &  $^{2}\Pi_{1/2}$ -- $^{2}\Pi_{3/2}$ J=5/2$^{+}$ -- 5/2$^{-}$ &   48.817 & 1.34 $\pm$ 0.14 &  &-22.8 $\pm$ 2.8\\
p-H$_{2}$O   & $4_{40}$ -- $3_{31}$ &  49.281 & 0.71 $\pm$ 0.13  & & -9.2 $\pm$ 1.6\\
o-H$_{2}$O & $4_{41}$ -- $3_{30}$ &  49.336& 1.82 $\pm$ 0.26 & 0.63 $\pm$ 0.14 & -17.1 $\pm$ 2.4& 48.1 $\pm$ 10.6\\
o-H$_{2}$O  &   $6_{34}$ -- $5_{23}$ &    49.390 &  & 0.99 $\pm$ 0.21 &  & 44.1 $\pm$ 9.3\\
p-H$_{2}$O   & $9_{28}$ -- $8_{17}$ &  50.634 &  & 0.19 $\pm$ 0.02 & & 35.4 $\pm$ 3.7\\
CO$^{(1)}$ & 52 -- 51  & 50.887 & & 0.71 $\pm$ 0.23 & & 1.1 $\pm$ 0.4\\
$[$OIII$]$  &     $^{3}$P$_{2}$ -- $^{3}$P$_{1}$ & 51.815 &  & 37.1 $\pm$ 0.50 &  & 7.2 $\pm$ 0.1 \\
p-H$_{2}$O   & $5_{33}$ -- $4_{22}$ &  53.137 & 0.24 $\pm$ 0.04 & 0.41  $\pm$ 0.03& -32.1 $\pm$ 5.4 & 36.5 $\pm$ 2.7\\
OH  & $^{2}\Pi_{1/2}$ -- $^{2}\Pi_{3/2}$ J= 3/2$^{+}$ -- 3/2$^{-}$ &   53.261 & 2.55 $\pm$ 0.32 & 1.10 $\pm$ 0.62& -20.7 $\pm$ 2.5 & 14.5 $\pm$ 8.1 \\
OH  & $^{2}\Pi_{1/2}$ -- $^{2}\Pi_{3/2}$ J= 3/2$^{-}$ -- 3/2$^{+}$  & 53.351 & 2.45 $\pm$ 0.26 & 0.26 $\pm$ 0.13 & -14.2 $\pm$ 1.5& 45.1 $\pm$ 19.9 \\
p-H2O   & $4_{31}$ -- $3_{22}$ &  56.324 & 1.09 $\pm$ 0.11 & 0.59 $\pm$ 0.25 & -5.1 $\pm$ 0.5& 44.9 $\pm$ 18.9 \\
$[$NIII$]$  &   $^{2}$P$_{3/2}$ -- $^{2}$P$_{1/2}$ & 57.329& &  5.97 $\pm$ 0.12 &  & 2.9 $\pm$ 0.1\\
p-H2O  & $4_{22}$ -- $3_{13}$ &  57.636& 2.15$^{a}$ $\pm$ 0.24 &  0.52 $\pm$ 0.19 & -15.5 $\pm$ 1.7& 48.9 $\pm$ 17.8\\
CO &  45 -- 44 &   58.547&  & 0.12 $\pm$ 0.037 &  & 6.7 $\pm$ 2.1\\
o-H$_{2}$O  & $4_{32}$ -- $3_{21}$ &   58.698 & 1.57 $\pm$ 0.13& 1.80 $\pm$ 0.14 & -29.9 $\pm$ 2.5& 37.3 $\pm$ 2.9\\
HDO & $3_{30}$ -- $2_{11}$  & 59.928 & & 0.15 $\pm$ 0.013 & & 37.5 $\pm$ 3.2\\
p-H$_{2}$O   & $4_{31}$ -- $4_{04}$ &  61.808 &  & 0.52 $\pm$ 0.02  & & 43.5 $\pm$ 1.7 \\
HDO & $10_{65}$ -- $10_{56}$  & 62.231 & & 0.02 $\pm$ 0.006 & & 4.5 $\pm$ 1.3\\
$[$OI$]$  &   $^{3}$P$_{1}$ -- $^{3}$P$_{2}$  & 63.184 &  & 65.33 $\pm$ 0.54 &  & 14.6 $\pm$ 0.1\\
p-H$_{2}$O & $8_{18}$ -- $7_{07}$ & 63.322 &  & 0.30 $\pm$ 0.12 &  &30.3 $\pm$ 12.1\\
o-H$_{2}$O & $8_{08}$ -- $7_{17}$   & 63.457 & &0.27 $\pm$ 0.045 & & 41.5 $\pm$ 6.9\\
OH   &   $^{2}\Pi_{3/2}$ J= 9/2$^{-}$ -- 7/2$^{+}$ &   65.132 & 0.44 $\pm$ 0.091&  0.84 $\pm$ 0.12 &-22.9 $\pm$ 4.7& 29.6 $\pm$ 4.2\\
o-H$_{2}$O   & $6_{25}$ -- $5_{14}$ & 65.166&  & 1.09 $\pm$ 0.13 & &26.4 $\pm$ 3.1 \\
OH  &  $^{2}\Pi_{3/2}$ J= 9/2$^{+}$ -- 7/2$^{-}$ &   65.279 & 0.39 $\pm$ 0.10 & 0.59 $\pm$ 0.20 & -15.6 $\pm$ 3.9& 34.9 $\pm$ 11.8\\
 o-H$_{2}$O & $7_{16}$ -- $6_{25}$ & 66.092 &  & 0.78 $\pm$ 0.16 & & 30.3 $\pm$ 6.2\\
o-H$_{2}$O  &  $3_{30}$ -- $2_{21}$  &   66.437& 1.51 $\pm$ 0.18 & 1.39 $\pm$ 0.28 &  -22.2 $\pm$ 2.6& 40.5 $\pm$ 8.1\\
p-H$_{2}$O  & $3_{31}$ -- $2_{20}$  &  67.089 & 1.40 $\pm$ 0.21 & 1.07 $\pm$ 0.20 & -12.7 $\pm$ 1.9& 42.5 $\pm$ 7.9 \\
H$_{2}^{18}$O  & $3_{30}$ -- $2_{21}$  &  67.191 & 0.19 $\pm$ 0.031 &   &  -12.9 $\pm$ 2.1&  \\
o-H$_{2}$O$^{(1)}$  &  $3_{30}$ -- $3_{03}$  &  67.269 & 0.33 $\pm$ 0.11 & 0.64 $\pm$ 0.11 & -22.5 $\pm$ 7.5& 32.5 $\pm$ 5.6\\
CO  & 39 -- 38 &   67.336 &  & 0.48 $\pm$ 0.14 & & 12.4 $\pm$ 3.6\\
CO  & 38 -- 37 &   69.074 & & 0.45 $\pm$ 0.14 &  & 15.5 $\pm$ 4.8\\
CO & 37 -- 36 &   70.907 &  & 0.70 $\pm$ 0.061 & & 10.1 $\pm$ 0.9\\
p-H$_{2}$O &   $5_{24}$ -- $4_{13}$ &  71.066 & 0.15 $\pm$ 0.039 & 0.89 $\pm$ 0.059 & -22.5 $\pm$ 5.8& 28.5 $\pm$ 1.8\\
OH  & $^{2}\Pi_{1/2}$ J= 7/2$^{-}$ -- 5/2$^{+}$ &    71.171 & & 0.33 $\pm$ 0.049 &  & 28.2 $\pm$ 4.2\\
OH   & $^{2}\Pi_{1/2}$ J= 7/2$^{+}$ -- 5/2$^{-}$ &    71.218 & & 0.78 $\pm$ 0.12 & & 27.9 $\pm$ 4.3\\
p-H$_{2}$O &  $7_{17}$ -- $6_{06}$ & 71.539 & & 1.39 $\pm$ 0.13 &  & 22.3 $\pm$ 2.1\\
o-H$_{2}$O  &  $7_{07}$ -- $6_{16}$ &  71.946 & & 1.42 $\pm$ 0.28 &  & 26.7 $\pm$ 5.2\\
CO  & 36 -- 35 &   72.843 & & 0.62 $\pm$ 0.13 & & 1.6 $\pm$ 0.3\\
HDO$^{(1)}$ & $9_{63}$ -- $10_{46}$ & 74.792 &  & 0.07 $\pm$ 0.02 & & 19.5 $\pm$ 5.6\\
CO & 35 -- 34 & 74.890 &  & 3.46 $\pm$ 0.40 &  & 0.1 $\pm$ 0.01\\
o-H$_{2}$O &$3_{21}$ -- $2_{12}$   & 75.380 & 0.88 $\pm$ 0.25 & 5.67 $\pm$ 0.21 & -31.5 $\pm$ 8.9 & 28.5 $\pm$ 1.1\\
o-H$_{2}$O &$5_{51}$ -- $5_{42}$   & 75.779 &  & 0.32 $\pm$ 0.04  & & 19.2 $\pm$ 2.4\\
H$_{2}^{18}$O &$3_{21}$ -- $2_{12}$ & 75.868 &  0.08  $\pm$ 0.01  & & -10.1 $\pm$ 0.6 & \\
o-H$_{2}$O &$5_{50}$ -- $5_{41}$ & 75.908 &  & 0.27 $\pm$ 0.01  &   &  20.1 $\pm$ 0.7 \\
CO  & 34 -- 33 &   77.059& & 0.60 $\pm$ 0.09 & & 4.9 $\pm$ 0.7\\
o-H$_{2}$O &  $4_{23}$ -- $3_{12}$ &   78.741 & & 6.18 $\pm$ 0.17 & & 26.9 $\pm$ 0.7\\
p-H$_{2}$O  & $6_{15}$ -- $5_{24}$ & 78.928 & & 0.43 $\pm$ 0.11 &  &16.5 $\pm$ 4.2\\
OH  &  $^{2}\Pi_{3/2}$ J= 1/2$^{-}$ -- 3/2$^{+}$ &  79.118 & & 3.89 $\pm$ 0.16 &  & 16.4 $\pm$ 0.7\\
OH  &  $^{2}\Pi_{3/2}$ J= 1/2$^{+}$ -- 3/2$^{-}$ &   79.182 & & 3.73 $\pm$ 0.38 &  & 19.1 $\pm$ 1.9\\
CO  &   33 -- 32 &   79.359 & & 2.32 $\pm$ 0.20 & & 3.7  $\pm$ 0.3\\
CO &    32 -- 31 &  81.806 & & 1.52 $\pm$ 0.09 & & 10.8 $\pm$ 0.6\\
o-H$_{2}$O &  $6_{16}$ -- $5_{05}$ &  82.030 & & 2.78 $\pm$ 0.13 & & 22.4 $\pm$ 1.1\\
o-H$_{2}$O &  $8_{36}$ -- $8_{27}$ &  82.974 & & 0.27 $\pm$ 0.06 & & 15.1 $\pm$ 3.2 \\
p-H$_{2}$O &  $6_{06}$ -- $5_{15}$ &  83.283 & & 1.29 $\pm$ 0.08 & & 31.1 $\pm$ 1.8\\
OH     &  $^{2}\Pi_{3/2}$ J= 7/2$^{+}$ -- 5/2$^{-}$ &    84.420 & & 1.72 $\pm$ 0.13 &  & 30.8 $\pm$ 2.3\\
o-H$_{2}$O$^{(1)}$ & $11_{48}$ -- $10_{55}$  & 84.514 &  & 0.15  $\pm$ 0.10 & & 2.1 $\pm$ 1.4\\
OH     &   $^{2}\Pi_{3/2}$ J= 7/2$^{-}$ -- 5/2$^{+}$ &   84.597 & 1.13 $\pm$ 0.38& 1.34 $\pm$ 0.26 &-21.4 $\pm$ 7.9& 36.2 $\pm$ 7.1 \\
o-H$_{2}$O$^{(2)}$ &   $7_{16}$ -- $7_{07}$ & 84.766 &  & 0.47 $\pm$ 0.05 & & 19.2 $\pm$ 1.9\\
H$_{2}^{17}$O$^{(1,2)}$ &   $7_{16}$ -- $7_{07}$ & 84.778 &  &   & &  \\
CO  &    30 -- 29 &   87.190 & & 1.58 $\pm$ 0.11 &  & 2.8 $\pm$ 0.2\\
$[$OIII$]$ &  $^{3}$P$_{1}$ -- $^{3}$P$_{0}$ & 88.356 & & 13.2 $\pm$ 0.29 &  & 9.8 $\pm$ 0.1\\
o-H$_{2}$O &   $3_{22}$ -- $2_{11}$ & 89.988 &  & 2.17 $\pm$ 0.17 & & 29.1 $\pm$ 2.3\\
CO & 29 -- 28 & 90.163 & & 1.58 $\pm$ 0.14 &  & 5.1 $\pm$ 0.4\\
p-H$_{2}$O    & $11_{57}$ -- $10_{64}$ &   90.643 & & 0.16 $\pm$ 0.03 & & 42.5 $\pm$ 8.0 \\
o-H$_{2}$O &   $6_{43}$ -- $6_{34}$ & 92.811 & & 0.49 $\pm$ 0.06 &  & 8.9 $\pm$ 1.1\\
CO & 28 -- 27 & 93.349 &  & 1.89 $\pm$ 0.11 & & 10.8 $\pm$ 0.6\\
o-H$_{2}$O & $7_{35}$ -- $7_{26}$  & 93.379 &  & 0.36  $\pm$ 0.05 & & 20.5 $\pm$ 2.9\\
p-H$_{2}$O    & $5_{42}$ -- $5_{33}$ &   94.206 & & 0.62 $\pm$ 0.02 & & 14.6 $\pm$ 0.5\\
o-H$_{2}$O &   $6_{25}$ -- $6_{16}$ & 94.643 &  & 0.52  $\pm$ 0.06 &  & 22.7 $\pm$ 0.3\\
o-H$_{2}$O &  $4_{41}$ -- $4_{32}$ & 94.703 &  & 0.72  $\pm$ 0.06 & & 27.1 $\pm$ 2.2\\
p-H$_{2}$O &  $5_{15}$ -- $4_{04}$ & 95.626 &  & 1.42  $\pm$ 0.10 & & 27.4 $\pm$ 1.3\\
p-H$_{2}$O &  $4_{41}$ -- $4_{32}$ & 95.883 &  & 0.67  $\pm$ 0.06 & & 29.2 $\pm$ 2.7\\
OH     &  $^{2}\Pi_{3/2}$ -- $^{2}\Pi_{1/2}$ J= 3/2$^{+}$ -- 5/2$^{-}$ &   96.314 &  & 0.88  $\pm$ 0.05 & & 10.5 $\pm$ 0.6  \\
OH     &  $^{2}\Pi_{3/2}$ -- $^{2}\Pi_{1/2}$ J= 3/2$^{-}$ -- 5/2$^{+}$ &  96.368 &  & 0.74  $\pm$ 0.03 & & 13.4 $\pm$ 0.6\\
CO   & 27 -- 26 &   96.773 &  & 3.82  $\pm$ 0.11 & & 9.2 $\pm$ 0.3\\
OH   &  $^{2}\Pi_{1/2}$ J= 5/2$^{-}$ -- 3/2$^{+}$ &  98.725  & &  3.61  $\pm$ 0.33 & & 38.8 $\pm$ 3.5\\
OH$^{(2)}$   &  $^{2}\Pi_{1/2}$ J= 5/2$^{+}$ -- 3/2$^{-}$ &  98.737  &  & & & 15.7 $\pm$ 3.5\\
o-H$_{2}$O &   $5_{05}$ -- $4_{14}$ & 99.492 &   & 5.61  $\pm$ 0.12 &  & 22.2 $\pm$ 0.5\\
H$_{2}^{18}$O$^{(1)}$ &   $5_{05}$ -- $4_{14}$ & 99.784 &   & 0.03 $\pm$0.01  &  & 24.6 $\pm$ 0.8\\
CO  & 26 -- 25 & 100.461 &  & 3.44  $\pm$ 0.06 &  & 11.5 $\pm$ 0.2\\
p-H$_{3}$O$^{+}$$^{(1)}$ & 2$_{1}^{-}$ -- 1$_{1}^{+}$ & 100.577 &
& 0.03
$\pm$ 0.01 & \\
o-H$_{3}$O$^{+}$$^{(1)}$ & 2$_{0}^{-}$ -- 1$_{0}^{+}$ & 100.869 &
& 0.09
$\pm$ 0.05 &  \\
o-H$_{2}$O & $5_{14}$ -- $4_{23}$ & 100.913 & &  3.05  $\pm$ 0.17 & & 21.1 $\pm$ 1.2 \\
p-H$_{2}$O & $2_{20}-1_{11}$ & 100.983 & &  2.35  $\pm$ 0.09 & & 29.9 $\pm$ 1.2\\
p-H$_{2}$O & $6_{24}$ -- $5_{33}$ & 101.210 & &  0.19  $\pm$ 0.04 &  & 14.8 $\pm$ 2.9\\
H$_{2}^{18}$O & $2_{20}$ -- $1_{11}$ & 102.008 &  0.22   $\pm$ 0.03& & -11.7 $\pm$ 1.3& \\
p-H$_{2}$O & $6_{42}$ -- $6_{33}$ & 103.916 &  & 0.08  $\pm$ 0.02 &  & 19.4 $\pm$ 5.6\\
p-H$_{2}$O & $6_{15}$ -- $6_{06}$ & 103.939 &  & 0.11  $\pm$ 0.02 &  &18.5 $\pm$ 3.7\\
o-H$_{2}$O & $6_{34}$ -- $6_{25}$ & 104.090 &  & 0.61  $\pm$ 0.11 & & 24.2 $\pm$ 4.3 \\
CO & 25 -- 24 & 104.445 &  &3.67  $\pm$ 0.10 & & 8.32 $\pm$ 0.2 \\
o-H$_{2}$O &  $2_{21}$ -- $1_{10}$ & 108.073 & & 3.22  $\pm$ 0.08 & & 29.1 $\pm$ 0.7\\
CO  & 24 -- 23 & 108.763 & & 3.28  $\pm$ 0.12 & & 9.7 $\pm$ 0.4\\
H$_{2}^{18}$O &  $2_{21}$ -- $1_{10}$ & 109.350 &0.08 $\pm$ 0.02 &  0.18 $\pm$ 0.013 &-8.1 $\pm$ 1.7& 48.9 $\pm$ 3.5\\
p-H$_{2}$O & $5_{24}$ -- $5_{15}$ & 111.626 &  & 0.43  $\pm$ 0.03 & & 26.7 $\pm$ 1.7\\
o-H$_{2}$O & $7_{43}$ -- $7_{34}$ & 112.511 &  & 0.09  $\pm$ 0.02 & & 11.6 $\pm$ 2.7\\
CO & 23 -- 22 & 113.458 &  & 4.82  $\pm$ 0.15 & & 10.4 $\pm$ 6.3\\
o-H$_{2}$O & $4_{14}$ -- $3_{03}$ & 113.537 &  & 5.10  $\pm$ 0.13 & & 22.6 $\pm$ 0.6\\
p-H$_{2}$O &  $5_{33}$ -- $5_{24}$ & 113.944 &  & 1.38  $\pm$ 0.06 & & 27.6 $\pm$1.3\\
H$_{2}^{18}$O &  $4_{14}$ -- $3_{03}$ & 114.297 &  & 0.16  $\pm$  0.01& & 44.6 $\pm$ 1.7\\
OH    & $^{2}\Pi_{3/2}$ -- $^{2}\Pi_{1/2}$ J= 7/2$^{-}$ -- 5/2$^{+}$ & 115.153 & & 0.18  $\pm$ 0.01 &  & 20.2 $\pm$ 1.6\\
CO  & 22 -- 21 & 118.581 &  & 6.58  $\pm$ 0.09 & & 12.5 $\pm$ 0.2\\
OH    & $^{2}\Pi_{3/2}$ J= 5/2$^{-}$ -- 3/2$^{+}$ & 119.234 &0.39 $\pm$ 0.07   & 1.76  $\pm$ 0.17 &-45.5 $\pm$ 8.4& 33.9 $\pm$ 3.3\\
OH    & $^{2}\Pi_{3/2}$ J= 5/2$^{+}$ -- 3/2$^{-}$ & 119.442 &  & 2.42  $\pm$ 0.15 & & 35.5 $\pm$ 2.2\\
$^{18}$OH$^{(1)}$    &$^{2}\Pi_{3/2}$ J= 5/2$^{+}$ -- 3/2$^{-}$  & 119.966 & 0.11 $\pm$ 0.04&  0.10 $\pm$ 0.025  & -25.8 $\pm$ 9.1&  51.4 $\pm$ 12.8\\
$^{18}$OH$^{(1)}$    &$^{2}\Pi_{3/2}$ J= 5/2$^{-}$ -- 3/2$^{+}$  & 120.172& 0.03 $\pm$ 0.01&0.031   $\pm$0.005  & -5.1 $\pm$ 0.9 & 28.9 $\pm$ 4.6\\
o-H$_{2}$O &$4_{32}$ -- $4_{23}$ & 121.719 &  & 2.28  $\pm$ 0.13 & & 28.1 $\pm$ 1.6\\
$[$NII$]$$^{(1)}$ & $^{3}$P$_{1}$ -- $^{3}$P$_{2}$ & 121.890   & & 0.93 $\pm$ 0.47 &  &\\
CO &  21 -- 20 & 124.193 &  & 7.89 $\pm$ 0.58 &  & 15.0 $\pm$ 1.1\\
o-NH$_{3}$ & 4$_{3}^{-}$ -- 3$_{3}^{+}$  & 124.647& & 0.08 $\pm$ 0.01 & & 40.8 $\pm$ 2.3\\
p-NH$_{3}$$^{(1)}$ & 4$_{2}^{-}$ -- 3$_{2}^{+}$  & 124.796 &  & 0.03 $\pm$ 0.01 & & 11.5 $\pm$ 0.5\\
o-NH$_{3}$ & 4$_{0}^{-}$ -- 3$_{0}^{+}$  & 124.913 & & 0.05 $\pm$ 0.01& & 8.6 $\pm$ 0.6\\
HDO & $5_{33}$ -- $5_{24}$   & 124.954& & 0.02 $\pm$ 0.004& & 9.1 $\pm$ 1.9\\
p-H$_{2}$O & $4_{04}$ -- $3_{13}$ & 125.354   & & 2.25  $\pm$ 0.09  & & 25.5 $\pm$ 1.1 \\
p-H$_{2}$O & $3_{31}$ -- $3_{22}$ & 126.713   &  & 0.62  $\pm$ 0.08   & & 30.5 $\pm$ 4.1\\
 o-H$_{2}$O &  $7_{25}$ -- $7_{16}$ & 127.882   & & 0.24  $\pm$ 0.03  & &26.4 $\pm$ 3.1\\
13CO$^{(1)}$ & 21 -- 20 & 129.891 & & 0.02 $\pm$ 0.01 & & 10.3 $\pm$ 3.9\\
CO & 20 -- 19 & 130.369 & & 6.19  $\pm$ 0.22 & & 16.1 $\pm$ 0.6\\
o-H$_{2}$O & $5_{14}$ -- $5_{05}$ & 134.935 & & 0.86  $\pm$ 0.05 &  & 25.4 $\pm$ 1.6\\
HDO & $4_{31}$ -- $4_{22}$  & 135.425 & & 0.08 $\pm$ 0.02 & & 17.5 $\pm$ 4.1 \\
o-H$_{2}$O & $3_{30}$ -- $3_{21}$ & 136.494 & & 0.71  $\pm$ 0.05 & & 30.9 $\pm$ 2.3\\
CO  & 19 -- 18 & 137.196 & & 5.22   $\pm$ 0.05 &  & 18.4 $\pm$ 0.2\\
p-H$_{2}$O &  $3_{13}$ -- $2_{02}$ &  138.527 & & 2.67  $\pm$ 0.03 & & 26.4 $\pm$ 0.3\\
p-H$_{2}$O & $4_{13}$ -- $3_{22}$ & 144.517 & & 1.15  $\pm$ 0.07 &  & 18.4 $\pm$ 0.2\\
CO  & 18 -- 17 & 144.784 & & 3.98  $\pm$ 0.26 &  & 15.5 $\pm$ 1.1\\
$[$OI$]$ & $^{3}$P$_{0}$ -- $^{3}$P$_{1}$ & 145.525 & & 1.17  $\pm$ 0.19 & & 28.5 $\pm$ 4.6\\
p-H$_{2}$O & $4_{31}$ -- $4_{22}$ & 146.919 & & 1.14  $\pm$ 0.06 & & 22.1 $\pm$ 1.1\\
CO & 17 -- 16 &  153.267 & & 8.66  $\pm$ 0.13 &  & 14.6 $\pm$ 0.2\\
p-H$_{2}$O & $3_{22}$ -- $3_{13}$ & 156.193 & & 1.23  $\pm$ 0.08 & & 32.5 $\pm$ 2.2 \\
o-H$_{2}$O & $5_{23}$ -- $4_{32}$ & 156.266 & & 1.09  $\pm$ 0.22 & & 16.7 $\pm$ 3.4 \\
$[$CII$]$ & $^{2}$P$_{3/2}$ -- $^{2}$P$_{1/2}$ & 157.741 &  & 2.23  $\pm$ 0.06 &  & 14.9 $\pm$ 0.4\\
p-H$_{2}$O & $5_{32}$ -- $5_{23}$ & 160.504 & & 0.45  $\pm$ 0.06 &  & 24.9 $\pm$ 3.2\\
CO &  16 -- 15 & 162.812 &  & 9.48  $\pm$ 0.26 &  & 12.7 $\pm$ 0.4\\
\onecolumn
OH & $^{2}\Pi_{3/2}$ J=3/2$^{+}$ -- 1/2$^{-}$ & 163.131 & & 0.81  $\pm$ 0.19 & & 17.4 $\pm$ 4.1\\
OH & $^{2}\Pi_{3/2}$ J=3/2$^{-}$ -- 1/2$^{+}$ & 163.397 &  & 1.66  $\pm$ 0.38 & & 17.7 $\pm$ 4.1 \\
p-NH$_{3}$ & 3$_{2}^{-}$ -- 2$_{2}^{+}$  & 165.596 & & 0.11 $\pm$ 0.02 & & 14.4 $\pm$ 2.3\\
p-H$_{2}$O & $6_{33}$ -- $6_{24}$ & 170.131 & & 0.27   $\pm$ 0.01 & & 43.5 $\pm$ 1.3\\
CO$^{(3)}$ &  15 -- 14 & 173.631 &  & 9.80 $\pm$ 0.69  &  &  \\
H$_{2}^{18}$O & $3_{03}$ -- $2_{12}$ & 174.379 & &1.01  $\pm$ 0.06  & &3.2 $\pm$ 1.1\\
H$_{2}^{17}$O$^{(1)}$ & $3_{03}$ -- $2_{12}$ & 174.491 & & 0.01 $\pm$ 0.004    & & 4.8 $\pm$ 1.9 \\
o-H$_{2}$O & $3_{03}$ -- $2_{12}$ & 174.626 & & 2.38 $\pm$ 0.19  & & 20.1 $\pm$ 1.6\\
o-H$_{2}$O & $4_{32}$ -- $5_{05}$  & 174.914 & & 0.16 $\pm$ 0.01   & & 23.5 $\pm$ 1.3\\
o-H$_{2}$O& $2_{12}$ -- $1_{01}$ & 179.527 & & 2.55 $\pm$ 0.31  & & 23.8 $\pm$ 2.9\\
o-H$_{2}$O & $2_{21}$ -- $2_{12}$ & 180.488 & & 1.03 $\pm$ 0.11  & & 23.8 $\pm$ 2.5\\
p-H$_{3}$O$^{+}$$^{(2)}$ & 1$_{1}^{-}$ -- 1$_{1}^{+}$ & 181.054 &
& 0.05
$\pm$ 0.01 & & 11.6 $\pm$ 0.3 \\
H$_{2}^{18}$O$^{(2)}$ & $2_{21}$ -- $1_{01}$ & 181.053 &  &    & &   \\
CO$^{(3)}$$^{(2,3)}$ &  14 -- 13 & 185.999 &  &7.58  $\pm$ 0.22  & &  \\
p-H$_{2}$O$^{(2,3)}$ & $4_{13}$ -- $4_{04}$ & 187.110 & &    & &   \\

\end{longtable}
\end{center}
\twocolumn
 These lines were detected by the Short Wavelength
Spectrometer (SWS) towards IRc2 and the inferred [S~{\sc iii}]
33.5/18.7 $\mu$m ratio was found to be $\sim$ 0.55 (van Dishoeck
et al. 1998). This ratio gives an $N_{e}$(S~{\sc iii}) of
$\approx$ 1950 cm$^{-3}$, which exceeds $N_{e}$(O~{\sc
iii}) by a factor of two, probably due to the higher critical densities of the [S~{\sc iii}] lines.\\
\begin{figure}
 \centering
    \includegraphics[width=9cm]{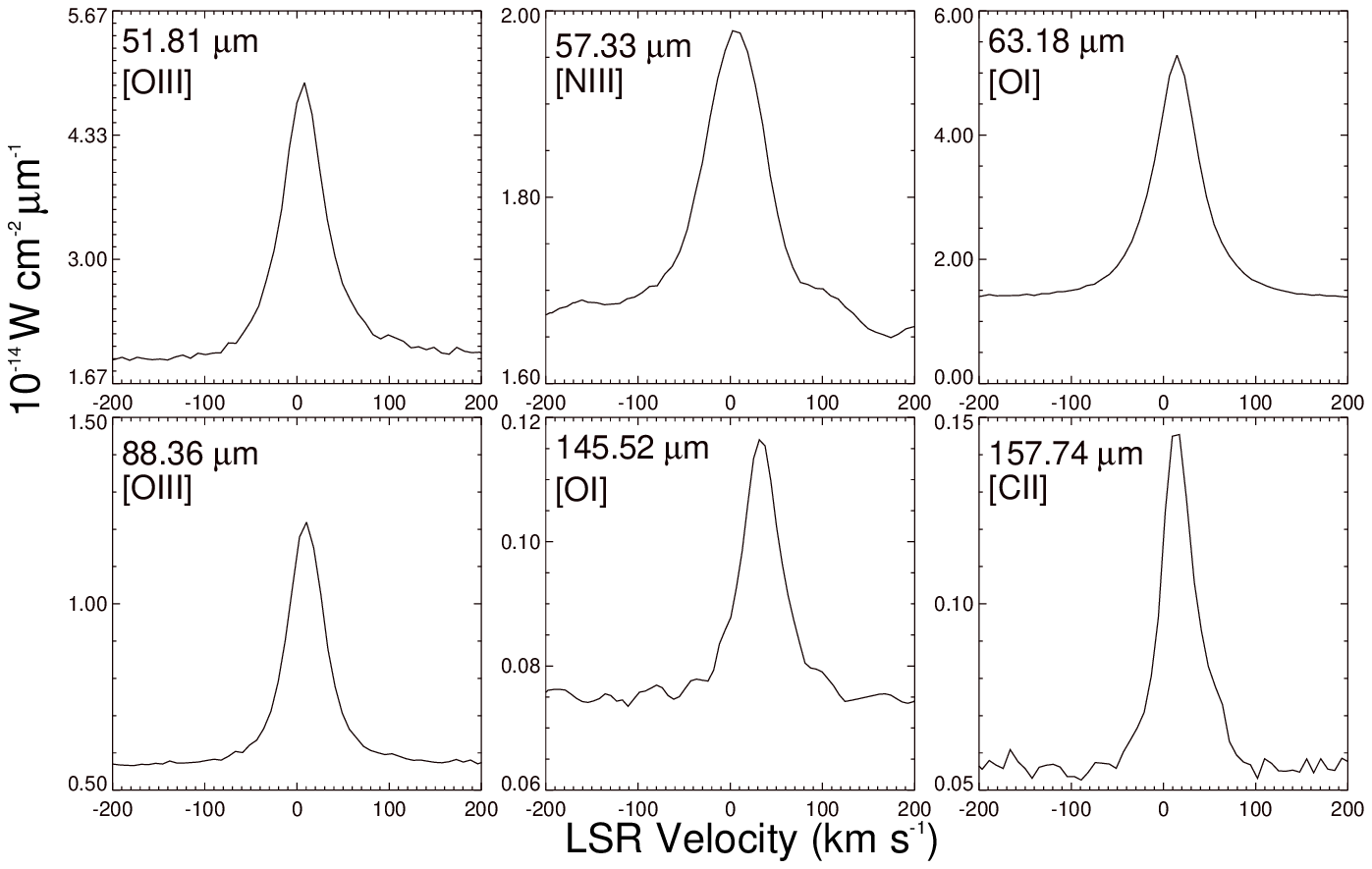}
    \caption{Transitions of  [O~{\sc iii}], [N~{\sc iii}], [O~{\sc i}] and [C~{\sc ii}] detected
    by the {\em ISO} LWS spectral survey towards Orion KL.
Note that the LSR velocity of the line peaks (except for the
[O~{\sc i}] line at 145.5 $\mu$m) is $\approx$ 10 km s$^{-1}$, in
agreement with the velocity of the quiescent gas
     ($\approx$ 9 km  s$^{-1}$). }
         \label{finas}
 \end{figure}
\begin{table*}
\caption{Detected fine-structure lines}             
\label{atomic}      
\centering                          
\begin{tabular}{c c c c}        
\hline\hline
Species  &  Transition  &  Rest Wavelength ($\mu$m) &  Flux (W cm$^{-2}$) \\    
\hline                        

 [O~{\sc iii}]  & $^{3}$P$_{0}$ -- $^{3}$P$_{1}$ & 51.814 & (3.71 $\pm$ 0.005)$\times$10$^{-16}$ \\

 [N~{\sc iii}]  & $^{2}$P$_{3/2}$ -- $^{2}$P$_{1/2}$ & 57.329  & (5.97 $\pm$ 0.12)$\times$10$^{-17}$ \\

 [O~{\sc i}]  & $^{3}$P$_{1}$ -- $^{3}$P$_{2}$ & 63.184 & (6.53 $\pm$ 0.05)$\times$10$^{-16}$ \\

 [O~{\sc iii}] & $^{3}$P$_{1}$ -- $^{3}$P$_{0}$ & 88.356 &  (1.52 $\pm$ 0.03)$\times$10$^{-16}$ \\

 [N~{\sc ii}] & $^{3}$P$_{1}$ -- $^{3}$P$_{2}$  & 121.889 & (9.3 $\pm$ 4.7)$\times$10$^{-19}$  \\

 [O~{\sc i}] & $^{3}$P$_{1}$ -- $^{3}$P$_{0}$ & 145.525 & (1.17 $\pm$ 0.19)$\times$10$^{-18}$ \\

 [C~{\sc ii}] & $^{2}$P$_{3/2}$ -- $^{2}$P$_{1/2}$ & 157.741 & (2.23 $\pm$ 0.06)$\times$10$^{-19}$ \\

\hline
\end{tabular}
\end{table*}

\begin{figure}
\centering
    \includegraphics{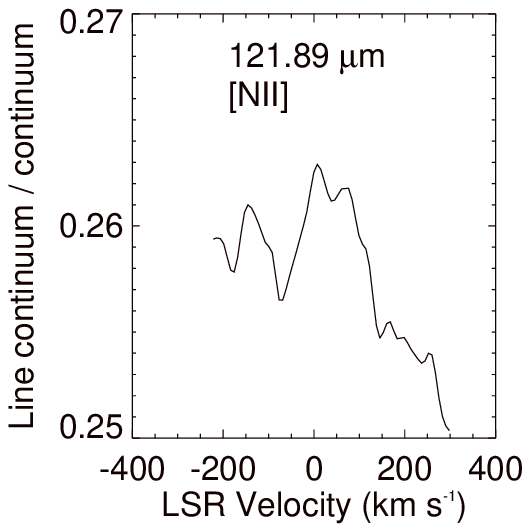}
    \caption{Tentative detection of the 122 $\mu$m line of [N~{\sc ii}].
    The broadness is due to hyperfine components in the
transition as can be seen in the detection of the [N~{\sc ii}] 122
$\mu$m line in Sgr B2 (Polehampton et al. 2006 in prep.)}
         \label{nitro}
 \end{figure}

\subsection{[O~{\sc i}] and [C~{\sc ii}]}
The [C~{\sc ii}] 157.7 $\mu$m and [O~{\sc i}] 63.2, 145.5 $\mu$m
lines are important coolants in photodissociation regions (PDRs),
whose heating is thought to be dominated by energetic
photoelectrons ejected from dust grains following FUV photon
absorption (Tielens \& Hollenbach 1985). As important coolants,
these lines can also originate
in shocks driven by jets and outflows from young stellar objects (Hollenbach \& McKee 1989).\\
The Orion nebula is an H~{\sc ii} region which is ionised by the
Trapezium stars. The luminosity of the ionising stars is about
10$^{5}$L$_{\odot}$, located approximately 0.15pc from the
molecular cloud leading to an incident FUV flux of about G
$\approx$ 10$^5$ G$_o$ (Tielens \& Hollenbach 1985). The fraction
of the FUV flux converted into line emission in these transitions
is generally of the order 10$^{-2}$ -- 10$^{-3}$ (Tielens \&
Hollenbach 1985) and is a function of both the gas temperature and
the
electron density. \\
The [C~{\sc ii}] 157.7 /[O~{\sc i}] 63.2 $\mu$m ratio in Orion KL
inferred from our observations is 3.42 $\times$ 10$^{-2}$, in good
agreement with Tielens \& Hollenbach's predictions. The [O~{\sc
i}] 63.2/145.5 $\mu$m ratio combined with the [O~{\sc i}] 63.2
$\mu$m/[C~{\sc ii}] 157.7 $\mu$m flux ratio can also be used to
derive the temperature and density of PDRs (Watson 1983). Using
the PDR temperature density plane defined in Liu et al. 2001, we
inferred a temperature of 300 K and a density of log $N_{H}$=5.4
(cm$^{-3}$). Considering this, it seems natural to postulate that
the lines originate in the PDR region. However, Orion KL could
also be an important shock region (Chernoff, Hollenbach \& McKee
1982) and a potential major contribution to [O~{\sc i}] emission
from dissociated
shocked gas should be taken into account.\\
Table ~\ref{compara_tielens} lists the surface brightnesses
measured with the 80$^{\prime\prime}$ LWS beam compared with those
observed towards $\theta^{1}$ Ori C (column 3) by the {\em KAO}
with a $\approx$ 45$^{\prime\prime}$ beam (Melnick, Gull, \&
Harwit 1979; Ellis \& Werner 1985; Werner et al. 1984; Stacey,
Smyers, Kurtz \& Harwit 1983; Russell et al. 1980; Phillips \&
Huggins 1981). Column 5 lists predicted line surface brightnesses
from the photodissociation model of Tielens et al. (1985) and
column 6 lists the predicted line surface brightnesses from the
dissociative shock
model of Hollenbach \& McKee (1989) (for $n_{0}$ = 10$^{5}$ cm$^{-3}$ and v = 30 -- 80 km s$^{-1}$).\\
We found that although our [O~{\sc i}] 63.18 $\mu$m and [O~{\sc
i}] 145.52 $\mu$m line intensities are in good agreement with the
shock model prediction, the observed [C~{\sc ii}] line intensity
is much higher than predicted by shock models. A PDR model can
reproduce the {\em ISO} LWS [O~{\sc i}] 63.2 $\mu$m and [C~{\sc
ii}] surface brightness levels within 35\%, although
over-predicting the LWS [O~{\sc i}] 145.5 $\mu$m emission by a
factor of 2.7. We note, however, that for the smaller beam {\em
KAO}
observations the PDR model can match the observed surface brightness levels for all three lines.\\

\begin{table*}

\centering                          
\begin{tabular}{c c c c c c}        
\hline
\hline
Species  &  Wavelength ($\mu$m) & {\em KAO} at $\theta^{1}$ Ori C  &  {\em ISO} LWS BN/KL &  PDR model & Shock model \\
 &  & (ergs cm$^{-2}$ s$^{-1}$ sr$^{-1}$) & (ergs cm$^{-2}$ s$^{-1}$ sr$^{-1}$) & (ergs cm$^{-2}$ s$^{-1}$
 sr$^{-1}$) & (ergs cm$^{-2}$ s$^{-1}$ sr$^{-1}$)\\
\hline                        

 [O~{\sc i}]  & 63.18 & 4--6$\times10^{-2}$ & 5.5$\times10^{-2}$ & 4.6$\times10^{-2}$  & 6.0$\times10^{-2}$\\

[O~{\sc i}] & 145.52 & 3--6$\times10^{-3}$ & 1.3$\times10^{-3}$  & 3.5$\times10^{-3}$ & 1.2$\times10^{-3}$\\

 [C~{\sc ii}] & 157.74 & 4--7$\times10^{-3}$ & 2.5$\times10^{-3}$  & 3.8$\times10^{-3}$ & $<$ 1$\times10^{-4}$ \\

\hline
\end{tabular}
 \caption{Comparison of the observed line intensities towards Orion KL
with the PDR model calculations of Tielens et al. (1984) and the
shock model of Hollenbach et al. (1989).}
\label{compara_tielens}             
\end{table*}

\subsection{Molecular species}
\subsubsection{Water lines}
High velocity gas was first detected at the centre of the Orion-KL
region as broad wings on `thermal' molecular lines in the
millimeter range and as high velocity maser features in the 22 GHz
line of H$_{2}$O (Genzel et al. 1981). These high velocity motions
may be caused by mass outflows from newly formed stars. Many
theoretical studies of the Orion region (e.g. Draine \& Roberge
1982; Chernoff et all. 1982; Neufeld \& Melnick 1987) have
concluded that the rich emission spectrum from thermally excited
water vapour should play a
significant role in cooling the gas. \\
 In conditions where
the temperature exceeds $\sim$ 400 K, OH and H$_{2}$O are rapidly
formed through the following reactions (Elitzur \& Watson 1978):
\begin{equation}
O + H_{2} \rightarrow OH + H
\end{equation}
\begin{equation}
 OH + H_{2} \rightarrow H_{2}O + H
\end{equation}

 At higher temperatures, OH is also formed via molecular oxygen
destruction:
\begin{equation}
O_{2} + H_{2} \rightarrow OH + OH
\end{equation}
eventually being processed into water again by reaction (2).
Consequently, OH and H$_{2}$O fractional abundances increase in
shocks and so does the overall rate of reaction. \\
Molecules and atoms behind a shock front act as cooling agents by
emitting infrared and microwave photons. Collisions between
molecules at high temperature populate the vibrational and
rotational levels of the molecules.
 Following these
reactions, rotational and vibrational water transitions are
predicted to occur, providing a ubiquitous tracer of shock-heated
gas.\\

\begin{figure*}
    \centering
    \includegraphics[width=14cm,height=22cm]{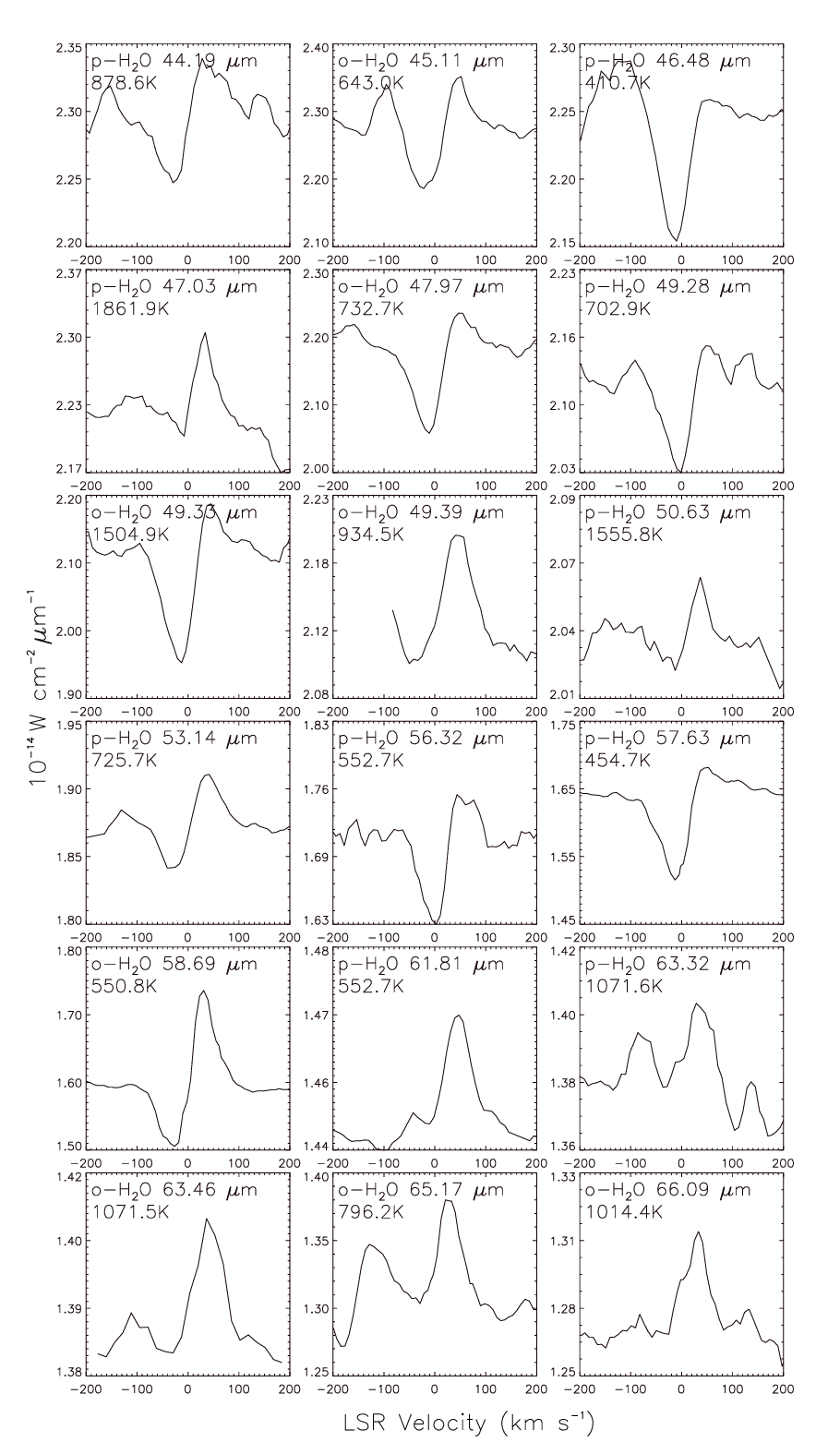}\\

    \caption{Water lines detected by the survey towards Orion KL}
    \label{agua1}
   \end{figure*}
   \addtocounter{figure}{-1}
\begin{figure*}
\centering
    \includegraphics[width=14cm,height=22cm]{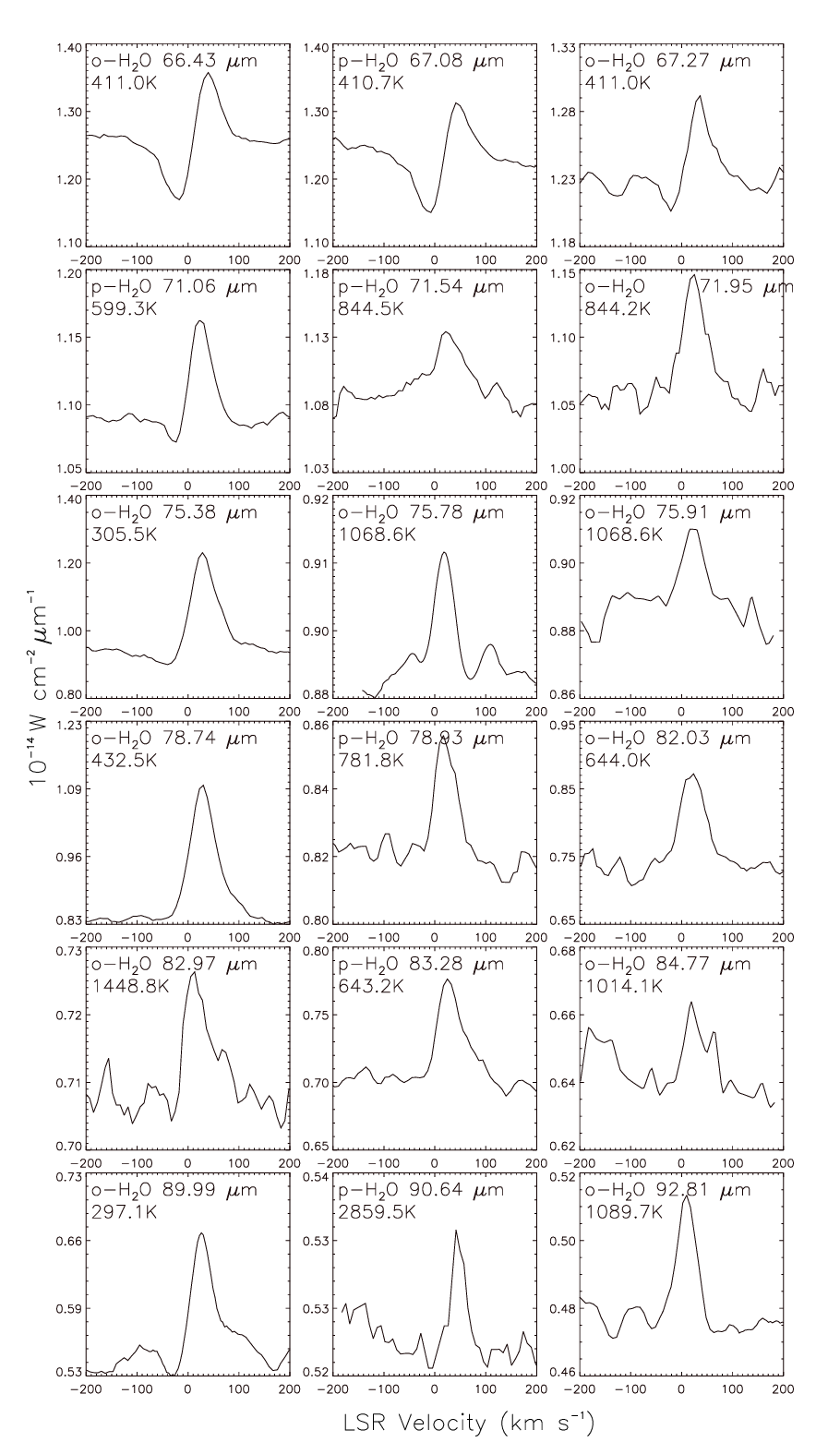}\\

         \caption{Continued}
         \label{agua2}
   \end{figure*}
   \addtocounter{figure}{-1}
\begin{figure*}
    \centering
    \includegraphics[width=14cm,height=22cm]{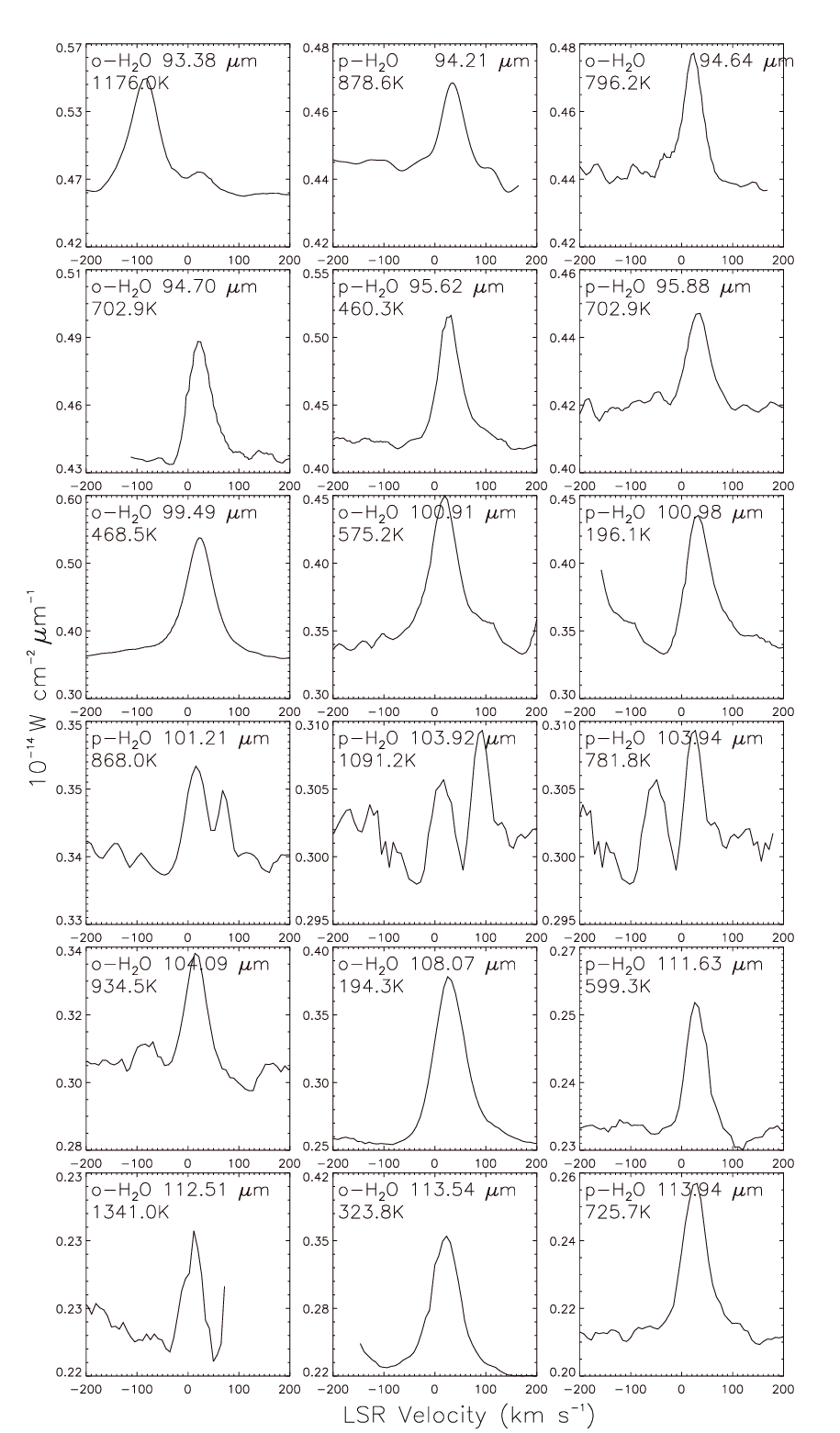}\\

     \caption{Continued}
    \label{agua3}
   \end{figure*}
   \addtocounter{figure}{-1}
\begin{figure*}
    \centering

    \includegraphics[width=14cm,height=22cm]{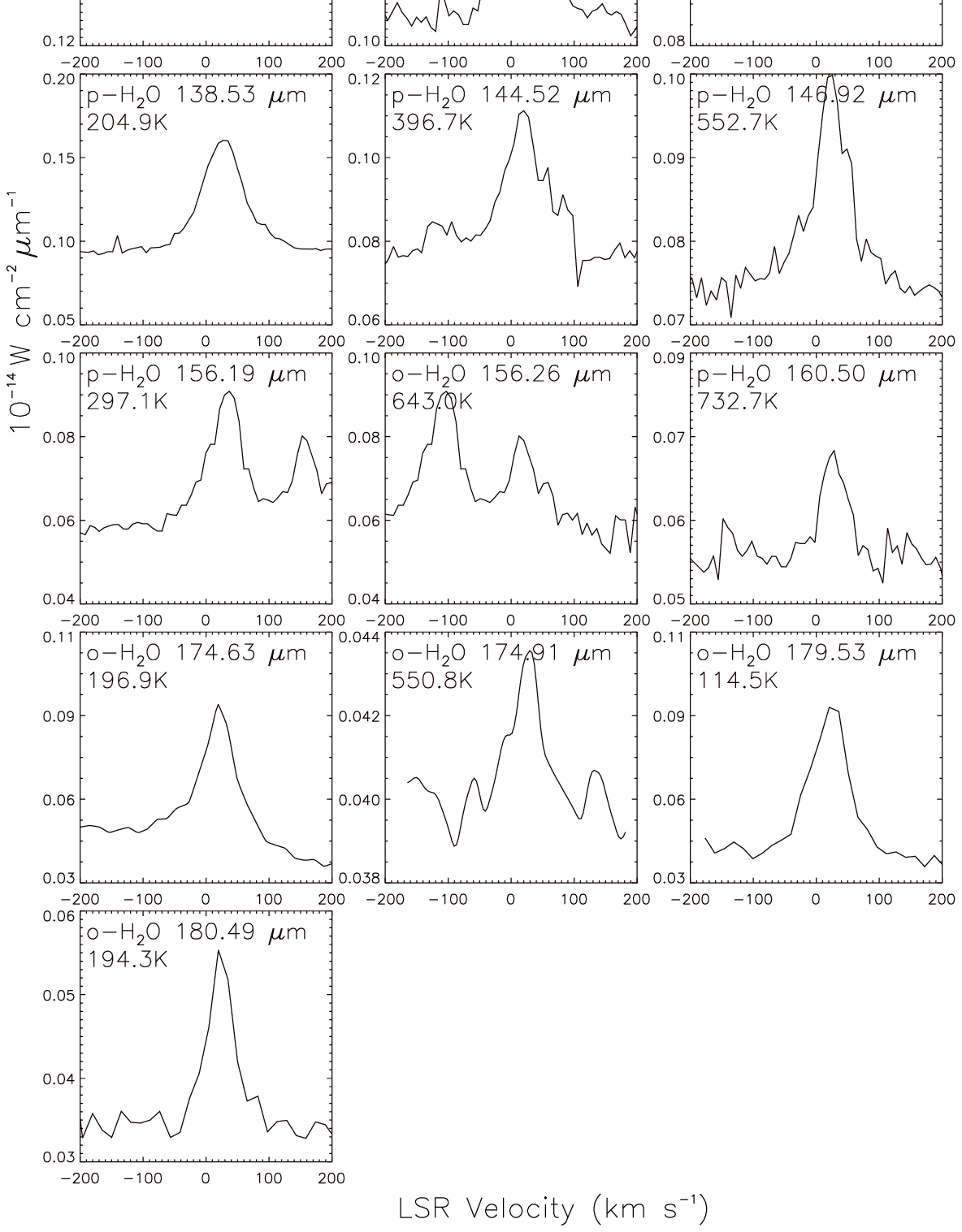}
    \caption{Continued}
    \label{agua3}
   \end{figure*}

The widespread nature of the water vapour around IRc2 has been
probed with maps at 183 GHz (Cernicharo et al. 1990, 1994,
Cernicharo \& Crovisier, 2005), the first time that its abundance
was estimated in the different large-scale components of Orion
IRc2. Harwit et al. (1998) analysed 8 lines of water observed by
LWS FP in L04 mode, concluding that these lines arise from a
molecular cloud subjected to a magnetohydrodynamic C-type shock.
From their modelling, they derived an H$_{2}$O/H abundance of 5
$\times$ 10$^{-4}$. However, the interpretation of these lines in
the $\approx$ 80$^{\prime\prime}$ LWS beam and the determination
of the water abundance in the different components of Orion
remains a long standing problem. This is basically due to two main
issues: the complexity of the different dynamical and chemical
processes that take place within the region encompassed by the LWS
beam, including outflows and several gas components, and the need
for new H$_{2}$O collisional rates appropriate for the
temperatures
prevailing in shocks. \\
Water lines appear in the survey as resolved (typically 70 km
s$^{-1}$ FWHM) with a total of 70 detected lines. The line
profiles range from predominantly P-Cygni at shorter wavelengths
to predominantly pure emission at longer wavelengths (see
Figure~\ref{agua1}). Radial velocities appear to be centred at
$\approx$ -- 15 km s$^{-1}$  in all absorption lines (shorter
wavelengths) consistent with the results found in the SWS range
(from 2 $\mu$m to 45 $\mu$m, van Dishoeck et al. 1998; Wright et
al. 2000). However, the pure emission lines of H$_{2}$O peak at
around $\approx$ +30 km s$^{-1}$ (see Section 4.4), where the
velocity of the quiescent gas is 9 km s$^{-1}$ (Cohen et al.
2006). The same behaviour is found for OH lines, where this has
been interpreted as evidence of an outflow at a velocity of $\geq$
25 $\pm$ 5 km s$^{-1}$ (see discussion in Section 4.3.3; also
Goicoechea et al. 2006).

\subsubsection{Water isotopes}
\begin{figure}
    \centering
    \includegraphics[width=9cm,height=6cm]{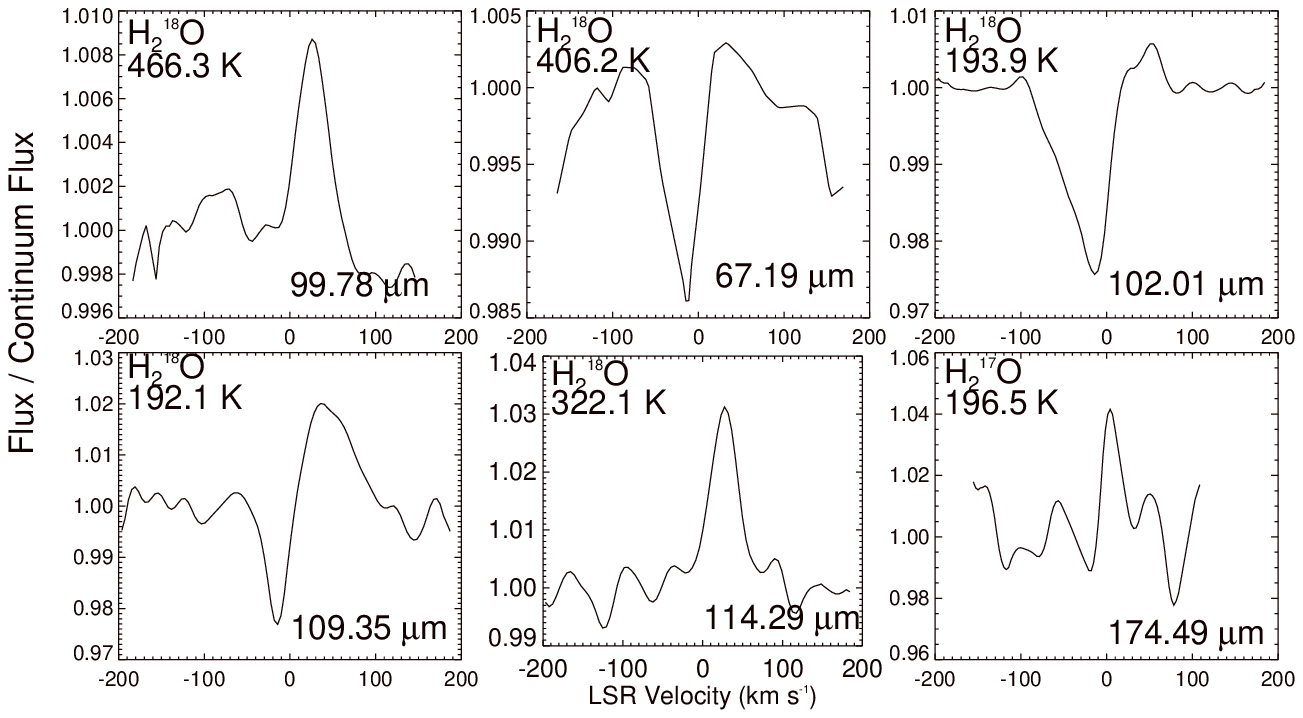}\\

    \caption{A selection of water isotope lines detected by the survey}
    \label{agua_isotopos}
   \end{figure}

\begin{figure}
    \centering
    \includegraphics[width=9cm,height=6cm]{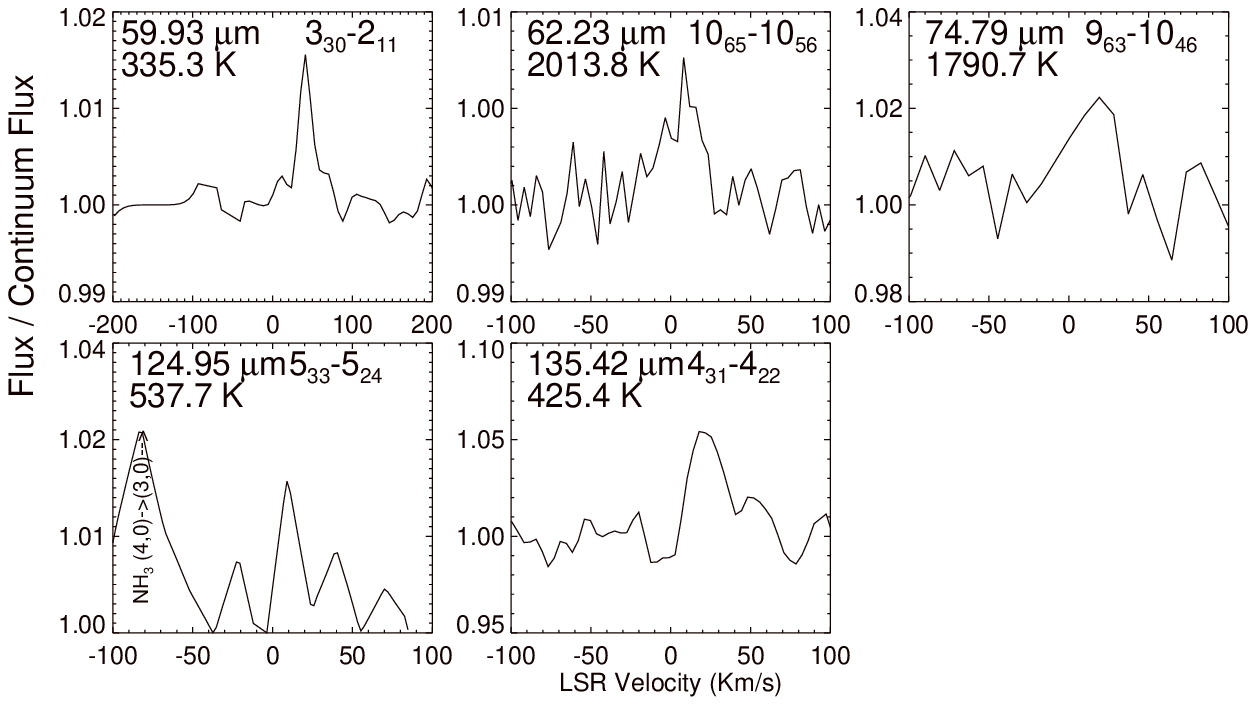}
    \caption{Tentative detections of five transitions of HDO observed towards Orion KL }
    \label{hdo}
\end{figure}

\begin{figure*}
\centering
    \includegraphics{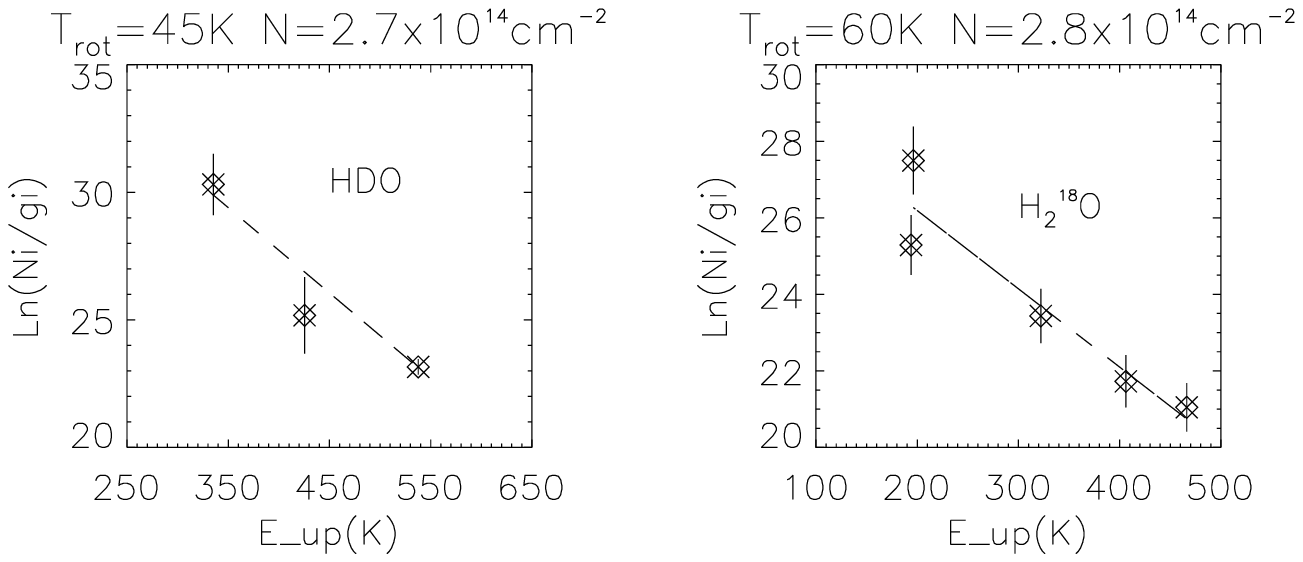}
    \caption{Rotational diagrams i.e. the natural logarithm of the column density in the $i$th stage, $N_{i}$, divided by the
    degeneracy $g_{i}$ versus the upper energy $E_{up}$ for the HDO and H$_{2}^{18}$O transitions detected by the survey
    towards Orion KL. Due to the high uncertainties of the high energy HDO lines, only the lowest energy transitions are considered.
    Note also that P-Cygni lines are excluded from the diagram.}
         \label{aguaiso_rota}
 \end{figure*}

Several transitions of the water isotopic variants H$_{2}^{17}$O,
H$_{2}^{18}$O and HDO are detected in the survey
(Figure~\ref{agua_isotopos} and Figure~\ref{hdo}). The detection
of isotopes is of special interest for abundance determinations
and the interpretation of the spatial origin of the lines. These
lines are excellent water tracers as they are more likely to be
optically thin, so that they can be used to derive the water
abundance via the adoption of a [$^{16}$O]/[$^{18}$O] isotopic ratio.\\
The detections of H$_{2}^{18}$O at 67.19 $\mu$m and 102.01 $\mu$m
are of particular interest (see Figure~\ref{agua_isotopos}). These
H$_{2}^{18}$O lines show a main absorption component that
contrasts with the equivalent H$_{2}^{16}$O transitions at 66.43
$\mu$m and 100.98 $\mu$m, which appear as a P-Cygni and an
emission line, respectively. The fact that both lines peak at
$\approx$ -- 10 km s$^{-1}$ LSR
can be interpreted as indicating an origin in the main outflow.\\
The H$_{2}^{18}$O rotational diagram (Figure~\ref{aguaiso_rota})
leads to a column density of $N_{col}$ $\approx$ 2.8 $\times$
10$^{14}$ cm$^{-2}$ and $T_{rot}$
= 60 K. If we adopt [$^{16}$O]/[$^{18}$O] = 500 (Bergin et al. 1998), the estimated total H$_{2}$O column density is $\approx$ 1.5 $\times$ 10$^{17}$ cm$^{-2}$.\\
However, the interpretation of the exact spatial origin of the
water lines is not straightforward and requires more sophisticated
models. A detailed study of these lines will be published in
future papers (Cernicharo et al. in preparation) and is beyond
the scope of this work.\\

A total of 5 far-infrared HDO lines are tentatively identified for
the first time towards Orion KL, from 3$_{30}\rightarrow 2_{11}$
at 59.93 $\mu$m to 4$_{31}\rightarrow 4_{22}$ at 135.42 $\mu$m
(see Figure~\ref{hdo}). The HDO rotational diagram gives a total
column density of $N_{col}$ $\approx$ 3 $\times$ 10$^{14}$
cm$^{-2}$. This value is significantly lower than that obtained
from millimeter wave observations (Turner et al. 1975, Beckman et
al.
 1982, Moore et al. 1986). However, if the main HDO emission
originated in the Hot Core, a low column density could be
explained if the emission from the core is blocked by the
intermediate-velocity gas. This hypothesis was previously pointed
out by Pardo et al. (2001) in their analysis of the first
detections of two submillimeter HDO lines from the KL region
(transitions 2$_{12}\rightarrow 1_{11}$ at 848.9 GHz and
1$_{11}\rightarrow
0_{00}$ at 893.6 GHz), where the total HDO column density was estimated to be in the range (4 -- 6) $\times$ 10$^{16}$ cm$^{-2}$. \\
They concluded that the HDO transitions detected at above 800 GHz
arise mainly from a very compact region (HPW 15$^{\prime\prime}$)
in the Plateau. Our lower column density result ($\approx$
10$^{14}$ cm$^{-2}$) could therefore be explained if the compact
HDO emission is diluted within the LWS beam. Special care has to
be taken in the interpretation of tentative detections as the
rotational diagram for these HDO lines could underestimate the
total column density. More sophisticated models are needed to aid
the interpretation.

\subsubsection{OH Lines}
Far-infrared line emission from excited OH rotational states was
first detected toward the embedded star-forming region in Orion KL
by Storey et al. ~\cite{storey} using the {\em KAO}. In these
early measurements, the $^{2}\Pi_{3/2}$ J=5/2 $\rightarrow$ 3/2
features at 119.23 $\mu$m and 119.44 $\mu$m were seen in emission.
The fact that the excitation temperature above the ground state
for J$_{up}$=5/2 is about 121 K and that the OH was observed
$\approx$ 30$^{\prime\prime}$ north of KL, led these authors to
assume that the emission they detected came from the shocked gas
region
surrounding BN/KL. \\
Watson et al.~\cite{watson85} reported {\em KAO} observations of
the two lowest lying pure rotational transitions in the
$^{2}\Pi_{3/2}$ electronic state of OH, at 119 $\mu$m and 84
$\mu$m. Melnick et al. ~\cite{melnick90} modelled these
detections, along with their own {\em KAO} detections of the
$^{2}\Pi_{1/2}$  J=3/2 $\rightarrow$ 1/2 line at 163 $\mu$m, the
$^{2}\Pi_{1/2}$ $\rightarrow$ $^{2}\Pi_{3/2}$ J=3/2$^{-}
\rightarrow$ 3/2$^{+}$ line at 53 $\mu$m and the $^{18}$OH
$^{2}\Pi_{3/2}$ J=5/2$^{+} \rightarrow$ 3/2$^{-}$ line at 119.44
$\mu$m. They found that
the best fit to the data required the presence of at least three components (see discussion below).\\

A total of 22 transitions of OH are detected by our survey,
ranging from the $^{2}\Pi_{1/2}-^{2}\Pi_{3/2}$
J=5/2$^{-}$$\rightarrow$5/2$^{+}$ transition at 48.7 $\mu$m to the
$^{2}\Pi_{3/2}$ J=3/2$^{-}$$\rightarrow$1/2$^{+}$ transition at
163.4 $\mu$m. These transitions range up to upper energy levels of
$\approx$ 618K. Their line profiles show a similar behaviour to
those of water, evolving from pure absorption or P-Cygni at
shorter wavelengths to pure emission at longer wavelengths. \\
Table~\ref{compara_oh} lists a comparison of the fluxes and
surface brightnesses of seven OH transitions measured with
different beam sizes by different observers. \\

\begin{figure*}
    \centering
    \includegraphics[width=14cm, height=22cm]{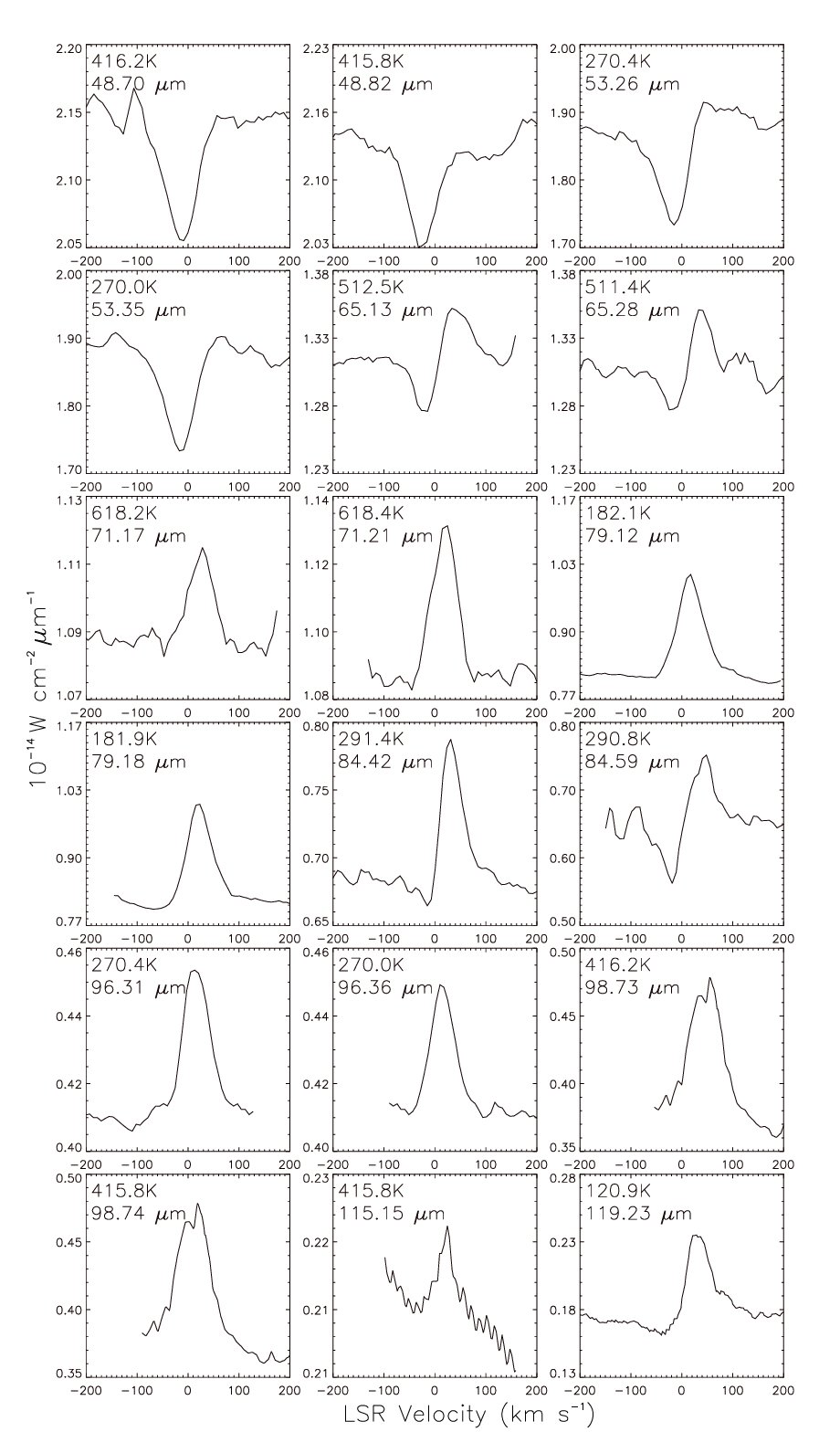}\\

    \caption{OH lines observed in the survey}
    \label{oh_levels}
\end{figure*}
\addtocounter{figure}{-1}
\begin{figure*}
    \centering
   \includegraphics[width=14cm, height=5cm]{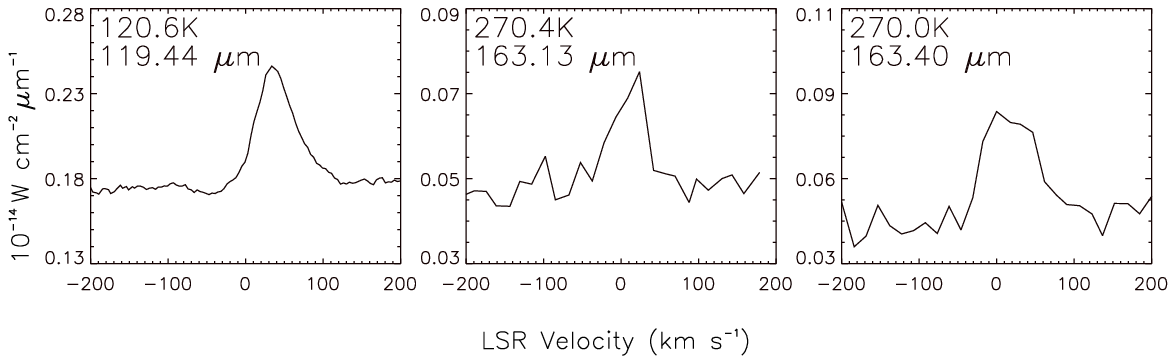}\\

    \caption{Continued}
    \label{oh_levels}
\end{figure*}
\begin{table*}

\centering
\renewcommand{\footnoterule}{}                        
\begin{tabular}{l c c c c c  }        
\hline \hline
Line  &  Wavelength  & FWHM Beam Size  &  Reference$^{(a)}$ & Flux  & Surface Br.   \\
 & ($\mu$m)& (arcsec)   & &(10$^{-17}$W cm$^{-2}$)$^{(b)}$  & (10$^{-3}$ergs cm$^{-2}$ s$^{-1}$sr$^{-1}$)\\

\hline                        

 $^{2}\Pi_{1/2}$ J= 3/2$^{-}$$\rightarrow$1/2$^{+}$ & 163.396 & 67 &  1&1.66 $\pm$ 0.38 & 2.04    \\

  & & 55 & 2& 1.3  & 1.6\\
 & & 60 & 2 & 1.2 $\pm$ 0.6 & 1.2\\

 $^{2}\Pi_{1/2}$ J= 3/2$^{+}$$\rightarrow$1/2$^{-}$ & 163.121 & 67 & 1 & 0.81 $\pm$ 0.19 & 0.99    \\

 & & 55 & 2 & 1.3 & 1.6\\
& & 60 & 2 & 1.2 $\pm$ 0.6 & 1.2\\
$^{2}\Pi_{3/2}$ J= 5/2$^{+}$$\rightarrow$3/2$^{-}$ & 119.441 & 78 & 1 & 2.42 $\pm$ 0.13 & 2.16 \\
 & & 44 & 4 & 0.7 $\pm$ 0.3 & 1.3 $\pm$ 0.5 \\
$^{2}\Pi_{3/2}$ J= 5/2$^{-}$$\rightarrow$3/2$^{+}$ & 119.234 & 78 & 1 &1.76 $\pm$ 0.17   & 1.57\\
 & & 44 & 4 & 0.83  & 1.5 \\
 & & 45 & 5 & 1.88 & 3.3 \\
$^{2}\Pi_{3/2}$ J= 7/2$^{-}$$\rightarrow$5/2$^{+}$ & 84.597 & 77 & 1 &(1.13 $\pm$ 0.38)$^{abs}$ & 1.01$^{abs}$  \\
 & & & & (1.34 $\pm$ 0.26)$^{em}$ &  1.19$^{em}$\\
 & & 30 & 4 & 0.56 & 2.1 \\
 & & 60 & 3 &1.4 $\pm$ 0.4  & 1.4 \\
$^{2}\Pi_{3/2}$ J= 7/2$^{+}$$\rightarrow$5/2$^{-}$ & 84.420 & 77 & 1 & 1.72 $\pm$ 0.13 & 1.54  \\
& & 30 & 4 & $<$0.5  & $<$1.9 \\
& & 60 & 3 &1.0 $\pm$ 0.3  &1.0\\
$^{2}\Pi_{1/2}$ J= 3/2$^{-}$$\rightarrow$$^{2}\Pi_{3/2}$ J=3/2$^{+}$ & 53.351 & 85 & 1 & (2.45 $\pm$ 0.26)$^{abs}$ & 1.86$^{abs}$  \\
 & & & & (0.26 $\pm$ 0.13)$^{em}$ & 0.19$^{em}$\\
 & & 40 & 5 & 3.00$^{abs}$ & 6.6$^{abs}$  \\
 & & & & 0.63$^{em}$ & 1.4$^{em}$\\

\hline \hline

$^{(a)}$ References :\\
1 Values  using {\em ISO} LWS\\
 data in this work\\
2  Melnick et al. 1987 \\
3 Viscuso et al. 1985 \\
4 Watson et al. 1985\\
5 Melnick et al. 1990\\
$^{(b)}$\\
$^{(abs)}$ P Cygni\\
absorption component \\
$^{(em)}$ P Cygni \\
emission component \\


\end{tabular}
 \caption{Comparison of OH line measurements obtained with different beam sizes and instruments. The uncertainty in the absolute fluxes and intensities is $\pm$ 30\% unless
 indicated otherwise}            
\label{compara_oh}

\end{table*}

A similar behaviour to that of the water line velocities is also
found for the OH detections, showing center velocities between --
10 km s$^{-1}$ and +30 km s$^{-1}$ (see Section 4.4) when the line
appear in absorption or emission respectively. Radiative transfer
modelling has been recently performed for the OH lines (Goicoechea
et al. 2006) and concludes that most of the newly detected excited
OH lines in the survey originate in a compact region (D $\approx$
25$^{\prime\prime}$), not resolved by the LWS, that forms part of
the low velocity plateau component. Average values of the physical
conditions in this region are estimated to be: $n(\rm {H}_{2}$)
$\approx$ 5 $\times$ 10$^{5}$ cm$^{-3}$, $T_{k}$ $\approx$ 100 K
and an abundance of $\chi$(OH) $\approx$ (0.5 -- 1) $\times$
10$^{-6}$.

\subsubsection{OH isotopes}
\begin{figure}
    \centering
    \includegraphics[width=9cm]{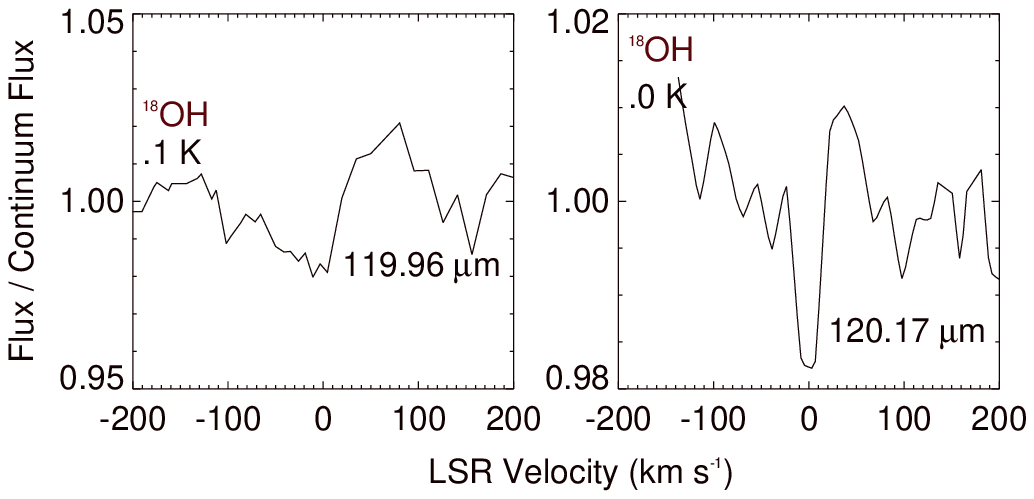}\\

    \caption{$^{18}$OH lines observed in the survey towards Orion KL. The left plot is the  $^{2}\Pi_{3/2}$
5/2$^{+}$$\rightarrow$3/2$^{-}$ transition at 119.96 $\mu$m and
the right plot is the $^{2}\Pi_{3/2}$
5/2$^{-}$$\rightarrow$3/2$^{+}$ transition at 120.17 $\mu$m.}
    \label{oh_isotopos}
   \end{figure}

The detection of $^{18}$OH at the sensitivity of the FP-LWS
observations was found to be very difficult. The weakness of the
lines makes them easily confused with noise and only the $^{18}$OH
$^{2}\Pi_{3/2}$ 5/2$^{+}$$\rightarrow$3/2$^{-}$ transition at
119.96 $\mu$m and the $^{2}\Pi_{3/2}$
5/2$^{-}$$\rightarrow$3/2$^{+}$ transition at 120.17 $\mu$m were
tentatively identified (see
Figure \ref{oh_isotopos}).\\
As with water, isotopic variants play an important role in the
interpretation of the OH lines. Both $^{18}$OH transitions
detected show a difference in their line profiles when compared
with their corresponding $^{16}$OH transitions. As the $^{18}$OH
optical depths are smaller than those measured for the $^{16}$OH
transitions, and since both line profiles are P-Cygni, the
$^{18}$OH lines indicate that OH is predominantly associated with
the Plateau outflow (see also Goicoechea et al. 2006).

\subsubsection{CO emission lines}
High-J CO transitions are well known among the most common tracers
of highly excited regions. OH and
H$_{2}$O can react with C$^{+}$ via the reactions:\\
\begin{equation}
C^{+}+OH \rightarrow CO^{+} + H
\end{equation}
\begin{equation}
C^{+}+H_{2}O \rightarrow HCO^{+} + H
\end{equation}
\begin{equation}
CO^{+} + H_{2} \rightarrow HCO^{+} + H
\end{equation}

\begin{equation}
HCO^{+} + e \rightarrow CO + H
\end{equation}
 These reactions are
sufficiently rapid to ensure that substantial amounts of CO are
produced in shocks (Pineau des For\^ets
et al. 1986, 1987)\\
 Watson et al.
\cite{watson80} made the first far-infrared detection of
interstellar CO with observations of the J=21 -- 20 and J=22 -- 21
transitions at 124 $\mu$m and 119 $\mu$m respectively, from the
BN/KL region. Using collisional excitation cross sections
calculated by Storey et al. \cite{storey}, they derived CO
rotational level populations for a number of temperatures and
densities. They showed that the observed line intensities could be
modelled by emission from two components: a 2000 K component with
$n(\rm {H}_{2}$)$\sim$1 $\times$ 10$^{6}$cm$^{-3}$ and a 400 --
1000 K component with $n(\rm {H}_{2}$)$\sim$5-2 $\times$
10$^{6}$cm$^{-3}$. More recent studies using {\em ISO} data
indicate that the CO emission can be modelled using three
temperature components, describing the plateau
and the ridge emission (Sempere et al. 2000, Maret et al. 2001)\\
\begin{figure*}
    \centering
    \includegraphics[width=14cm,height=22cm]{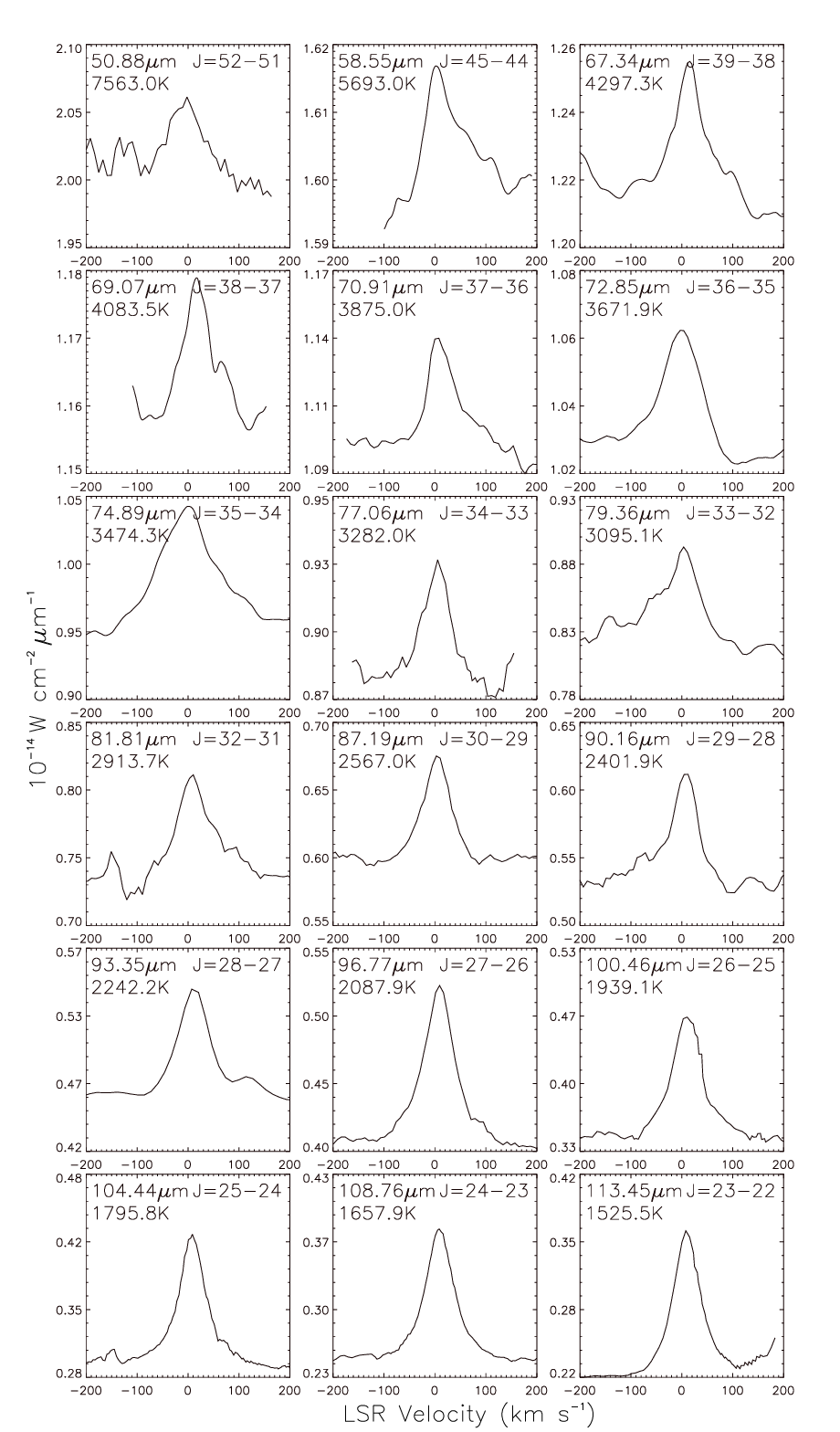}\\
    \caption{CO lines detected by the survey}
         \label{co1}
   \end{figure*}
\addtocounter{figure}{-1}

\begin{figure*}
    \centering
    \includegraphics{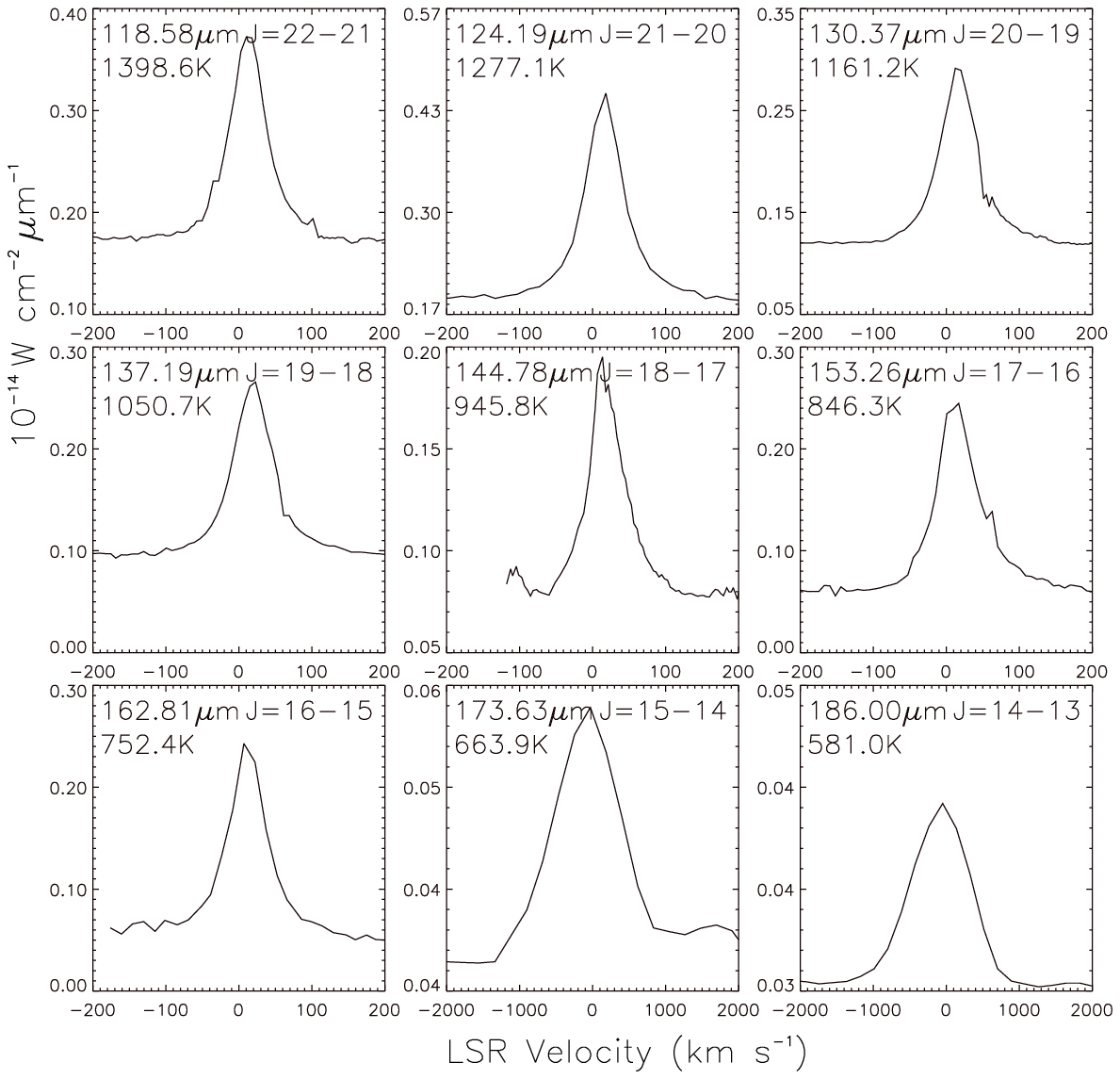}\\
    \caption{continued}
         \label{co2}
   \end{figure*}
 CO detections originating form J$_{up}$=14 to J$_{up}$=52 are
identified by our survey in a total of 26 detected emission lines.
Line profiles are shown in Figure~\ref{co1} and the measured line
fluxes and intensities are compared in Table~\ref{cotable} with
those measured by Storey et al.~\cite{storey} and Watson et
al.\cite{watson85} using the {\em KAO}. Note that the field of
view for {\em KAO}'s observations was 60 arcsec for the J$_{up}$ =
17, 16 lines and $\sim$44 arcsec for the J$_{up}$ = 21, 22, 26,
27, 30 and 31 lines. An interesting result from
Table~\ref{cotable} is that almost identical surface brightnesses
were measured for the J=17--16 and J=16--15 transitions when
observed with the 60$^{\prime\prime}$ {\em KAO} beam and the
80$^{\prime\prime}$ LWS beam. However for the higher J CO lines we
find that the surface brightness in the 44$^{\prime\prime}$ {\em
KAO} beam is about 60\% higher than in the 80$^{\prime\prime}$ LWS
beam, while the integrated fluxes in the two beam sizes are
similar, indicating that the source size may be smaller than 44
arcsec for
these transitions.\\
 CO is also an important tracer of H$_2$,
since it is the second most abundant molecule in the interstellar
medium after molecular hydrogen. Despite theoretical and
observational uncertainties in the use of a canonical
$N$(CO)/$N$(H$_{2}$) ratio (Williams et al. 1984; van Dishoeck et
al. 1992; Sakamoto et al. 1996), theoretical studies of the
CO/H$_{2}$ abundance ratio have concluded that on large scales it can be considered to be constant (Taylor et al. 1993).\\

\begin{figure*}
    \centering
    \includegraphics[width=15cm]{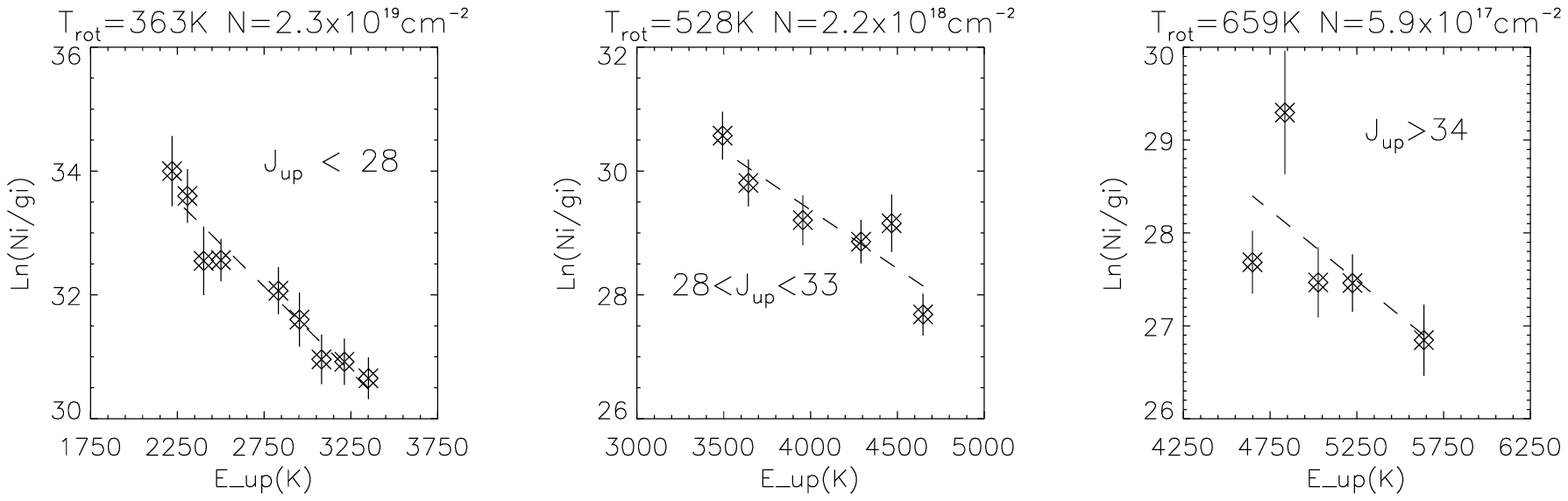}
    \caption{Rotational diagrams i.e. the natural logarithm of the column density in the $i$th stage, $N_{i}$, divided by the
    degeneracy $g_{i}$ versus the upper energy $E_{up}$ for the observed CO emission towards Orion KL using {\em ISO} LWS-FP data.}
         \label{co_rotas}
   \end{figure*}

In Figure~\ref{co_rotas} we show rotational diagrams for three
different CO J-ranges, finding three different rotational
temperatures. Below, we compare our results with the model of
Sempere et al. \cite{sempere} which was developed using a
radiative transfer model fit to {\em ISO} data obtained at low
spectral resolution (grating mode), together with selected
observations at high spectral resolution. In this model:
\begin{itemize}
\item The CO emission from J=18 to 33 can be explained by a two
temperature component model of the plateau region; the inner
region reproduces the emission from J=33 to 28 and the colder gas
contributes to the lower J lines. The high J transitions (J$>$ 34)
reveal the presence of a very hot gas component ( $T$ $\approx$
1500--2000K).
 \item They assumed
densities of 10$^{7}$ cm$^{-3}$ and a temperature of 400 K for the
inner region and 10$^{6}$ cm$^{-3}$, 300K for the external part of
the plateau. \item The resulting column densities in Sempere et
al's model were: $N$(CO)=10$^{19}$ cm$^{-2}$ and $N$(CO)=3.5
$\times$ 10$^{18}$ cm$^{-2}$ for the inner and outer plateau, and
$N$(CO)=10$^{17}$ cm$^{-2}$ for the hot gas component.\\
The results from our rotational diagrams shows consistency with
the different temperature components predicted by Sempere et al's
model. Our results confirm the presence of hot gas ($T$ $\approx$
660 K; Figure~\ref{co_rotas}) which differs by $\approx$ 300 K
from the warm component ($T$ $\approx$ 360 K, see
Figure~\ref{co_rotas}). However, we found that this warm component
has a higher CO column density than that found by Sempere et al.,
with a column density of the order of $\approx$ 2.3 $\times$
10$^{19}$ cm$^{-2}$. According to our results, the inner part of
the Plateau component is traced by the warm gas which emits the CO
transitions with J $<$ 28 with a rotational temperature of
$\approx$ 360 K.
\end{itemize}

\subsubsection{CO isotopic variants}
CO isotopic variants are hardly detected in the survey. Only one
line of $^{13}$CO (the J = 21--20 transition at 129.89 $\mu$m) is
tentatively detected with a flux of 2 $\pm$ 1$\times$10$^{-19}$ W
cm$^{-2}$.

\begin{table*}
\begin{center}
    \leavevmode
    \footnotesize
    \begin{tabular}[h]{lrcccc}
      \hline \\[-5pt]
      Transition & $\lambda$ ($\mu$m)      &  LWS Flux$^{a}$  & LWS Intensity$^{a}
      $ &{\em KAO} Flux$^{b}$  & {\em KAO} Intensity$^{b}$        \\[-5pt]
       & & & & &  \\

        &  & [10$^{-17}$W cm$^{-2}$] & [10$^{-3}$ergs$^{-1}$cm$^{-2}$sr$^{-1}$] &[10$^{-17}$W cm$^{-2}$] &
    [10$^{-3}$ergs$^{-1}$cm$^{-2}$sr$^{-1}$] \\
    \hline \\[-5pt]

      J=34--33  & 77.05  & 0.60 & 0.52  & 0.43 & 0.86 \\
      J=30--29  & 87.19  & 1.58 & 1.37  & 1.6 & 2.41\\
      J=27--26  & 96.77  & 3.82 & 3.47  & 4.3 & 6.47 \\
      J=26--25  & 100.46 & 3.44 & 3.13  & 1.9 & 3.80 \\
      J=22--21  & 118.58 & 6.58 & 5.87  & 5.3 & 11.0\\
      J=21--20  & 124.19 &  7.89    &  8.87     & 6.8 & 10.2\\
      J=17--16  & 153.27 & 8.66 & 10.35  &7.0$^{c}$ & 10.5\\
      J=16--15  & 162.81 & 9.48 & 9.88 &6.4$^{c}$ & 9.63\\

      \hline \\

      \end{tabular}

  \end{center}
 \begin{tabular}{l }

 $(a)$ Line fluxes  and intensities averaged over a 80$^{\prime\prime}$ beam observed by the {\em ISO} LWS in Fabry-P\'erot
       mode.\\
   $(b)$ Line fluxes  and intensities averaged over a 44$^{\prime\prime}$ beam measured by Watson et al. (1981)
    using the UC Berkeley tandem Fabry-P\'erot\\
   spectrometer.\\
   $(c)$ Storey et al. (1981) using a 60$^{\prime\prime}$ beam.\\

   \end{tabular}
    \caption{LWS Fabry-P\'erot CO detections towards Orion KL, compared with those of Watson et al. (1981)}
    \label{cotable}
\end{table*}


\subsection{Line kinematics}
\begin{figure*}
\centering
    \includegraphics[width=10cm,height=20cm]{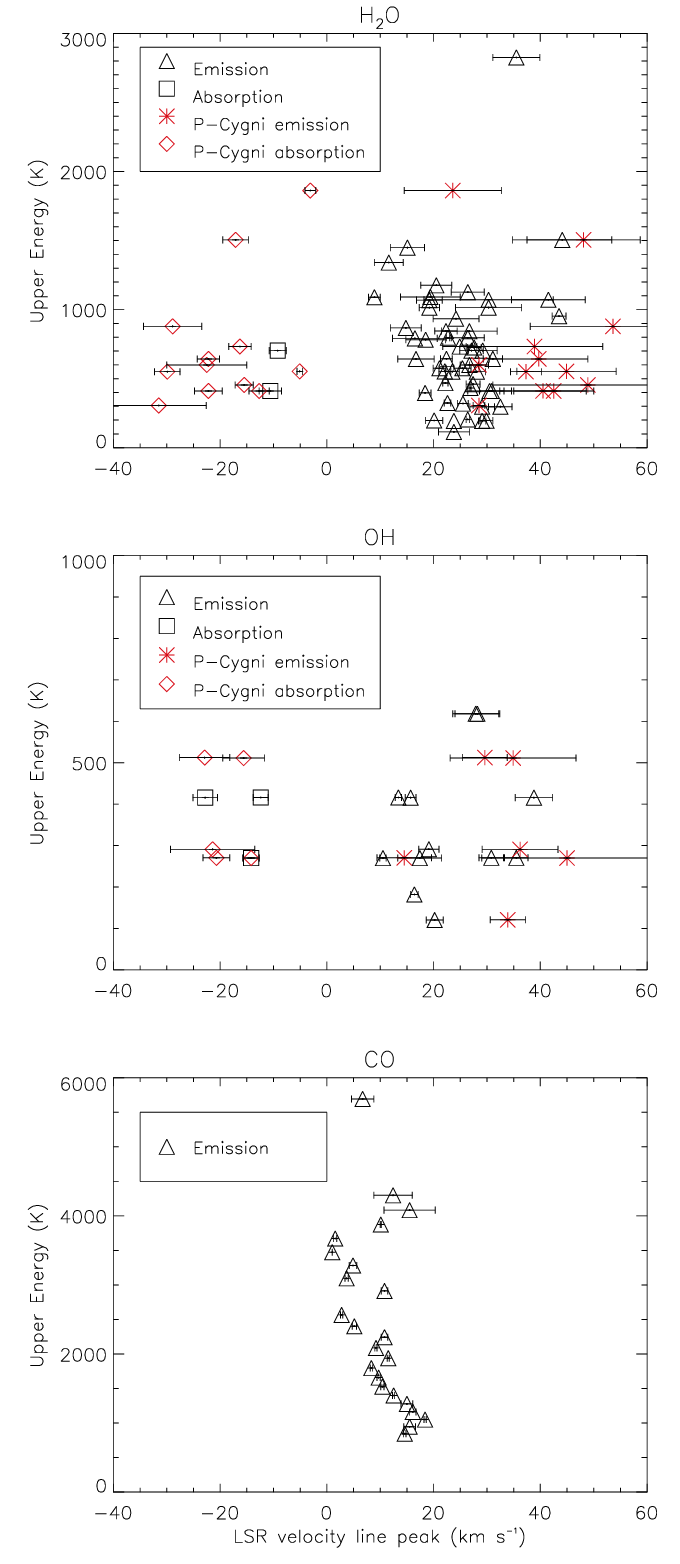}
    \caption{LSR peak velocities of H$_{2}$O, OH and CO lines observed
    towards Orion KL with the {\em ISO} LWS-FPs.
    Note that the LSR velocity of the quiescent gas is $\approx$ 9 km s$^{-1}$ LSR (Cohen et al. 2006). Errors
    are from goodness of fit estimations.}
         \label{velo}
 \end{figure*}

Although kinematical properties are diluted in the large LWS beam,
resolved line profiles and velocity peaks can trace the overall
dynamical gas properties. Figure~\ref{velo} shows the upper
transition energy of the main molecular detections of H$_{2}$O, OH
and CO, as a function of the emission and absorption LSR line
velocity peaks. For the H$_{2}$O and OH lines, both pure emission
or absorption and P-Cygni peaks are plotted. The velocity of the
quiescent gas is 9.0 $\pm$ 0.5 km s$^{-1}$(Cohen et al. 2006).
Subtracting this value from the velocity peaks gives an indication
of the exact blue- or red-shift of the lines.
 Several conclusions can be deduced
from Figure~\ref{velo}:
\begin{itemize}
\item H$_{2}$O and OH radial velocities trace the velocity
distribution of the same expanding gas. Absorption lines are
centred between --15 and --30 km s$^{-1}$ and emission lines
between +15 and +45 km s$^{-1}$. This confirms previous analyses
of these lines (Harwit et al. 1998; Cernicharo et al. 1999;
Goicoechea et al. 2006) which interpreted them as evidence for
outflows driven by the star formation activity of Orion KL.
 \item
 For H$_{2}$O and OH, the absorption and emission components of
 the P-Cygni lines tend to be at more extreme velocities than
 those of the `pure' absorption or emission lines. Detailed
 radiative transfer modelling is underway to attempt to match this
 behaviour.

\item The CO radial velocities indicate a different excitation
mechanism to that of H$_{2}$O and OH, with emission peaks centred
between (5 -- 15) km s$^{-1}$. Considering an average velocity
uncertainty of $\pm$10 km s $^{-1}$, this range is consistent with
the velocity of the quiescent gas ($\approx$ 9 km s$^{-1}$). As
very high energy CO transitions are detected, the lines may
originate in the hot and warm quiescent post-shocked gas.

\end{itemize}

\subsection{H$_{3}$O$^{+}$ line detections}

H$_{3}$O$^{+}$ is  one of the key species in the interstellar
chemistry of oxygen. This saturated molecular ion leads to the
 formation of OH and H$_{2}$O by dissociative recombination, with an imprecisely known
 branching ratio (Bates 1986; Sternberg 1995). It can be formed from H$_{3}^{+}$ by the reaction:
 \begin{equation}
H_{3}^{+} + H_{2}O \rightarrow H_{3}O^{+} + H_{2}
\end{equation}
or via the reactions:
\begin{equation}
H_{3}^{+} + O \rightarrow OH^{+} + H_{2}
\end{equation}
\begin{equation}
OH^{+} + H_{2} \rightarrow H_{2}O^{+} + H
\end{equation}
\begin{equation}
H_{2}O^{+} + H_{2} \rightarrow H_{3}O^{+} + H
\end{equation}

Its formation is more likely to occur via reactions 9--11 (Wootten
et al. 1986). Destruction of H$_{3}$O$^{+}$ occurs via electron
recombination:
\begin{equation}
H_{3}O^{+} + e^{-} \rightarrow H_{2}O + H
\end{equation}
\begin{equation}
H_{3}O^{+} + e^{-} \rightarrow OH + H_{2}
\end{equation}

leading to the production of H$_{2}$O and OH. \\
H$_{3}$O$^{+}$ has a pyramidal structure; inversion transitions
are produced when the oxygen atom tunnels through the plane of the
hydrogen atoms. The ground state inversion splitting is $\approx$
55 cm$^{-1}$ (Liu et al. 1985). This large value makes the
fundamental transitions of the $\nu_{2}$ mode lie at submillimeter
and far-infrared wavelengths (Wootten et al. 1986, Bogey et al.
1985). H$_{3}$O$^{+}$ was first detected at 365 GHz in OMC-1 and
Sgr B2 by Wootten et al. (1991) and
 more recently in the far-infrared towards Sgr B2 (Goicoechea and Cernicharo, 2001; Polehampton et al. 2006).
 Wootten et al. (1991) modeled the excitation of the
 para-H$_{3}$O$^{+}$ transition at 365 GHz, finding abundances
 $\chi$(H$_{3}$O$^{+}$) $\approx$ 1 $\times$ 10$^{-9}$
 $\rightarrow$ 5 $\times$ 10$^{-9}$.\\

\begin{figure}
    \centering
    \includegraphics[width=9cm]{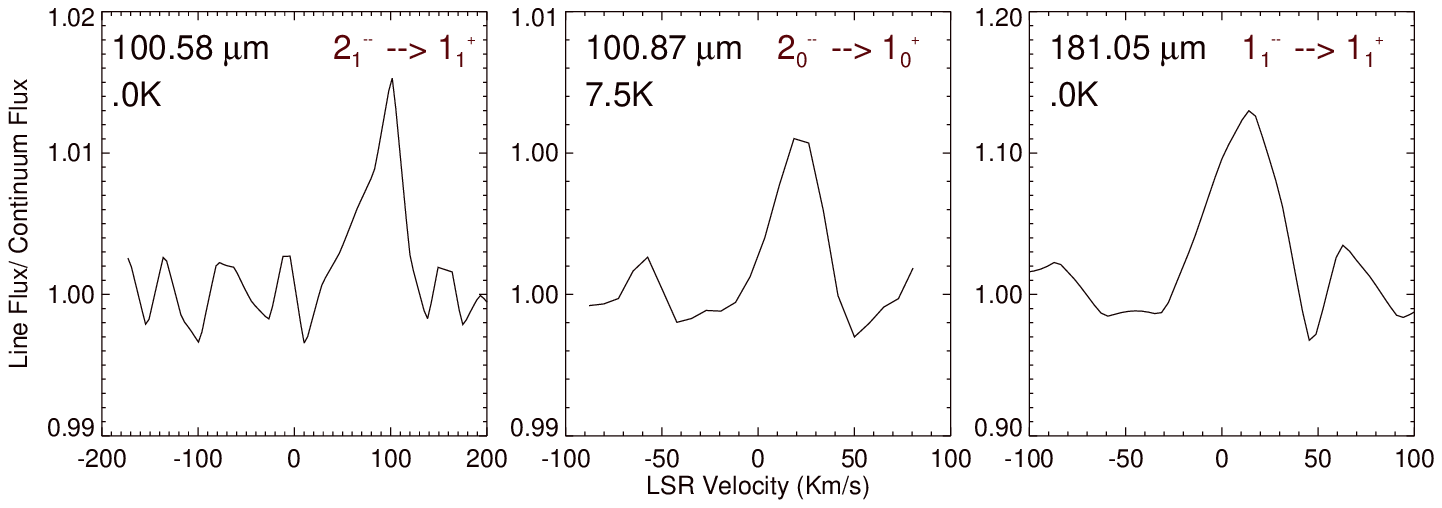}
    \caption{H$_{3}$O$^{+}$ lines observed towards Orion KL }
    \label{h3o+}
\end{figure}

The lowest rotational levels of the ortho and para ladders of the
$\nu_{2}$ ground-state H$_{3}$O$^{+}$ inversion mode
(0$^{+}\rightarrow$0$^{-}$) are detected in the survey. These are
the rotational inversions 2$_{1}^{-}$$\rightarrow$1$_{1}^{+}$ at
100.58 $\mu$m and 2$_{0}^{-}$$\rightarrow$1$_{0}^{+}$ at 100.87
$\mu$m and the pure inversion 1$_{1}^{-}$$\rightarrow$1$_{1}^{+}$
transition at 181.05 $\mu$m (see Figure~\ref{h3o+}). The optical
depths of these transitions estimated with the RADEX code are
$\tau$ $>$ 10. We used as input the column density values inferred
by Wootten et al. (1991; $\approx$ 10$^{14}$ cm$^{-2}$),
$T_{K}$=80 K and $n_{\rm {H_{2}}}$=10$^{5}$ cm$^{-3}$. The
rotational diagram method cannot therefore be applied to derive
physical parameters. Further modelling of optically thick lines
including radiative excitation by the FIR dust continuum emission
is needed and will be published in forthcoming papers.

\subsection{NH$_{3}$ line detections}

\begin{figure}
    \centering
    \includegraphics[width=9cm,height=8cm]{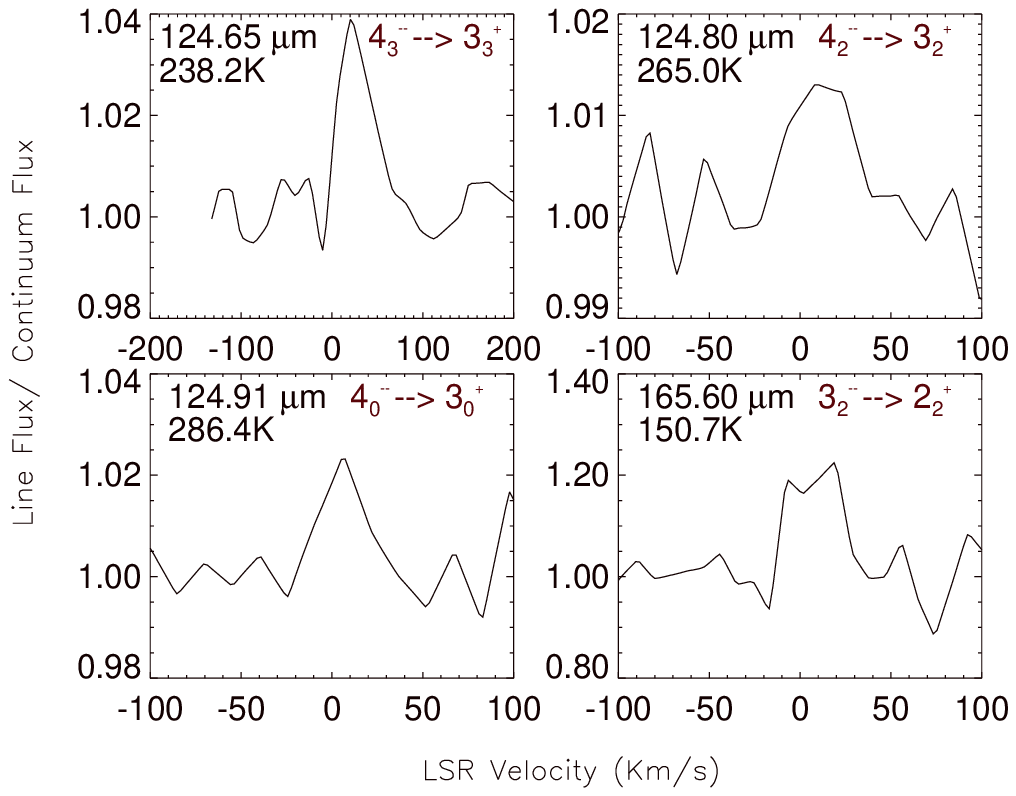}

    \caption{NH$_{3}$ lines observed towards Orion KL}
    \label{amonio}
\end{figure}

\begin{figure}
 \centering
    \includegraphics[width=7cm,height=6cm]{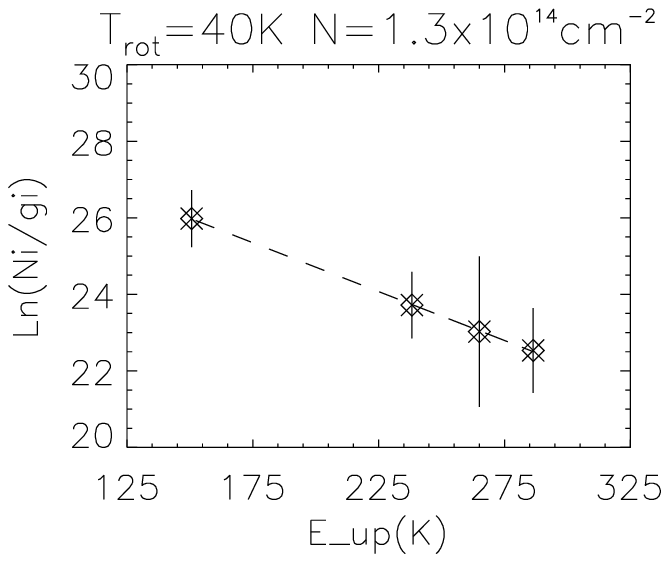}
    \caption{Rotational diagram for the NH$_{3}$ lines detected towards Orion KL.}
         \label{amonia_rota}
 \end{figure}

Like H$_{3}$O$^{+}$, the NH$_{3}$ molecule has a pyramidal
structure. Inversion transitions are produced when the nitrogen
atom tunnels through the plane of the hydrogen atoms (analogous to
the tunnelling of the O atom in H$_{3}$O$^{+}$), with the
significant difference that the inversion splitting $\approx$ 1
cm$^{-1}$, much smaller than that of H$_{3}$O$^{+}$ (Ho \& Townes,
1983).
\\NH$_{3}$ was first detected
at far-IR wavelengths by the {\em KAO} towards Orion-KL (Townes et
al. 1983), in the rotational $\nu_{2}$=1; 4$_{3}^{-}$ --
3$_{3}^{+}$ line at 124.6 $\mu$m. The fact that this line is very
optically thick ($\tau \approx$ 10$^{3}$) and that it is seen in
emission, led the authors to consider that this level is
collisionally excited, probably originating in the Hot Core
component.\\
The 4$_{3}^{-}$ -- 3$_{3}^{+}$ transition at 124.6 $\mu$m is also
detected by the LWS, as well as the following transitions:
4$_{2}^{-}$ -- 3$_{2}^{+}$ 124.8 $\mu$m, 4$_{0}^{-}$ --
3$_{0}^{+}$ 124.9 $\mu$m and 3$_{2}^{-}$ -- 2$_{2}^{+}$ 165.6
$\mu$m (see
Figure~\ref{amonio}). \\
We estimated the optical depths of these transitions with the
RADEX code, using as input the inferred column density of
10$^{14}$ cm$^{-2}$, $T_{K}$=80K and $n_{\rm {H_{2}}}$=10$^{5}$
cm$^{-3}$. For these conditions, only the 3$_{2}^{-}$ --
2$_{2}^{+}$ transition at 165.6 $\mu$m is moderately optically
thick ($\tau$ = 1.4). Figure~\ref{amonia_rota} shows the
rotational diagram of NH$_{3}$, with an inferred column density of
1.3 $\times$ 10$^{14}$ cm$^{-2}$ and a rotational temperature of
40 K. However, if the NH$_{3}$ column density is $ \geq$ 10$^{15}$
cm$^{-2}$ the code predicts high optical depths and our rotational
calculation could be underestimated. Estimations of the NH$_{3}$
abundance using more sophisticated models will be published in
forthcoming papers.

\begin{table*}
\centering
\begin{tabular}{c c c c c c }
\hline\hline
Molecule & Method & $N_{col}$ & $T_{rot}$ & Lines & Note  \\
& & cm$^{-2}$ & K\\
 \hline
CO & 1 & (1-3) $\times$ 10$^{19}$& 360 &15 & Inner Plateau   \\
CO& 1 & (1-4) $\times$ 10$^{18}$& 530 &  5& Plateau\\
CO& 1 & (2-7) $\times$ 10$^{17}$& 660 & 5& Hot Gas\\
$^{13}$CO & & & &1 \\
OH & 2& (2.5-5.1)$^{(a)}$ $\times$ 10$^{16}$ & & 22 & Low velocity Plateau \\
H$_{2}$O & 3    & (1-3) $\times$ 10$^{17}$ & 60 &70 & \\
HDO & 1 & (1-5) $\times$ 10$^{14}$  & 45& 5 & \\
H$_{2}^{18}$O & 1 & (2-5) $\times$ 10$^{14}$ & 60 &  5& \\
H$_{2}^{17}$O &  &  &&  1& \\
NH$_{3}$ &1 &(1-5) $\times$ 10$^{14}$  &40 &4 \\
H$_3$O$^{+}$ & & & &3 & \\
 \hline
 \hline
\end{tabular}
\caption{ Rotational temperatures and beam averaged column
densities N$_{col}$, estimated using LTE rotational diagram method
or using non-Local radiative transfer models, for the different
molecular species identified in the far-infrared survey towards
Orion KL with {\em ISO} LWS-FP data.Column 2 list the method used
to estimate the column density; [1] rotational diagram method,[2]
non-Local radiative transfer models, [3] via the isotopic
ratio.$^{(a)}$ From Goicoechea et al. 2006}
 \label{moleculas}
\end{table*}

\subsection{Unidentified and weak features}

Some isotopic lines that were detected by the LWS in L04 mode were
very difficult to detect in the L03 scans, due to differences
between repeated scans.
 This affected the detection of weak transitions. However, some unidentified lines were detected after
carrying out an exhaustive analysis, in which the lines were seen
in at least two different observations plotted together in order
to discriminate between glitches and spurious features. The
observations were also analyzed before performing smoothing
corrections, to ensure that no features were missed in this step.
Table~\ref{unidentifies} gives a list of the unidentified features
found in the survey. The line fluxes were derived using similar
fits to those used for identified lines.\\
Amongst the candidate detections are several rotational
transitions of the H$_{2}$O vibrational bending mode $\nu$2 (44.09
and 84.36 $\mu$m). There are also many rotational transitions in
the far-IR of slightly asymmetrical species that could be
contributing to the spectrum but may not have been detected at the
FP resolution, such as HNO, HNCO or HOCO$^{+}$ and the low-energy
bending modes of carbon chains.

\section{Conclusions}
\begin{table*}
\centering                          
\begin{tabular}{c c c  c }        
\hline
Wavelength observed & Flux &absorption & emission  \\
($\mu$m) & (10$^{-18}$W cm$^{-2}$) &  &\\
 \hline
 44.09 $\pm$ 7.5 $\times$ 10$^{-3}$ & 1.54 $\pm$ 0.98  &  & X   \\
 46.29 $\pm$ 5.5 $\times$ 10$^{-4}$ & 1.46 $\pm$ 0.31  & & X  \\
48.48 $\pm$ 5.5 $\times$ 10$^{-4}$ & 6.36 $\pm$ 1.82 &  & X  \\
48.75 $\pm$ 1.1 $\times$ 10$^{-3}$ & 2.58 $\pm$ 0.83  &  & X \\
52.29 $\pm$ 5.7 $\times$ 10$^{-4}$ & 4.14$\pm$ 1.61  & & X \\
52.57 $\pm$ 9.5 $\times$ 10$^{-4}$ & 6.07 $\pm$ 2.01  & & X \\
58.02 $\pm$ 8.5 $\times$ 10$^{-3}$ & 1.44 $\pm$ 0.38  & X \\
58.24 $\pm$ 5.5 $\times$ 10$^{-3}$ & 1.67 $\pm$ 0.53 & &X \\
63.78 $\pm$ 3.6 $\times$ 10$^{-4}$ & 5.06 $\pm$ 1.06  & & X \\
64.34 $\pm$ 9.5 $\times$ 10$^{-3}$ & 0.87 $\pm$ 0.27  & & X  \\
66.63 $\pm$ 8.5 $\times$ 10$^{-4}$ & 2.03 $\pm$ 0.74  & & X \\
67.14 $\pm$ 4.5 $\times$ 10$^{-3}$ & 1.55 $\pm$ 0.75  & & X \\
68.61 $\pm$ 8.8 $\times$ 10$^{-3}$ & 1.98 $\pm$ 0.47 &   &X \\
84.36 $\pm$ 5.5 $\times$ 10$^{-3}$ & 1.57 $\pm$ 0.93  & & X \\
91.13 $\pm$ 7.5 $\times$ 10$^{-3}$ &  0.36 $\pm$ 0.15 &   & X \\
91.97 $\pm$ 3.2 $\times$ 10$^{-4}$ & 4.08 $\pm$ 0.10 &   & X \\
101.78 $\pm$ 4.1 $\times$ 10$^{-4}$ & 0.31 $\pm$ 0.10  & & X  \\
104.39 $\pm$ 5.6 $\times$ 10$^{-3}$ & 0.68 $\pm$ 0.37  & & X  \\
108.30 $\pm$ 2.2 $\times$ 10$^{-4}$ & 0.32 $\pm$ 0.05  & & X  \\
112.82 $\pm$ 5.4 $\times$ 10$^{-3}$ & 0.29 $\pm$ 0.14  & & X \\
114.96 $\pm$ 7.5 $\times$ 10$^{-3}$ & 0.27 $\pm$ 0.093  & & X \\
115.68 $\pm$ 2.5 $\times$ 10$^{-3}$ & 0.41 $\pm$ 0.22  & & X  \\
116.27 $\pm$ 6.9 $\times$ 10$^{-4}$ & 0.89 $\pm$ 0.24 & & X  \\
116.60  $\pm$ 9.5 $\times$ 10$^{-3}$ & 0.49 $\pm$ 0.35 & & X \\
122.37 $\pm$ 4.4 $\times$ 10$^{-4}$ & 1.21 $\pm$ 0.027 & X  \\
123.80  $\pm$ 2.7 $\times$ 10$^{-4}$&  7.40 $\pm$ 0.088 & & X\\
126.64 $\pm$ 8.5 $\times$ 10$^{-3}$ & 0.19 $\pm$ 0.056 & &  X \\
131.44 $\pm$ 9.5 $\times$ 10$^{-4}$ & 1.23 $\pm$ 0.45  & &  X \\
133.34 $\pm$ 3.2 $\times$ 10$^{-3}$ & 0.76 $\pm$ 0.36 & &  X \\
136.61 $\pm$ 2.3 $\times$ 10$^{-3}$ & 0.69 $\pm$ 0.20 & & X  \\
137.50  $\pm$ 8.4 $\times$ 10$^{-3}$ & 0.99 $\pm$ 0.48  & & X \\
139.54 $\pm$ 4.8 $\times$ 10$^{-3}$ & 1.66 $\pm$ 0.065  & & X \\
140.60 $\pm$ 8.6 $\times$ 10$^{-3}$ & 0.33 $\pm$ 0.15  & & X \\
144.47 $\pm$ 5.9 $\times$ 10$^{-4}$ & 4.11 $\pm$ 0.86  & & X  \\

 \hline \hline
\end{tabular}
\caption{List of unidentified and weak features found on the
survey towards Orion KL with {\em ISO} high resolution spectra
Fabry-P\'erot mode. The unidentified line selection criteria was
based on the features that were seen in at least two different
observations plotted together in order to discriminate between
glitches and spurious features. The observations were also
analysed before performing smoothing corrections, to assure that
no features were missed in this step.}
 \label{unidentifies}
\end{table*}
A high spectral resolution 44--188 $\mu$m line survey towards
Orion KL has been carried out with the {\em ISO} LWS in
Fabry-P\'erot mode. A total of 152 lines are identified and a
further 34 lines remain to be identified. A basic analysis of the
molecular detections was carried out by deriving rotational
temperatures and column densities (see summary in
Table~\ref{moleculas}) and by comparison with previous
measurements and published models.
 We found that:
 \begin{itemize}
 \item
The spectrum is dominated by the main molecular coolants: H$_2$O,
CO and OH together with the forbidden lines [O~{\sc i}], [O~{\sc
iii}], [N~{\sc iii}] and [C~{\sc ii}].
 \item The analysis of the
[O~{\sc i}] and [C~{\sc ii}] fine structure lines indicates that a
PDR model can reproduce the observed [O~{\sc i}] 63.2 $\mu$m and
[C~{\sc ii}] 157.7 $\mu$m surface brightness levels although it
over-predicts the [O~{\sc i}] 145.5 $\mu$m emission by a factor of
2.7.
 \item The water and OH
P-Cygni profiles, along with the kinematical line properties,
confirm that most of the detected emission is associated with gas
expanding in the outflow from the KL cluster.
 \item The observed molecular emission is consistent with
previous models where several main components were clearly
distinguished; the CO detections confirm the differentiation into
physically distinct components, with column densities ranging from
$\approx$ 10$^{19}$ to $\approx$ 10$^{17}$ cm$^{-2}$ and
temperatures from 350 to 650 K. \item HDO and H$_3$O$^+$ are
tentatively detected for the first time in the far-infrared range
towards Orion KL. The derived HDO column density ($\approx$
10$^{14}$ cm$^{-2}$) is lower than the values (10$^{16}$
cm$^{-2}$) obtained at millimeter and submillimeter wavelengths.
Due to the high optical depths of the lines, more sophisticated
non-local radiative transfer models are needed to estimate the
H$_3$O$^+$ and NH$_3$ column densities.
\end{itemize} The 80 arcsec {\em ISO} LWS beam
size encompasses a large range of physical conditions, ranging
from quiescent cool gas to outflows that alter the chemistry of
the region. Consequently, the exact interpretation of the line
survey requires modelling. We plan to model the chemistry and
dynamics of the main components using a coupled chemical-radiative
transfer code.

\section*{Acknowledgements}
      The authors would like to thank Glenn White and the anonymous referee for
      their suggestions and helpful comments.
      The {\em ISO} Spectral Analysis Package (ISAP) is a joint
      development by the LWS and SWS Instrument Teams and Data
      Centres. Contributing institutes are CESR, IAS, IPAC, MPE,
      RAL and SRON. LIA ia a joint development of the {\em ISO}-LWS
      Instrument Team at Rutherford Appleton Laboratories (RAL,
      UK- the PI institute) and the Infrared Processing and
      Analysis Center (IPAC/Caltech, USA).\\
      JRG was supported by a Marie Curie Intra-European Fellowship
      under contract MEIF-CT-2005-515340 within the 6th European Community Framework programme.

\end{document}